\documentclass[preprint,nonatbib]{elsarticle}

\usepackage{lineno,hyperref}
\usepackage{latexsym}
\usepackage{graphicx}
\usepackage{multicol,multirow}
\usepackage{amsmath,amssymb,amsfonts}
\usepackage{mathrsfs}
\usepackage{amsthm}
\usepackage{subcaption}
\usepackage{upgreek}
\usepackage{color}
\usepackage[dvipsnames]{xcolor}

\newtheorem*{remark}{Remark}
\DeclareMathOperator*{\argmin}{arg\,min}

\usepackage{algorithm}
\usepackage{algpseudocode}

\newcommand{\bmat}[1]{\begin{bmatrix}#1\end{bmatrix}}

\definecolor{Black}{rgb}{0,0,0}
\definecolor{Blue}{rgb}{0,0,1}
\definecolor{Red}{rgb}{1,0,0}
\definecolor{Green}{rgb}{0,1,0}
\definecolor{Cyan}{rgb}{0,0.72,0.92}
\definecolor{Amethyst}{rgb}{0.6,0.4,0.8}
\definecolor{Bronze}{rgb}{0.8,0.5,0.2}
\definecolor{Violet}{rgb}{0.54,0.17,0.89}

\newcommand{\RA}[1]{{\textcolor{Red}{#1}}}  % edits related to the 1st reviewer's comments
\newcommand{\RB}[1]{{\textcolor{Blue}{#1}}} % edits related to the 2nd reviewer's comments
\newcommand{\RC}[1]{{\textcolor{Violet}{#1}}} % all other edits

\renewcommand{\textcolor}[2]{#2} % Change the textcolor to black globally
\journal{Journal of Computational Physics}

\makeatletter
\let\c@author\relax
\makeatother
\usepackage[style=ieee, citestyle=numeric-comp, backend=biber]{biblatex}
\begin{document}

\begin{frontmatter}

\title{gLaSDI: Parametric Physics-informed Greedy Latent Space Dynamics Identification}
\date{}
% \tnotetext[mytitlenote]{Fully documented templates are available in the elsarticle package on \href{http://www.ctan.org/tex-archive/macros/latex/contrib/elsarticle}{CTAN}.}

\author[1]{Xiaolong He\corref{corresponding_author}}
\cortext[corresponding_author]{Corresponding author}
\ead{xih251@eng.ucsd.edu}
\author[2]{Youngsoo Choi}
\author[3]{William D. Fries}
\author[4]{Jonathan L. Belof}
\author[1]{Jiun-Shyan Chen}

\address[1]{Department of Structural Engineering, University of California, San Diego, La Jolla, CA, 92093, USA}
\address[2]{Center for Applied Scientific Computing, Lawrence Livermore National Laboratory, Livermore, CA, 94550, USA}
\address[3]{Applied Mathematics, School of Mathematical Sciences, University of Arizona, Tucson, AZ, 85721, USA}
\address[4]{Materials Science Division, Physical and Life Science Directorate, Lawrence Livermore National Laboratory, Livermore, CA, 94550, USA}

% \address{Radarweg 29, Amsterdam}
% \fntext[myfootnote]{Since 1880.}

% %% or include affiliations in footnotes:
% \author[mymainaddress,mysecondaryaddress]{Elsevier Inc}

% \author[mysecondaryaddress]{Global Customer Service\corref{mycorrespondingauthor}}
% \cortext[mycorrespondingauthor]{Corresponding author}

% \address[mymainaddress]{1600 John F Kennedy Boulevard, Philadelphia}
% \address[mysecondaryaddress]{360 Park Avenue South, New York}

\begin{abstract}
A parametric adaptive physics-informed greedy Latent Space Dynamics Identification (gLaSDI) method is proposed for accurate, efficient, and robust data-driven reduced-order modeling of high-dimensional nonlinear dynamical systems. 
In the proposed gLaSDI framework, an autoencoder discovers intrinsic nonlinear latent representations of high-dimensional data, while dynamics identification (DI) models capture local latent-space dynamics.  
An interactive training algorithm is adopted for the autoencoder and local DI models, which enables identification of simple latent-space dynamics and enhances accuracy and efficiency of data-driven reduced-order modeling. 
To maximize and accelerate the exploration of the parameter space for the optimal model performance, an adaptive greedy sampling algorithm integrated with a physics-informed residual-based error indicator and random-subset evaluation is introduced to search for the optimal training samples on the fly.
Further, to exploit local latent-space dynamics captured by the local DI models for an improved modeling accuracy with a minimum number of local DI models in the parameter space, a $k$-nearest neighbor convex interpolation scheme is employed.
The effectiveness of the proposed framework is demonstrated by modeling various nonlinear dynamical problems, including Burgers equations, nonlinear heat conduction, and radial advection. 
The proposed adaptive greedy sampling outperforms the conventional predefined uniform sampling in terms of accuracy.
Compared with the high-fidelity models, gLaSDI achieves \RB{17 to 2,658}$\times$ speed-up with 1 to 5$\%$ relative errors.
\end{abstract}
\end{frontmatter}

% \linenumbers

% ===============================================================================
\section{Introduction}\label{sec:introduction}
% ------------ Significance of high-fidelity physical simulations and difficulties --------------
Physical simulations have played an increasingly significant role in developments of engineering, science, and technology.
The widespread applications of physical simulations in digital twins systems \cite{jones2020characterising,liu2021review} is one recent example. 
Many physical processes are mathematically modeled by time-dependent nonlinear partial differential equations (PDEs).
As it is difficult or even impossible to obtain analytical solutions for many highly complicated problems, various numerical methods have been developed to approximate the analytical solutions.
However, due to the complexity and the domain size of problems, high-fidelity forward physical simulations can be computationally intractable even with high performance computing, which prohibits their applications to problems that require a large number of forward simulations, such as design optimization \cite{wang2007large,white2020dual}, optimal control \cite{choi2015practical}, uncertainty quantification \cite{smith2013uncertainty,biros2011large}, and inverse analysis \cite{biros2011large,galbally2010non}.

% ------------ Linear/Nonlinear subspace ROM --------------
To achieve accurate and efficient physical simulations, data can play a key role. For example, various physics-constrained data-driven model reduction techniques have been developed, such as the projection-based reduced-order model (ROM), in which the state fields of the full-order model (FOM) are projected to a linear or nonlinear subspace so that the dimension of the state fields is significantly reduced. 
Popular \textit{linear} projection techniques include the proper orthogonal decomposition (POD) \cite{berkooz1993proper}, the reduced basis method \cite{patera2007reduced}, and the balanced truncation method \cite{safonov1989schur}, while autoencoders \cite{demers1993non,hinton2006reducing} are often applied for \textit{nonlinear} projection \cite{kim2022fast,kim2020efficient,lee2020model}. 
The linear-subspace ROM (LS-ROM) has been successfully applied to various problems, such as nonlinear heat conduction \cite{hoang2021domain}, Lagrangian hydrodynamics \cite{copeland2022reduced,cheung2022local,lauzon2022sopt}, nonlinear diffusion equations \cite{hoang2021domain,fritzen2018algorithmic}, Burgers equations \cite{lauzon2022sopt,choi2020sns,choi2019space,carlberg2018conservative}, convection-diffusion equations \cite{mclaughlin2016stabilized,kim2021efficient}, Navier-Stokes equations \cite{stabile2018finite, iliescu2014variational}, Boltzmann transport problems \cite{hughes2020discontinuous,choi2021space}, fracture mechanics \cite{chen2015model,he2019decomposed}, molecular dynamics \cite{lee2013proper,lee1a2013rbf}, fatigue analysis under cycling-induced plastic deformations \cite{kaneko2021hyper}, topology optimization \cite{gogu2015improving,choi2019accelerating}, structural design optimization \cite{mcbane2021component,choi2020gradient}, etc. 
Despite successes of the classical LS-ROM in many applications, it is limited to the assumption that intrinsic solution space falls into a low-dimensional subspace, which means the solution space has a small Kolmogorov $n$-width. 
This assumption is not satisfied in advection-dominated systems with sharp gradients, moving shock fronts, and turbulence, which prohibits the applications of the LS-ROM approaches for these systems. 
On the other hand, it has been shown that nonlinear-subspace ROMs based on autoencoders outperforms the LS-ROM on advection-dominated systems \cite{kim2022fast,fries2022lasdi}.

% ------------ Extension of linear-subspace ROM for advection-dominated systems --------------
\RC{Several strategies have been developed to extend LS-ROM for addressing the challenge posed by advection-dominated systems, which can be mainly categorized into Lagrangian-based approaches and methods based on a transport-invariant coordinate frame.
In the first category of approaches, Lagrangian coordinate grids are leveraged to build a ROM that propagates both the wave physics and the coordinate grid in time \cite{lu2021dynamic,lu2020lagrangian,mojgani2017lagrangian}.
Although these methods work well, their applicability is limited by the requirement of full knowledge of the governing equations for obtaining the Lagrangian grid.
The second category of strategies is based on transforming the system dynamics to a moving coordinate frame by adding a time-dependent shift to the spatial coordinates such that the system dynamics are absent of advection, such as the shifted POD method \cite{reiss2018shifted} and the implicit feature tracking algorithm based on a minimal-residual ROM \cite{mirhoseini2021model}.
Despite the effectiveness of these methods, the high computational costs prohibit their applications in practice.}

% ------------ Intrusive ROM vs non-intrusive, interpretability --------------
Most aforementioned physics-constrained data-driven projection-based ROMs are \textit{intrusive}, which require plugging the reduced-order solution representation into the discretized system of governing equations. Although the intrusive brings many benefits, such as extrapolation robustness, a requirement of less training data, and high accuracy, the implementation of the intrusive ROMs requires not only sufficient understanding of the numerical solver of the high-fidelity simulation but also access to the source code of the numerical solver.

% ------------ Nonintrusive ROM, without latent-space dynamics learning --------------
In contrast, \textit{non-intrusive} ROMs are purely data-driven. It requires neither access to the source code nor the knowledge of the high-fidelity solver. Many non-intrusive ROMs are constructed based on interpolation techniques that provide nonlinear mapping to relate inputs to outputs. Among various interpolation techniques, such as Gaussian processes \cite{qian2006building,tapia2018gaussian}, radial basis functions \cite{daniel2007hydraulic,huang2015hull}, Kriging \cite{han2013improving,han2012hierarchical}, neural networks (NNs) have been most popular due to their strong flexibility and capability supported by the universal approximation theorem \cite{bock2019review}. NN-based surrogates have been applied to various physical simulations, such as fluid dynamics \cite{kutz2017deep}, particle simulations \cite{paganini2018calogan}, bioinformatics \cite{min2017deep}, deep Koopman dynamical models \cite{morton2018deep}, porous media flow \cite{kadeethum2021framework,kadeethum2021continuous,kadeethum2022non,kadeethum2022reduced}, etc. However, pure black-box NN-based surrogates lack \textit{interpretability} and suffer from unstable and inaccurate generalization performance. 
\RC{For example, Swischuk, et al. \cite{swischuk2019projection} compared various ML models, including NNs, multivariate polynomial regression, $k$-nearest neighbors (\RB{k-NNs}), and decision trees, used to learn nonlinear mapping between input parameters and low-dimensional representations of solution fields obtained by POD projection.
The numerical examples of this study shows that the highly flexible NN-based model performs worst, which highlights the importance of choosing an appropriate ML strategy by considering the bias-variance trade off, especially when the training data coverage of the input space is sparse.}

% ------------ Nonintrusive ROM based on latent-space dynamics learning --------------
In recent years, several ROM methods have been integrated with latent-space learning algorithms. Kim, et al. \cite{kim2019deep} proposed a DeepFluids framework in which the autoencoder was applied for nonlinear projection and a latent-space time integrator was used to approximate the evolution of the solutions in the latent space. 
Xie, et al. \cite{xie2019non} applied the POD for linear projection and a multi-step NN to propagate the latent-space dynamical solutions. 
Hoang, et al. \cite{hoang2022projection} applied the POD to compress space-time solution space to obtain space-time reduced-order basis and examined several surrogate models to map input parameters to space-time basis coefficients, including multivariate polynomial regression, \RB{k-NNs}, random forest, and NNs. 
Kadeethum, et al. \cite{kadeethum2022non} compared performance of the POD and autoencoder compression along with various latent space interpolation techniques, such as radial basis function and artificial neural networks.
However, the latent-space dynamics models of these methods are complex and lack interpretability.

To improve the interpretability and generalization capability, it is critical to identify the underlying equations governing the latent-space dynamics. 
Many methods have been developed for the identification of interpretable governing laws from data, including symbolic regression that searches both parameters and the governing equations simultaneously \cite{koza1994genetic,schmidt2009distilling}, parametric models that fit parameters to equations of a given form, such as the sparse identification of nonlinear dynamics (SINDy) \cite{brunton2016discovering}, \RC{and operator inference \cite{peherstorfer2016data,qian2020lift,benner2020operator}.}
Cranmer et al. \cite{cranmer2020discovering} applied graph neural networks to learn sparse latent representations and symbolic regression with a genetic algorithm to discover explicit analytical relations of the learned latent representations, which enhances efficiency of symbolic regression to high-dimensional data \cite{cranmer2020pysr}.
Instead of genetic algorithms, other techniques have been applied to guide the equation search in symbolic regression, such as gradient descent \cite{sahoo2018learning,kusner2017grammar} and Monte Carlo Tree Search with asymptotic constraints in NNs \cite{li2019neural}.

Champion, et al. \cite{champion2019data} applied an autoencoder for nonlinear projection and SINDy to identify simple ordinary differential equations (ODEs) that govern the latent-space dynamics. The autoencoder and the SINDy model were trained interactively to achieve simple latent-space dynamics.
However, the proposed SINDy-autoencoder method is not parameterized and generalizable.
Bai and Peng \cite{bai2021non} proposed parametric non-intrusive ROMs that combine the POD for linear projection and regression surrogates to approximate dynamical systems of latent variables, including support vector machines with kernel functions, tree-based methods, \RB{k-NNs}, vectorial kernel orthogonal greedy algorithm (VKOGA), and SINDy. 
The ROMs integrated with VKOGA and SINDy deliver superior cost versus error trade-off.
\RC{Additionally, various non-intrusive ROMs have been developed based on POD-based linear projection with latent space dynamics captured by polynomials through operator inference \cite{peherstorfer2016data, geelen2022operator, guo2022bayesian, geelen2022localized, mcquarrie2021non, qian2020lift, swischuk2020learning, benner2020operator, jain2021performance, mcquarrie2021data, peherstorfer2020sampling, khodabakhshi2022non, yildiz2021learning}.
For example, Qian, et al. \cite{qian2020lift} introduced a lifting map to transform non-polynomial physical dynamics to quadratic polynomial dynamics and then combined POD-based linear projection with operator inference to identify quadratic reduced models for dynamical systems.
Due to the limitation of the POD-based linear projection, these non-intrusive ROMs have difficulties with advection-dominated problems.}

\begin{figure}[htp]
    \centering
    \includegraphics[width=1\linewidth]{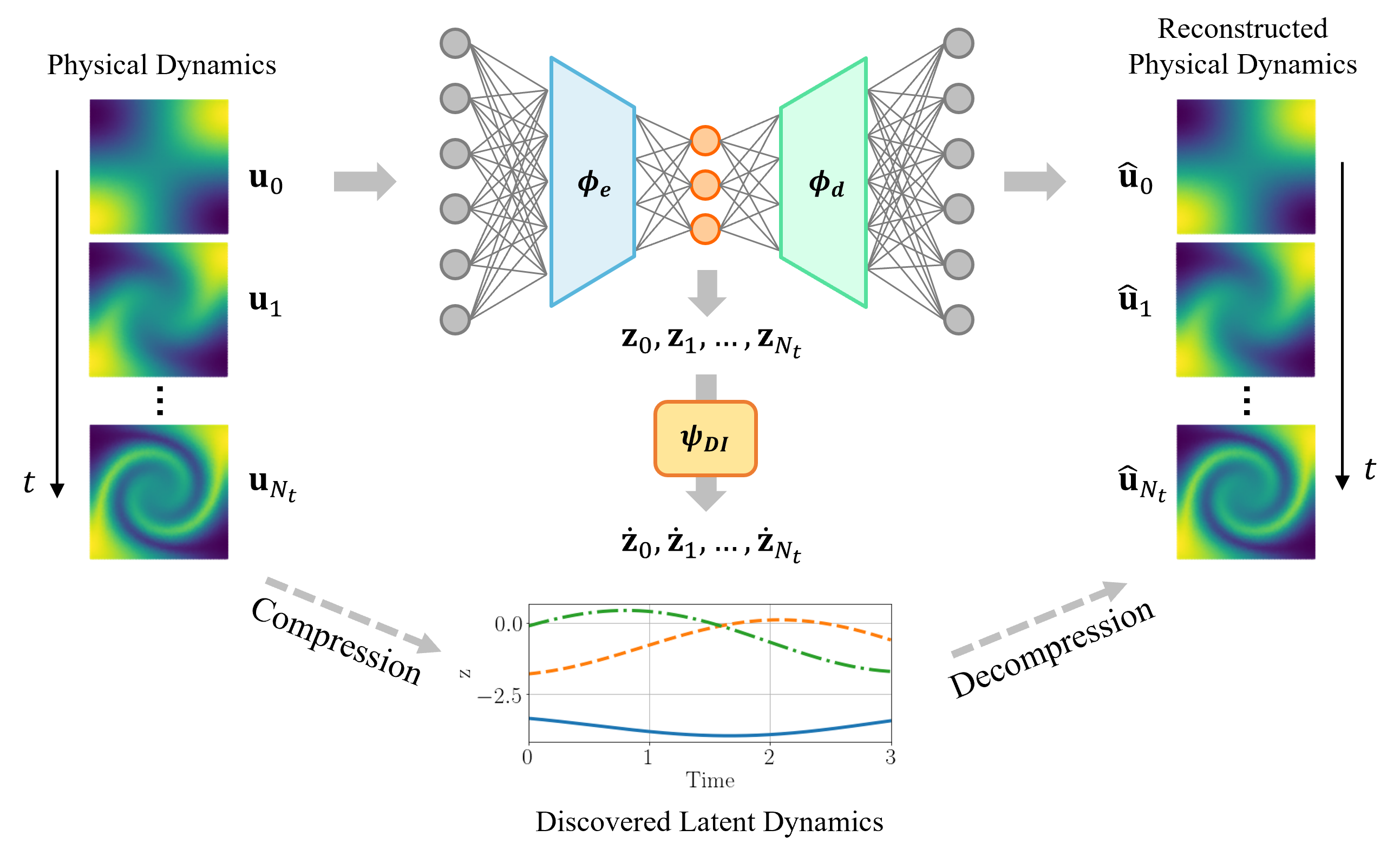}
    \caption{\RB{Schematics of the latent space dynamics identification (LaSDI) algorithm.} The autoencoder performs nonlinear projection and discovers intrinsic latent representations of high-fidelity solutions, while the dynamics identification (DI) model with strong interpretability approximates the ordinary differential equations that govern the latent-space dynamics.}
    \label{fig.lasdi}
\end{figure}

\RC{To address this challenge, Issan and Kramer \cite{issan2022predicting} recently proposed a non-intrusive ROM based on shifted operator inference by transforming the original coordinate frame of dynamical systems to a moving coordinate frame in which the dynamics are absent of translation and rotation.}
Fries, et al. \cite{fries2022lasdi} proposed a parametric latent space dynamics identification (LaSDI) framework in which an autoencoder was applied for nonlinear projection and a set of parametric dynamics identification (DI) models were introduced in the parameter space to identify local latent-space dynamics, as illustrated in Fig. \ref{fig.lasdi}.
\RB{The LaSDI framework can be viewed as a generalization of aforementioned non-intrusive ROMs built upon latent-space dynamics identification, since it allows linear or nonlinear projection and enables latent-space dynamics to be captured by flexible DI models based on general nonlinear functions.}
However, since a sequential training procedure was adopted for the autoencoder and the DI models, the lack of interaction between the autoencoder and the DI models leads to a strong dependency on the complexity and quality of the latent-space dynamics on the autoencoder architecture, which could pose challenges to the subsequent training of the DI models and thus affect the model performances.
Most importantly, all the above-mentioned approaches rely on predefined training samples, such as uniform or Latin hypercube sampling that may not be optimal in terms of the number of samples for achieving the best model performance in the prescribed parameter space. As the generation of the simulation data can be computationally expensive, it is important to minimize the number of samples.

% ------------ proposed method --------------
In this study, we propose a parametric adaptive greedy latent space dynamics identification (gLaSDI) framework for accurate, efficient, and robust physics-informed data-driven reduced-order modeling. To maximize and accelerate the exploration of the parameter space for optimal performance, an adaptive greedy sampling algorithm integrated with a physics-informed residual-based error indicator and random-subset evaluation is introduced to search for the optimal and minimal training samples on the fly. 
The proposed gLaSDI framework contains an autoencoder for nonlinear projection to discover intrinsic latent representations and a set of local DI models to capture local latent-space dynamics \RB{in the parameter space}, which is further exploited by an efficient \RB{k-NN} convex interpolation scheme.
\RB{The concept of gLaSDI is similar to the active learning algorithms that are allowed to choose training data from which it learns \cite{settles2009active,melville2004diverse}. 
The DI models in gLaSDI can be viewed as the active learners in active learning, which are selected to maximize ``diversity" of learners in the parameter space and minimize prediction errors.}
The autoencoder training and dynamics identification in the gLaSDI take place interactively to achieve an optimal identification of simple latent-space dynamics. 
\RB{The effectiveness and enhanced performance of the proposed gLaSDI framework is demonstrated by modeling various nonlinear dynamical problems with a comparison with the LaSDI framework \cite{fries2022lasdi}.}

% ------------ outline --------------
The remainder of this paper is organized as follows. 
The governing equations of dynamical systems is introduced in Section \ref{sec:govern_eqn}.
In Section \ref{sec:glasdi}, the ingredients of the proposed gLaSDI framework, the mathematical formulations, the training and testing algorithms are introduced.
In Section \ref{sec:result}, the effectiveness and capability of the proposed gLaSDI framework are examined by modeling various nonlinear dynamical problems, including Burgers equations, nonlinear heat conduction, and radial advection. 
The effects of various factors on model performance are investigated, including the number of nearest neighbors for convex interpolation, the latent-space dimension, the complexity of the DI models, and the size of the parameter space. 
A performance comparison between uniform sampling, i.e., LaSDI, and the physics-informed greedy sampling, i.e., gLaSDI, is also presented.
Concluding remarks and discussions are summarized in Section \ref{sec:conclusion}.

\section{Governing equations of dynamical systems}\label{sec:govern_eqn}
A parameterized dynamical system characterized by a system of ordinary differential equations (ODEs) is considered
\begin{subequations}\label{eq.govern_eqn}
    \begin{align}
    & \frac{d\mathbf{u}(t; \boldsymbol{\mu})}{dt} = \mathbf{f}(\mathbf{u},t; \boldsymbol{\mu}), \quad t \in [0,T], \\
    & \mathbf{u}(0; \boldsymbol{\mu}) = \mathbf{u}_0(\boldsymbol{\mu})
    \end{align}
\end{subequations}
where $T \in \mathbb{R}_{+}$ is the final time; 
$\boldsymbol{\mu} \in \mathcal{D} \subseteq \mathbb{R}^{n_{\mu}}$ is the parameter in a parameter domain $\mathcal{D}$.
$\mathbf{u}(t; \boldsymbol{\mu}): [0,T] \times \mathcal{D} \rightarrow \mathbb{R}^{N_u}$ is the parameterized time-dependent solution to the dynamical system; 
$\mathbf{f}: \mathbb{R}^{N_u} \times [0,T] \times \mathcal{D} \rightarrow \mathbb{R}^{N_u}$ denotes the velocity of $\mathbf{u}$; 
$\mathbf{u}_0$ is the initial state of $\mathbf{u}$. 
Eq. (\ref{eq.govern_eqn}) can be considered as a semi-discretized equation of a system of partial differential equations (PDEs) with a spatial domain $\Omega \subseteq \mathbb{R}^d, d \in \mathbb{N}(3)$, and $\mathbb{N}(N) := \{1,...,N\}$. Spatial discretization can be performed by numerical methods, such as the finite element method. 

A uniform time discretization is considered in this study, with a time step size $\Delta t \in \mathbb{R}_{+}$ and $t_n = t_{n-1} + \Delta t$ for $n \in \mathbb{N}(N_t)$ where $t_0 = 0$, $N_t \in \mathbb{N}$. Various explicit or implicit time integration schemes can be applied to solve Eq. (\ref{eq.govern_eqn}). For example, with the implicit backward Euler time integrator, the \RB{approximate} solutions to Eq. (\ref{eq.govern_eqn}) can be obtained by solving the following nonlinear system of equations
\begin{equation}\label{eq.backward_euler}
    \mathbf{u}_n = \mathbf{u}_{n-1} + \Delta t \mathbf{f}_n,
\end{equation}
where $\mathbf{u}_n := \mathbf{u}(t^n; \boldsymbol{\mu})$, and $\mathbf{f}_n := \mathbf{f}(\mathbf{u}(t^n; \boldsymbol{\mu}), t^n; \boldsymbol{\mu})$. The residual function of Eq. (\ref{eq.backward_euler}) is expressed as
\begin{equation}\label{eq.residual}
    \mathbf{r}(\mathbf{u}_n; \mathbf{u}_{n-1}, \boldsymbol{\mu}) = \mathbf{u}_n - \mathbf{u}_{n-1} - \Delta t \mathbf{f}_n.
\end{equation}

 Solving the dynamical system of equation (Eq. (\ref{eq.govern_eqn})) could be computationally expensive, especially when the solution dimension ($N_u$) is large and the computational domain ($\Omega$) is geometrically complex. In this work, an efficient and accurate data-driven reduced-order modeling framework based on physics-informed greedy latent space dynamics identification is proposed, which will be discussed in details in the next section.

\section{gLaSDI}\label{sec:glasdi}
The ingredients of the proposed physics-informed parametric adaptive greedy latent space dynamics identification (gLaSDI) framework are introduced in this section, including autoencoders, dynamics identification (DI) models, $k$-nearest neighbors (\RB{k-NN}) convex interpolation, and an adaptive greedy sampling procedure with a physics-informed error indicator. 
An autoencoder is trained to discover an intrinsic nonlinear latent representation of high-dimensional data from dynamical PDEs, while training cases are sampled on the fly and associated local DI models are trained simultaneously to capture localized latent-space dynamics. 
An interactive training procedure is employed for the autoencoder and local DI models, which is referred to as the \RB{\textit{interactive Autoencoder-DI training}}, as shown in Fig. \ref{fig.glasdi}. 
The interaction between the autoencoder and local \RB{(in the parameter space)} DI models enables the identification of simple and smooth latent-space dynamics and therefore accurate and efficient data-driven reduced-order modeling. 
\RB{In the following, Sections \ref{sec:autoencoder} - \ref{sec:knn_convex} review the basics of autoencoders, the dynamics identification method, and the convex interpolation scheme, respectively, which are fundamental to the physics-informed adaptive greedy sampling algorithm introduced in Section \ref{sec:greedy}. Section \ref{sec:algorithm} summarizes the algorithms for gLaSDI training and testing.}

The physics-informed adaptive greedy sampling can be performed on either a continuous parameter space ($\mathcal{D}$) or a discrete parameter space ($\mathcal{D}^h \subseteq \mathcal{D}$). 
In the following demonstration, a discrete parameter space ($\mathcal{D}^h$) is considered.
$\mathbb{D} \subseteq \mathcal{D}^h$ denotes a set of $N_{\mu}$ selected training sample points.

Let us consider a training sample point $\boldsymbol{\mu}^{(i)} \in \mathcal{D}^h$, $i \in \mathbb{N}(N_{\mu})$ and $\mathbf{u}_n^{(i)} \in \mathbb{R}^{N_u}$ as the solution at the $n$-th time step of the dynamical system in Eq. (\ref{eq.govern_eqn}) with the training sample point $\boldsymbol{\mu}^{(i)}$. 
The solutions at all time steps are arranged in a snapshot matrix denoted as $\mathbf{U}^{(i)} = [\mathbf{u}_0^{(i)}, ..., \mathbf{u}_{N_t}^{(i)}] \in \mathbb{R}^{N_u \times (N_t + 1)}$. Concatenating snapshot matrices corresponding to all training sample points gives a full snapshot matrix $\mathbf{U} \in \mathbb{R}^{N_u \times (N_t + 1) N_{\mu}}$
\begin{equation}\label{eq.full_snapshot}
    \mathbf{U} = \big[ \mathbf{U}^{(1)}, ... , \mathbf{U}^{(N_{\mu})} \big].
\end{equation}

% ===================================================================================================
\subsection{Autoencoders for nonlinear dimensionality reduction}\label{sec.autoencoders}\label{sec:autoencoder}
An autoencoder \cite{demers1993non,hinton2006reducing} is a special architecture of deep neural networks (DNNs) designed for dimensionality reduction or representation learning.
As shown in Fig. \ref{fig.lasdi}, an autoencoder consists of an encoder function $\boldsymbol{\phi}_e(\cdot;\boldsymbol{\theta}_{\text{enc}}) : \mathbb{R}^{N_u} \rightarrow \mathbb{R}^{N_z}$ and a decoder function $\boldsymbol{\phi}_d(\cdot;\boldsymbol{\theta}_{\text{dec}}): \mathbb{R}^{N_z} \rightarrow \mathbb{R}^{N_u}$, such that
\begin{subequations}\label{eq.autoencoder}
    \begin{align}
        & \mathbf{z}_n^{(i)} = \boldsymbol{\phi}_e(\mathbf{u}_n^{(i)};\boldsymbol{\theta}_{\text{enc}}), \\
        & \hat{\mathbf{u}}_n^{(i)} = \boldsymbol{\phi}_d(\mathbf{z}_n^{(i)};\boldsymbol{\theta}_{\text{dec}})
    \end{align}
\end{subequations}
where $\mathbf{u}_n^{(i)} \in \mathbb{R}^{N_u}$ denotes the solution of a sampling point $\boldsymbol{\mu}^{(i)} \in \mathcal{D}^h$, $i \in \mathbb{N}(N_{\mu})$, at the $n$-th time step; 
$N_z \ll N_u$ is the latent dimension, $\boldsymbol{\theta}_{\text{enc}}$ and $\boldsymbol{\theta}_{\text{dec}}$ are trainable parameters of the encoder and the \RA{decoder}, respectively; $\hat{\mathbf{u}}_n^{(i)} \in \mathbb{R}^{N_u}$ is the output of the autoencoder, a reconstruction of the original input $\mathbf{u}_n^{(i)}$.
With the latent dimension $N_z$ much smaller than the input dimension $N_u$, the encoder $\boldsymbol{\phi}_e$ is trained to compress the high-dimensional input $\mathbf{u}_n^{(i)}$ and learn a low-dimensional representation, denoted as a latent variable $\mathbf{z}_n^{(i)} \in \mathbb{R}^{N_z}$,
whereas the decoder $\boldsymbol{\phi}_d$ reconstructs the input data by mapping the latent variable back to the high-dimensional space, as illustrated in Fig. \ref{fig.lasdi}. 

Let us denote $\mathbf{Z}^{(i)} = [\mathbf{z}_0^{(i)}, ..., \mathbf{z}_{N_t}^{(i)}] \in \mathbb{R}^{N_z \times (N_t + 1)}$ as the matrix of latent variables at all time steps of the sampling point $\boldsymbol{\mu}^{(i)}$ and $\mathbf{Z} = \big[ \mathbf{Z}^{(1)}, ... , \mathbf{Z}^{(N_{\mu})} \big] \in \mathbb{R}^{N_z \times (N_t + 1) N_{\mu}}$ as the full latent variable matrix of all sampling points in the parameter space. 
The corresponding reconstructed full snapshot matrix is denoted by $\hat{\mathbf{U}} = \big[ \hat{\mathbf{U}}^{(1)}, ... , \hat{\mathbf{U}}^{(N_{\mu})} \big] \in \mathbb{R}^{N_u \times (N_t + 1) N_{\mu}}$, where $\hat{\mathbf{U}}^{(i)} = [\hat{\mathbf{u}}_0^{(i)}, ..., \hat{\mathbf{u}}_{N_t}^{(i)}] \in \mathbb{R}^{N_u \times (N_t + 1)}$, $i \in \mathbb{N}(N_{\mu})$, obtained from Eq. (\ref{eq.autoencoder}). 
The optimal trainable parameters of the autoencoder ($\boldsymbol{\theta}_{\text{enc}}$ and $\boldsymbol{\theta}_{\text{dec}}$ from Eq. (\ref{eq.autoencoder})) are obtained by minimizing the loss function:
\begin{equation}\label{eq.autoencoder_loss}
    \mathcal{L}_{recon} := || \mathbf{U} - \hat{\mathbf{U}} ||_{L_2}^2.
\end{equation}

\RB{A standard autoencoder often consists of symmetric architectures for the encoder and the decoder, but it is not required. In this study, a standard autoencoder is adopted and the encoder architecture is used to denote the autoencoder architecture for simplicity.} For example, the encoder architecture 6-4-3 denotes that there are 6, 4, and 3 artificial neurons in the input layer, the hidden layer, and the embedding layer that outputs the latent variables, respectively. As such, the decoder architecture in this case is 3-4-6 and the corresponding autoencoder architecture is 6-4-3-4-6.
===================================================================================================
\subsection{Latent-space dynamics identification}\label{sec:dyn_ident}
Instead of learning complex physical dynamics of high-dimensional data, a dynamics identification (DI) model is introduced to capture dynamics of the autoencoder-discovered low-dimensional representation associated with the high-dimensional data. 
Therefore, the problem of the high-dimensional dynamical system in Eq. (\ref{eq.govern_eqn}) is reduced to
\begin{equation}\label{eq.dynamics_identification}
    \frac{d\mathbf{z}(t; \boldsymbol{\mu})}{dt} = \boldsymbol{\psi}_{DI}(\mathbf{z},t; \boldsymbol{\mu}), \quad t \in [0,T],
\end{equation}
where $\mathbf{z}(t; \boldsymbol{\mu}) = \boldsymbol{\phi}_e(\mathbf{u}(t; \boldsymbol{\mu})) \in \mathbb{R}^{N_z}$ and $\boldsymbol{\phi}_e$ denotes the encoder function introduced in Section \ref{sec.autoencoders}. The dynamics of $\mathbf{u}$ can be reconstructed by using the decoder function: $\hat{\mathbf{u}}(t; \boldsymbol{\mu}) = \boldsymbol{\phi}_d(\mathbf{z}(t; \boldsymbol{\mu}))$.

Given the discrete latent variable matrix $\mathbf{Z}^{(i)} = [\mathbf{z}_0^{(i)}, ..., \mathbf{z}_{N_t}^{(i)}] \in \mathbb{R}^{N_z \times (N_t + 1)}$ of a sampling point $\boldsymbol{\mu}^{(i)}$, $i \in \mathbb{N}(N_{\mu})$, the governing dynamical function $\boldsymbol{\psi}_{DI}$ is approximated by a  user-defined library of candidate basis functions $\boldsymbol{\Theta}(\mathbf{Z}^{(i)T})$, expressed as
\begin{equation}\label{eq.dynamics_identification_discrete}
     \dot{\mathbf{Z}}^{(i)T} \approx \dot{\hat{\mathbf{Z}}}^{(i)T} = \boldsymbol{\Theta}(\mathbf{Z}^{(i)T}) \boldsymbol{\Xi}^{(i)},
\end{equation}
where $\boldsymbol{\Theta}(\mathbf{Z}^{(i)T}) = [\boldsymbol{b}_1(\mathbf{Z}^{(i)T}),\boldsymbol{b}_2(\mathbf{Z}^{(i)T}),...,\boldsymbol{b}_{N_b}(\mathbf{Z}^{(i)T})] \in \mathbb{R}^{(N_t+1)\times N_l}$ has ${N_b}$ candidate basis functions to capture the latent space dynamics, e.g., polynomial, trigonometric, and exponential functions, \RA{with $\boldsymbol{b}_i(\mathbf{Z}^{T}) \in \mathbb{R}^{(N_t+1) \times N_{l_i}}$; $N_l  = \sum_i^{N_b} N_{l_i} $ denotes the number of columns of the library matrix.
Note that $N_{l_i}$ is determined by the form of the basis function \cite{brunton2016discovering,fries2022lasdi}.
For example, if $b_i$ is an \textit{exponential} function, then 
\begin{equation}
    \boldsymbol{b}_i(\mathbf{Z}^{T}) = 
    \bmat{\exp(z_1(t_0)) & \hdots &  \exp(z_{N_z}(t_0)) \\
    \vdots & \ddots & \vdots  \\
    \exp(z_1(t_{N_t})) & \hdots &  \exp(z_{N_z}(t_{N_t}))}.
\end{equation}
If $b_i$ is a \textit{quadratic polynomial}, then 
\begin{equation}
    \boldsymbol{b}_i(\mathbf{Z}^{T}) = 
    \bmat{z_1^2(t_0) & z_1(t_0) z_2(t_0) & \hdots & z_2^2(t_0) & \hdots & z_{N_z}^2(t_0) \\
    \vdots & \vdots & \ddots & \vdots & \ddots & \vdots \\
    z_1^2(t_{N_t}) & z_1(t_{N_t}) z_2(t_{N_t}) & \hdots & z_2^2(t_{N_t}) & \hdots & z_{N_z}^2(t_{N_t})}.
\end{equation}
}
$\boldsymbol{\Xi}^{(i)}=[\boldsymbol{\xi}_1^{(i)}, \boldsymbol{\xi}_2^{(i)}, ..., \boldsymbol{\xi}_{N_z}^{(i)}] \in \mathbb{R}^{N_l \times N_z}$ is an associated coefficient matrix. 

In gLaSDI, the autoencoder and the DI model are trained simultaneously and interactively to identify simple and smooth latent-space dynamics.
Note that $\dot{\mathbf{Z}}^{(i)T} = [\dot{\mathbf{z}}_0^{(i)T}, ..., \dot{\mathbf{z}}_{N_t}^{(i)T}]^T \in \mathbb{R}^{(N_t + 1) \times N_z}$ in Eq. (\ref{eq.dynamics_identification_discrete}) can be obtained by applying the chain rule and automatic differentiation (AD) \cite{paszke2017automatic} to the encoder network, i.e.,
\begin{equation}\label{eq.zdot}
    \dot{\mathbf{z}}_n^{(i)} = \big(\nabla_{\mathbf{u}} \mathbf{z}_n^{(i)} \big) \dot{\mathbf{u}}_n^{(i)} = 
    \nabla_{\mathbf{u}} \boldsymbol{\phi}_e(\mathbf{u}_n^{(i)}) \dot{\mathbf{u}}_n^{(i)}, \quad n = 0,...,N_t.
\end{equation}
\RA{Enforcing the consistency on the predicted latent-dynamics, $\dot{\mathbf{Z}}$ and $\dot{\hat{\mathbf{Z}}}$, allows simple and smooth dynamics to be identified, which are determined by the selection of the basis functions in the DI model.
Therefore, }the following loss function is constructed, which imposes a constraint on the trainable parameters of the encoder ($\boldsymbol{\theta}_{\text{enc}}$ via Eq. (\ref{eq.zdot})) and the DI models ($\{\boldsymbol{\Xi}^{(i)}\}_{i \in \mathbb{N}(N_{\mu})}$ via Eq. (\ref{eq.dynamics_identification_discrete})),
% An optimal coefficient matrix of each sampling point, \YC{i.e., $\boldsymbol{\Xi}^{(i)}$ for $\boldsymbol{\mu}^{(i)}$,} is obtained by minimizing the following loss function
\begin{equation}\label{eq.loss_zdot}
    \mathcal{L}_{\dot{\mathbf{z}}} := || \dot{\mathbf{Z}} - \dot{\hat{\mathbf{Z}}} ||_{L_2}^2,
\end{equation}
with
\begin{subequations}\label{eq.loss_zdot2}
    \begin{align}
        \dot{\mathbf{Z}} & = \big[ \dot{\mathbf{Z}}^{(1)}, ... , \dot{\mathbf{Z}}^{(N_{\mu})} \big] \in \mathbb{R}^{N_z \times (N_t + 1) N_{\mu}}\\
        \dot{\hat{\mathbf{Z}}} &= \big[ \dot{\hat{\mathbf{Z}}}^{(1)}, ... , \dot{\hat{\mathbf{Z}}}^{(N_{\mu})} \big] \in \mathbb{R}^{N_z \times (N_t + 1) N_{\mu}}.
    \end{align}
\end{subequations}
The loss function in $\dot{\mathbf{z}}$ also enables identification of simple latent-dynamics when simple DI model is prescribed, which will be demonstrated in Section \ref{sec:result_2Dburger}-\ref{sec:result_heat}.
Note that the local DI models are considered to be point-wise (see the detailed description about point-wise and region-based DI models in \cite{fries2022lasdi}), which means each local DI model is associated with a distinct sampling point in the parameter space. Hence, each sampling point has an associated DI coefficient matrix.

\begin{figure}[htp]
    \centering
    \includegraphics[width=1\textwidth]{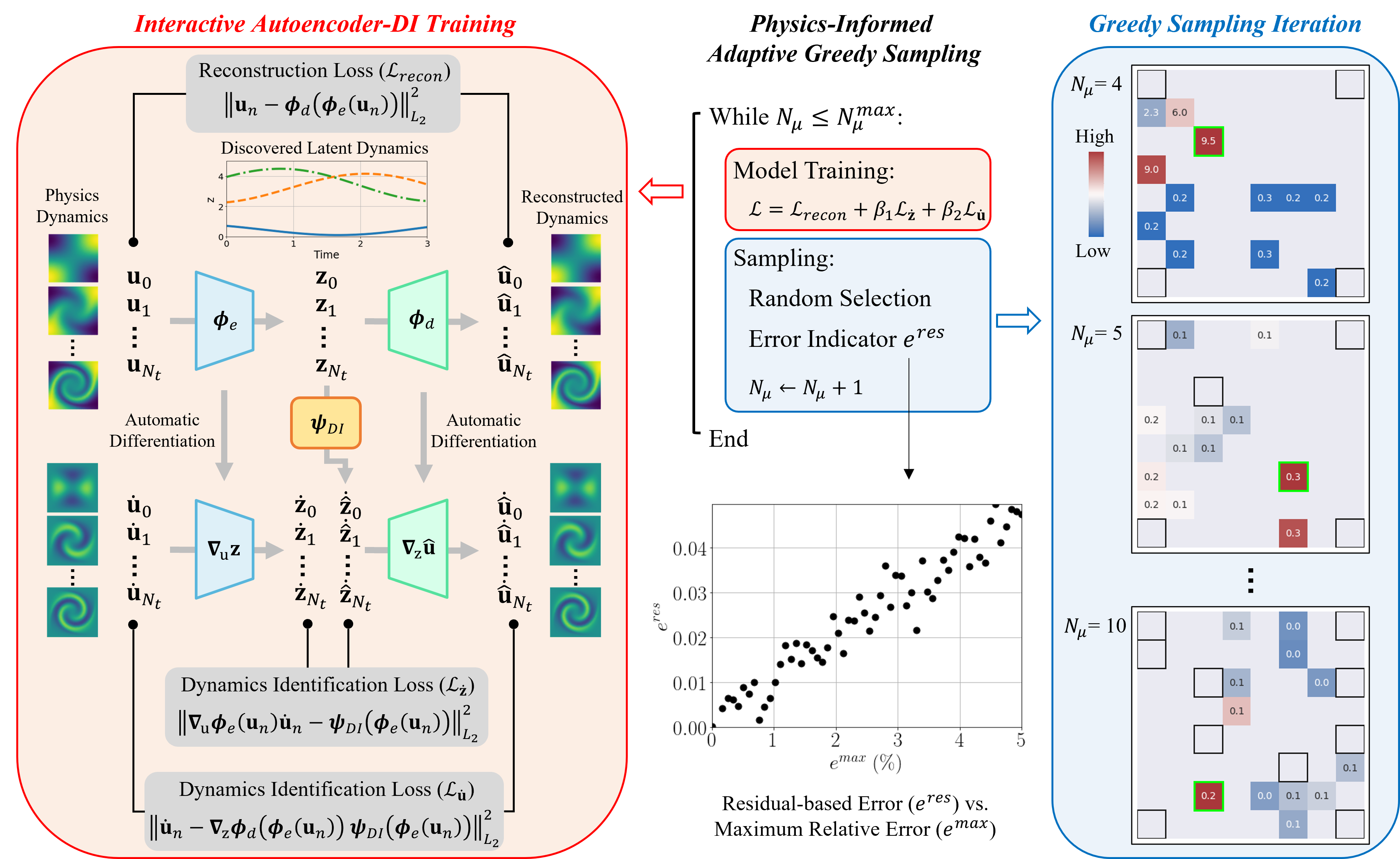}
    \caption{\RB{Schematics of the gLaSDI algorithm. The positive correlation between the residual-based error indicator $e^{res}$ and the maximum relative error $e^{max}$ indicates the computationally efficient $e^{res}$ can replace the computationally expensive $e^{max}$. Each black-filled circle represents one evaluated sample.}}
    \label{fig.glasdi}
\end{figure}

To further enhance the accuracy of the physical dynamics predicted by the decoder, a loss function is constructed to ensure the \RA{consistency between} the predicted dynamics gradients and the gradients of the solution data, in addition to the reconstruction loss function in Eq. (\ref{eq.autoencoder_loss}). The enhancement is demonstrated in Section \ref{sec:result_1Dburger_case3}.
The predicted dynamics gradients, $\dot{\hat{\mathbf{u}}}$, can be calculated from $\mathbf{u}$ by following the path: $\mathbf{u} \rightarrow \mathbf{z} \rightarrow \dot{\hat{\mathbf{z}}} \rightarrow \dot{\hat{\mathbf{u}}}$, as illustrated in Fig. \ref{fig.glasdi}, and applying the chain rule and AD to the decoder network, 
\begin{subequations}\label{eq.udot}
    \begin{align}
        \dot{\hat{\mathbf{u}}}_n^{(i)} 
        & = \frac{\partial \hat{\mathbf{u}}_n^{(i)}}{\partial \mathbf{z}_n^{(i)}}  \cdot \frac{\partial \mathbf{z}_n^{(i)} }{\partial t} \\
        & = \nabla_{\mathbf{z}} \boldsymbol{\phi}_d \big( \boldsymbol{\phi}_e(\mathbf{u}_n^{(i)} ) \big) \cdot \boldsymbol{\psi}_{DI}(\mathbf{z}_n^{(i)}) \\
        & = \nabla_{\mathbf{z}} \boldsymbol{\phi}_d \big( \boldsymbol{\phi}_e(\mathbf{u}_n^{(i)} ) \big) \cdot \boldsymbol{\Theta}(\boldsymbol{\phi}_e(\mathbf{u}_n^{(i)})^T) \boldsymbol{\Xi}^{(i)}. \quad n = 0,...,N_t,
    \end{align}
\end{subequations}
which involves all trainable parameters, including the encoder ($\boldsymbol{\theta}_{\text{enc}}$), the decoder ($\boldsymbol{\theta}_{\text{dec}}$), and the DI models ($\{\boldsymbol{\Xi}^{(i)}\}_{i \in \mathbb{N}(N_{\mu})}$).
The loss function in $\dot{\mathbf{u}}$ is defined as
\begin{equation}\label{eq.loss_udot}
    \mathcal{L}_{\dot{\mathbf{u}}} := || \dot{\mathbf{U}} - \dot{\hat{\mathbf{U}}} ||_{L_2}^2.
\end{equation}
with 
\begin{subequations}\label{eq.loss_udot2}
    \begin{align}
        \dot{\mathbf{U}} & = \big[ \dot{\mathbf{U}}^{(1)}, ... , \dot{\mathbf{U}}^{(N_{\mu})} \big] \in \mathbb{R}^{N_u \times (N_t + 1) N_{\mu}}\\
        \dot{\hat{\mathbf{U}}} &= \big[ \dot{\hat{\mathbf{U}}}^{(1)}, ... , \dot{\hat{\mathbf{U}}}^{(N_{\mu})} \big] \in \mathbb{R}^{N_u \times (N_t + 1) N_{\mu}}.
    \end{align}
\end{subequations}

Therefore, the loss function of the \RB{\textit{interactive autoencoder-DI training}} consists of three different loss terms, i.e., the reconstruction loss of the autoencoder in Eq. (\ref{eq.autoencoder_loss}), the DI loss in $\dot{\mathbf{z}}$ in Eq. (\ref{eq.loss_zdot}), and the DI loss in $\dot{\mathbf{u}}$ in Eq. (\ref{eq.loss_udot}). They are combined as a linear combination:
\begin{equation}\label{eq.total_loss}
    \mathcal{L} = \mathcal{L}_{recon} + \beta_1 \mathcal{L}_{\dot{\mathbf{z}}}  + \beta_2 \mathcal{L}_{\dot{\mathbf{u}}},
\end{equation}
where $\beta_1$ and $\beta_2$ denote the regularization parameters to balance the scale and contributions from the loss terms. A schematics of the \RB{\textit{interactive autoencoder-DI training}} is shown in Fig. \ref{fig.glasdi}.

Compared with a global DI model that captures global latent-space dynamics of all sampling points in the parameter space $\mathcal{D}^h$, each local DI model in the proposed framework is associated with one sampling point and thus captures local latent-space dynamics more accurately. Further, an efficient $k$-nearest neighbor (\RB{k-NN}) convex interpolation scheme, which will be introduced in the next subsection, is employed to exploit the local latent-space dynamics captured by the local DI models for an improved prediction accuracy, which will be demonstrated in Section \ref{sec:result_1Dburger}.
% with a minimum number of local DI models in the parameter space. 

% ===================================================================================================
\subsection{k-nearest neighbors convex interpolation}\label{sec:knn_convex}
To exploit the local latent-space dynamics captured by the local DI models for enhanced parameterization and efficiency, a \RB{k-NN} convexity-preserving partition-of-unity interpolation scheme is employed.
The interpolation scheme utilizes Shepard function \cite{shepard1968two} or inverse distance weighting, which has been widely used in data fitting and function approximation with positivity constraint \cite{babuvska1997partition,wendland2004scattered,he2014topology,he2021deep}.
Compared with other interpolation techniques, such as the locally linear embedding \cite{roweis2000nonlinear,he2020physics} and the radial basis function interpolation \cite{fries2022lasdi}, which require optimization to obtain interpolation weights and thus more computational cost, the employed Shepard interpolation is more efficient while preserving convexity.

Given a testing parameter $\boldsymbol{\mu} \in \mathcal{D}$, the DI coefficient matrix $\boldsymbol{\Xi}$ is obtained by a convex interpolation of coefficient matrices of its $k$-nearest neighbors (existing sampling points), expressed as
\begin{equation}\label{eq.convex_interp}
    \boldsymbol{\Xi}_{interp} = \mathcal{I} \left( \{\Psi^{(i)}(\boldsymbol{\mu}); \boldsymbol{\Xi}^{(i)}\}_{i \in \mathcal{N}_k(\boldsymbol{\mu})} \right) = \sum_{i \in \mathcal{N}_k(\boldsymbol{\mu})} \Psi^{(i)}(\boldsymbol{\mu}) \boldsymbol{\Xi}^{(i)},
\end{equation}
where $\boldsymbol{\Xi}_{interp}$ is the interpolated DI coefficient matrix of the testing parameter $\boldsymbol{\mu}$, $\mathcal{N}_k(\boldsymbol{\mu})$ is the index set of the $k$-nearest neighbor points of $\boldsymbol{\mu}$ selected from $\mathbb{D} \subseteq \mathcal{D}^h$ that contains the parameters of the training samples, and $\boldsymbol{\Xi}^{(i)}$ is the coefficient matrix of the sampling point $\boldsymbol{\mu}^{(i)}$. The selection of the $k$-nearest neighbors is based on the \RB{Mahalanobis} distance between the testing parameter and the training parameters, \RB{$||\boldsymbol{\mu}-\boldsymbol{\mu}^{(i)}||_{\mathbf{S}}$}. The interpolation functions are defined as
\begin{equation}\label{eq.shep_shape}
    \Psi^{(i)}(\boldsymbol{\mu}) = \frac{\phi(\boldsymbol{\mu}-\boldsymbol{\mu}^{(i)})}{\sum_{j=1}^{k} \phi(\boldsymbol{\mu}-\boldsymbol{\mu}^{(j)})},
\end{equation}
where $k$ is the number of nearest neighbors. In Eqs. (\ref{eq.convex_interp}) and (\ref{eq.shep_shape}), $\phi$ is a positive kernel function representing the weight on the coefficient set $\{\boldsymbol{\Xi}^{(i)}\}_{i \in \mathcal{N}_k(\boldsymbol{\mu})}$, and
$\mathcal{I}$ denotes the interpolation operator that constructs shape functions with respect to $\boldsymbol{\mu}$ and its neighbors.
It should be noted that these functions satisfy a partition of unity, i.e., $\sum_{i \in \mathcal{N}_k(\boldsymbol{\mu})} \Psi^{(i)}(\boldsymbol{\mu})=1$ for transformation objectivity. Furthermore, they are convexity-preserving when the kernel function $\phi$ is a positive function.
Here, an inverse distance function is used as the kernel function
% \begin{equation}\label{eq.shep_kern}
%     \phi(\boldsymbol{\mu}-\boldsymbol{\mu}^{(i)}) = \frac{1}{|| \boldsymbol{\mu}-\boldsymbol{\mu}^{(i)} ||_{L_2}^2}.
% \end{equation}
\begin{equation}\label{eq.shep_kern}
    \RB{\phi(\boldsymbol{\mu}-\boldsymbol{\mu}^{(i)}) = \frac{1}{|| \boldsymbol{\mu}-\boldsymbol{\mu}^{(i)} ||_{\mathbf{S}}^2},}
\end{equation}
\RB{where $|| \boldsymbol{\mu}-\boldsymbol{\mu}^{(i)} ||_{\mathbf{S}} = \sqrt{(\boldsymbol{\mu}-\boldsymbol{\mu}^{(i)})^T \mathbf{S}^{-1} (\boldsymbol{\mu}-\boldsymbol{\mu}^{(i)})}$ measures the Mahalanobis distance (multivariate distance) between $\boldsymbol{\mu}$ and $\boldsymbol{\mu}^{(i)}$.
The covariance matrix $\mathbf{S}$ is estimated from the sampled parameters and accounts for the scale and the correlation between the variables along each direction of the parameter space.
For uncorrelated variables with a similar scale, the Mahalanobis distance is equivalent to the Euclidean ($L_2$) distance.
In the numerical examples of Section \ref{sec:result}, the parameter variables along each direction of the parameter space are uncorrelated with a similar scale and therefore the Euclidean distance was applied.}
If the testing parameter point overlaps with one of the nearest neighbor points, i.e., $\boldsymbol{\mu} = \boldsymbol{\mu}^{(j)}$, $j \in \mathcal{N}_k(\boldsymbol{\mu})$, then $\Psi^{(j)}(\boldsymbol{\mu}) = 1$ and $\Psi^{(i)}(\boldsymbol{\mu}) = 0$, $\forall i \in \mathcal{N}_k(\boldsymbol{\mu})$ and $i \neq j$, resulting in $\boldsymbol{\Xi}_{interp} = \boldsymbol{\Xi}^{(j)}$, which is expected.

\begin{figure}[htp]
\centering
    \begin{subfigure}{0.4\textwidth}
        \centering
        \includegraphics[width=1\linewidth]{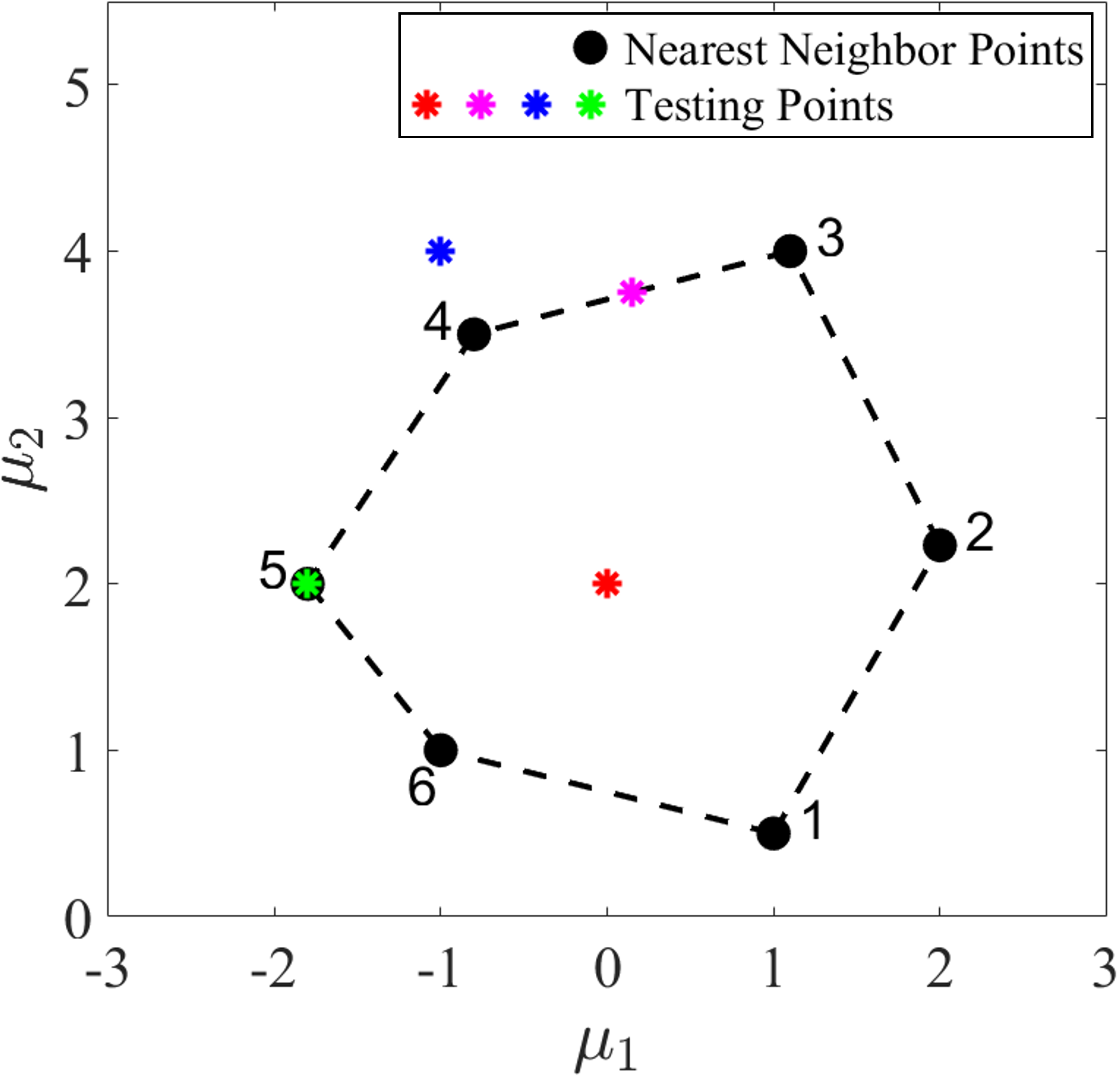}
        \caption{}
    \end{subfigure}
    \begin{subfigure}{0.43\textwidth}
        \centering
        \includegraphics[width=1\linewidth]{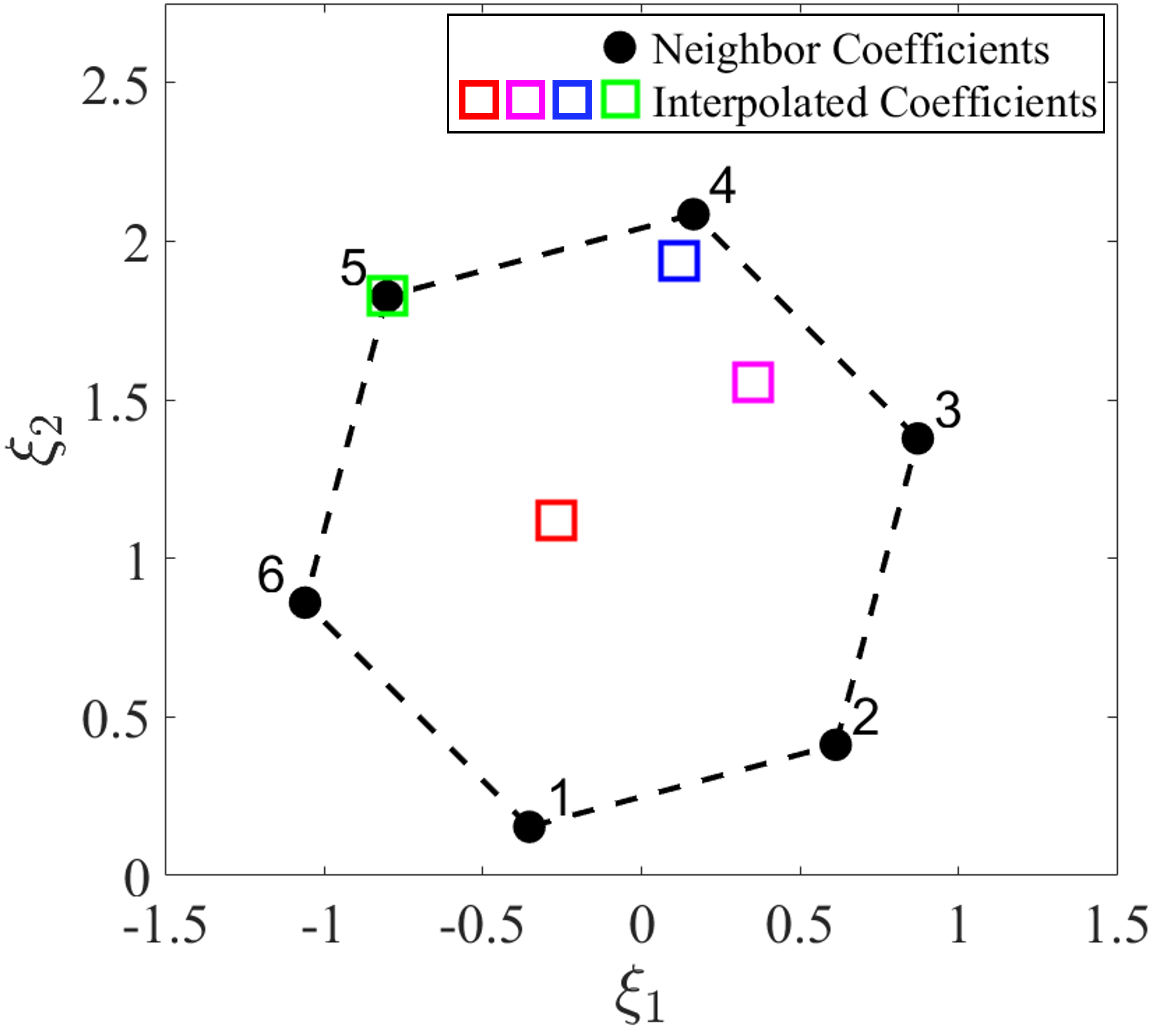}
        \caption{}
    \end{subfigure}
\caption{Demonstration of the convexity-preserving interpolation in Eq. (\ref{eq.convex_interp}): (a) The four asterisks denote the testing parameter points while the black solid dots denote the nearest neighbor parameter points of the testing parameter points. The black dash line depicts a locally convex hull formed by the nearest neighbor points. The interpolation functions are obtained using Eqs. (\ref{eq.shep_shape}-\ref{eq.shep_kern}) based on the distance between the testing points and nearest neighbor points. (b) The black solid dots denote the coefficients associated with the nearest neighbor parameter points in (a), which are used to interpolate the coefficients (squares) associated with the testing points in (a) by using the convex interpolation scheme in Eq. (\ref{eq.convex_interp}). The black dash line depicts a locally convex hull formed by the nearest neighbors' coefficients.}\label{fig.shep}
\end{figure}

\RB{Note that the interpolation functions are constructed in the parameter space, while the interpolation is performed in the DI coefficient space.}
For better visualization and demonstration in a two-dimensional \RB{coefficient} domain, we consider convex interpolation of two components from the \RB{DI} coefficient matrix, i.e., $\xi_1$ and $\xi_2$, as shown in Fig. \ref{fig.shep}(b). 
Fig. \ref{fig.shep}(a) shows four testing parameter points denoted by asterisks and their 6 nearest neighbor parameter points denoted by solid black dots.
The testing parameter points are at different locations relative to the convex hull (depicted by the black dash line) formed by the nearest neighbor points.
The interpolation functions are obtained using Eqs. (\ref{eq.shep_shape}-\ref{eq.shep_kern}) based on the distance between the testing points and the nearest neighbor points \RB{in the parameter space}, which are then used to interpolate the \RB{DI} coefficients of the testing points from the \RB{DI} coefficients of the nearest neighbors by using Eq. (\ref{eq.convex_interp}), as shown in Fig. \ref{fig.shep}(b). 
It can be seen that the interpolated \RB{DI} coefficients are all located within the convex hull (depicted by the black dash line) formed by the nearest neighbors' \RB{DI} coefficients, showing the desired convexity-preserving capability, which allows existing local DI models to be leveraged for prediction of latent-space dynamics of testing parameters. 
When the testing parameter point (the green asterisk) overlaps with the nearest neighbor point 5, the \RB{DI} coefficient of the testing point is identical to that of the nearest neighbor point 5.
% The convex interpolation is simple and efficient as the interpolation functions in Eq. (\ref{eq.shep_shape}) can be constructed easily in the parameter space.

\RB{\begin{remark}
The likelihood that the DI coefficient matrices ($\boldsymbol{\Xi}$) have a perfect negative correlation is very low. 
Further, the interpolation is weighted by the inverse of the distance between the nearest neighbors and the testing parameter point of interest, and the number of nearest neighbors involved in the interpolation is often greater than two to take the best advantage of the learned DI models. 
Therefore, a degenerate situation is not likely to happen.
\end{remark}}

% ===================================================================================================
\subsection{Physics-informed adaptive greedy sampling}\label{sec:greedy}
% As described in Section \ref{sec:dyn_ident}, a DI model captures local dynamics associated with a parameter $\boldsymbol{\mu} \in \mathbb{D}$. Such local dynamics could be different from that of a different parameter $\boldsymbol{\mu}' \in \mathbb{D}$. Constructing DI models at each newly queried parameter $\boldsymbol{\mu}'$ is computationally inefficient as it requires expensive computation of high-fidelity solution $\mathbf{U}$ at the parameter $\boldsymbol{\mu}'$.

% In the proposed algorithm, predefined parameter points in the parameter space for training are not required, but rather the training parameter points are sampled from the parameter space on-the-fly during training.
In the proposed algorithm, the training sample points are determined on the fly by a physics-informed adaptive greedy sampling algorithm to maximize parameter space exploration and achieve optimal model performance.
To this end, a sampling procedure integrated with an error indicator is needed. The most accurate error indicator would be actual relative error that is computed with high-fidelity solutions. However, requiring high-fidelity solution is computationally expensive, leading to undesirable training cost and time. Therefore, we adopt a physics-informed residual-based error indicator, which does not require high-fidelity solutions. The residual-based error indicator, defined in Eq.~\eqref{eq.residual_norm}, has a positive correlation with the maximum relative error and can be efficiently computed based only on predicted gLaSDI solutions. For example, see Figure~\ref{fig.glasdi}.
The physics-informed residual-based error indicator is integrated into an adaptive greedy sampling algorithm with a multi-level random-subset evaluation strategy.
Various termination criteria for the adaptive greedy sampling are discussed in the following subsection.

\subsubsection{Adaptive greedy sampling procedure}\label{sec:greedy_procedure}
To address the issues of parameter dependency of local latent-space dynamics efficiently and effectively, an adaptive greedy sampling procedure is applied to construct a database $\mathcal{DB}=\{\mathbf{U}^{(i)}\}_{i=1}^{N_{\mu}}$ on the fly during offline training, which corresponds to a set of sampled parameters $\mathbb{D}=\{\boldsymbol{\mu}^{(i)}\}_{i=1}^{N_{\mu}}$ from the discrete parameter space $\mathcal{D}^h$, $N_{\mu} < N_{\mathcal{D}} = |\mathcal{D}^h|$;
$\mathbf{U}^{(i)} = [\mathbf{u}_0^{(i)}, ..., \mathbf{u}_{N_t}^{(i)}] \in \mathbb{R}^{N_u \times (N_t + 1)}$ is the high-fidelity solution of the parameter $\boldsymbol{\mu}^{(i)}$. 
% $N_{\mathcal{DB}}$ denotes the size of the database $\mathcal{DB}$, equal to the number of sampled parameters, $|\mathbb{D}|$. 
% $N_{\mathcal{D}}$ is the total number of parameters in the \XH{discrete parameter space $\mathcal{D}^h$}. 
% The \XH{physics-informed adaptive} greedy sampling procedure is performed offline and \XH{yields} a \XH{small} database $\mathcal{DB}$ to satisfy a certain criterion, which will be discussed in the following subsection. 

The database is first initialized with a small set of parameters located, e.g., at the corners of the boundaries or at the center of the parameter space. 
To enhance sampling reliability and quality, the model training is performed before greedy sampling, as illustrated in Fig. \ref{fig.glasdi}. 
Therefore, the adaptive greedy sampling is performed after every $N_{up}$ epochs of training, which means a new training sample is added to the training database for every $N_{up}$ epochs.
At the $v$-th sampling iteration, a set of candidate parameters are considered and the parameter that maximizes an error indicator, $e \big( \mathbf{U}(\boldsymbol{\mu}), \hat{\mathbf{U}}(\boldsymbol{\mu}) \big)$,  
is selected. The definition of the error indicator is introduced in the next section. The greedy sampling procedure is summarized as
\begin{subequations}\label{eq.greedy_database}
    \begin{align}
        \boldsymbol{\mu}^* & = \underset{\boldsymbol{\mu} \in \mathbb{D}^{'}}{\operatorname{argmax}} \ 
        e \big( \mathbf{U}(\boldsymbol{\mu}), \hat{\mathbf{U}}(\boldsymbol{\mu}) \big), \\
        \mathcal{DB}_v & = \{\mathcal{DB}_{v-1}, \mathbf{U}^{*}) \}, \\
        \mathbb{D}_v & = \{\mathbb{D}_{v-1}, \boldsymbol{\mu}^* \},
    \end{align}
\end{subequations}
where $\mathbb{D}^{'} \subseteq \mathcal{D}^h$ denotes a set of $N^{'}\leq N_{\mathcal{D}}$ candidate parameters and $\mathbb{D}^{'} \cap \mathbb{D}_{v-1} = \emptyset$; $\mathbb{D}_{v-1}$ contains the parameters associated with $\mathcal{DB}_{v-1}$;
$\boldsymbol{\mu}^*$ denotes the selected parameter and $\mathbf{U}^*$ is the corresponding high-fidelity solution.
The iterations of greedy sampling continue until a certain criterion is reached, which will be discussed in the following subsection.
% either a prescribed tolerance of the error indicator or a maximum number of iterations is reached.

\subsubsection{Physics-informed residual-based error indicator}\label{sec:error_indicator}
Given an approximate gLaSDI solution, $\hat{\mathbf{U}}(\boldsymbol{\mu}) = [\hat{\mathbf{u}}_0 (\boldsymbol{\mu}), ..., \hat{\mathbf{u}}_{N_t} (\boldsymbol{\mu}) ]$, of the corresponding high-fidelity true solution, $\mathbf{U}(\boldsymbol{\mu}) = [\mathbf{u}_0 (\boldsymbol{\mu}), ..., \mathbf{u}_{N_t} (\boldsymbol{\mu})]$, 
the maximum relative error, $e^{max} \big( \mathbf{U}(\boldsymbol{\mu}), \hat{\mathbf{U}}(\boldsymbol{\mu}) \big)$ is defined as 
% \begin{equation}\label{eq.max_relative_error}
%     e^{max} \big( \mathbf{U}(\boldsymbol{\mu}), \hat{\mathbf{U}}(\boldsymbol{\mu}) \big) 
%     = \frac{\text{max}\{ ||\mathbf{u}_n(\boldsymbol{\mu}) - \hat{\mathbf{u}}_n(\boldsymbol{\mu})||_{L_2} \}_{n=0}^{Nt}}{\sum_{n=0}^{Nt}||\mathbf{u}_n(\boldsymbol{\mu})||_{L_2}/(N_t+1)}.
% \end{equation}
\begin{equation}\label{eq.max_relative_error}
    e^{max} \big( \mathbf{U}(\boldsymbol{\mu}), \hat{\mathbf{U}}(\boldsymbol{\mu}) \big) 
    = \underset{n \in \mathbb{N}_t}{\text{max}} \Bigg( \frac{||\mathbf{u}_n(\boldsymbol{\mu}) - \hat{\mathbf{u}}_n(\boldsymbol{\mu})||_{L_2}}{||\mathbf{u}_n(\boldsymbol{\mu})||_{L_2}} \Bigg).
\end{equation}
The maximum relative error as an error indicator provides the most accurate guidance to the greedy sampling procedure,
However, the evaluation of $e^{max} \big( \mathbf{U}(\boldsymbol{\mu}), \hat{\mathbf{U}}(\boldsymbol{\mu}) \big)$ is computationally inefficient because of the requirement of the high-fidelity true solution.
To ensure effective and efficient greedy sampling, the error indicator needs to satisfy the following criteria: 
(i) It must be positively correlated with the maximum relative error measure, as demonstrated in Fig. \ref{fig.glasdi}; 
(ii) The evaluation is computationally efficient, i.e., it must not involve any high-fidelity solution. 
A computationally feasible error indicator based on the residual of the governing equation is employed in this study, defined as
\begin{equation}\label{eq.residual_norm}
    e^{res} \big( \hat{\mathbf{U}}(\boldsymbol{\mu}) \big) 
    = \frac{1}{N_{ts}+1} \sum_{n=0}^{N_{ts}}||\mathbf{r}(\hat{\mathbf{u}}_n; \hat{\mathbf{u}}_{n-1}, \boldsymbol{\mu})||_{L_2}
\end{equation}
where the residual function $\mathbf{r}(\hat{\mathbf{u}}_n; \hat{\mathbf{u}}_{n-1}, \boldsymbol{\mu})$ is defined in Eq. (\ref{eq.residual}) and $N_{ts} < N_t$. The residual error indicator satisfies the aforementioned two conditions. For example, note that the evaluation of the error indicator requires only the predicted gLaSDI solutions and the fact that we use $N_{ts} < N_t$ further enhances computational efficiency of the error indicator. In this paper, $N_{ts}/N_t \approx 0.1$ is used. Furthermore, the adopted error indicator is positively correlated with the maximum relative error, which is demonstrated in Figure~\ref{fig.glasdi}. Note also that it is physics-informed as it is based on the residual of the discretized governing equations, which embeds physics (Eq. \RA{\eqref{eq.residual}}). 

Finally, the next parameter to be sampled is determined by
\begin{equation}\label{eq.optim_param}
    \boldsymbol{\mu}^* = \underset{\boldsymbol{\mu} \in \mathbb{D}^{'}}{\operatorname{argmax}} \ 
    e^{res} \big( \hat{\mathbf{U}}(\boldsymbol{\mu}) \big).
\end{equation}

\subsubsection{Termination criteria}\label{sec:termination_criteria}
% During training, the adaptive greedy sampling is performed after every $N_{up}$ epochs of training, which means a new training sample is added to the training set for every $N_{up}$ epochs.
The adaptive greedy sampling procedure is terminated until one of the following criteria is reached: 
(i) a prescribed maximum number of sampling points (local DI models), 
(ii) the maximum allowable training iterations, or
(iii) a prescribed target tolerance of the maximum relative error. 

As the training cost increases with the number of sampling points and the number of training iterations, criteria (i) and (ii) are considered if training efficiency is preferable. 
On the other hand, if one expects the optimal model performance, criterion (iii) is more suitable as it offers a guidance of the model accuracy in the parameter space. 

Since the maximum relative error is estimated by the residual-based error indicator, as described in Section \ref{sec:error_indicator}, criterion (iii) only provides an estimated model performance. To alleviate this issue, we exploit the ratio between the maximum relative error and the residual-based error indicator to approximate the correct target relative error. 
For example, at $v$-th sampling iteration,
the model is evaluated to obtain the maximum relative errors and the residual-based errors of all sampled parameters:
\begin{subequations}
    \begin{align}
        \mathbf{E}_v^{max} & = \{ e^{max} \big( \mathbf{U}(\boldsymbol{\mu}), \hat{\mathbf{U}}(\boldsymbol{\mu}) \big) \}_{\boldsymbol{\mu} \in \mathbb{D}_v}, \label{eq.max_rel_err_vector}\\
        \mathbf{E}_v^{res} & = \{ e^{res} \big( \hat{\mathbf{U}}(\boldsymbol{\mu}) \big) \}_{\boldsymbol{\mu} \in \mathbb{D}_v}, \label{eq.res_err_vector}
    \end{align}
\end{subequations}
where $e^{max} \big( \mathbf{U}(\boldsymbol{\mu}), \hat{\mathbf{U}}(\boldsymbol{\mu}) \big)$ and 
$e^{res} \big( \hat{\mathbf{U}}(\boldsymbol{\mu}) \big)$
are calculated from Eq. (\ref{eq.max_relative_error}) and Eq. (\ref{eq.residual_norm}), respectively. 
Note that $\mathbf{E}_v^{max}$ can be obtained because the database $\mathcal{DB}_{v}$ contains high-fidelity solutions of all sampled parameters in $\mathbb{D}_v$. 
Then, linear correlation coefficients $(k^*, b^*)$ between $\mathbf{E}_v^{res}$ and $\mathbf{E}_v^{max}$ are obtained by
\begin{equation}\label{eq.corr_eqn}
    (k^*, b^*) = \underset{k,b} \argmin
    || \mathbf{E}_v^{max} - \big( k \cdot \mathbf{E}_v^{res} + b \big) ||^2_{L_2}.
\end{equation}
Finally, the estimated maximum relative error is obtained by 
\begin{equation}\label{eq.est_emax}
    e_{v}^{max} = k^* \cdot \text{max}(\mathbf{E}_v^{res}) + b^*.
\end{equation}

As the correlation between $\mathbf{E}_v^{res}$ and $\mathbf{E}_v^{max}$ of all sampled parameters in $\mathbb{D}_v$ is used to estimate the maximum relative error $e_{v}^{max}$, it improves the termination guidance of the greedy sampling procedure in order to achieve the target $tol$, leading to more reliable reduced-order models. This  provides some level of confidence in the accuracy of the trained gLaSDI.

\subsubsection{Multi-level random-subset evaluation}\label{sec:greedy_subset}
To further accelerate the greedy sampling procedure \RB{and guarantee a desirable accuracy}, a two-level random-subset evaluation strategy is adopted in this study. \RB{At the $v$-th sampling iteration, a subset of parameters, $\mathbb{D}^{'} \subseteq \mathcal{D}^h$ ($\mathbb{D}^{'} \cap \mathbb{D}_{v-1} = \emptyset$), is randomly selected from the parameter space.} A small subset size $N_{subset} = |\mathbb{D}^{'}|$ is considered in the first-level random-subset selection so that the error evaluation of the parameters in the subset is efficient. When the tolerance of the error indicator $tol$ is reached the greedy sampling procedure moves to the second-level random-subset evaluation, where the subset size $N_{subset}$ doubles. \RB{A bigger subset size is chosen to increase the probability of achieving the desirable accuracy.} The greedy sampling procedure continues until the prescribed termination criterion is reached, see Algorithm~\ref{alg:glasdi_sampling} for more details.

% ===================================================================================================
\subsection{\RA{gLaSDI off-line and on-line stages}}\label{sec:algorithm}
The ingredients mentioned in earlier sections are integrated into the proposed greedy latent-space dynamics identification model. The training procedure of the gLaSDI model is summarized in Algorithm \ref{alg:glasdi_train}.
\begin{algorithm}
  \caption{Training of the gLaSDI model}\label{alg:glasdi_train}
  \textbf{Input}: 
  An initial parameter set $\mathbb{D}_0 \subseteq \mathcal{D}^h$ and 
  the associated database $\mathcal{DB}_0$; 
  an initial random subset size $N_{subset}$; 
  a greedy sampling frequency $N_{up}$; 
  the number of nearest neighbors $k$ for \RB{k-NN} convex interpolation; and
  one of the three terminal criteria:
  \begin{itemize}
      \item a target tolerance of the maximum relative error $tol$ \item the maximum number of sampling points $N_{\mu}^{max}$
      \item a maximum number of training epochs $N_{epoch}$
  \end{itemize}
  
  Note that $N_{epoch}$ is often used together with the error tolerance to avoid excessive training iterations in the case where the prescribed tolerance may not be achieved.
  
  \textbf{Output}: 
  gLaSDI sampled parameter set $\mathbb{D}_{v}$ and
  the associated database $\mathcal{DB}_{v}$;
  autoencoder parameters; $\boldsymbol{\theta}_{enc}$ and $\boldsymbol{\theta}_{dec}$;
  DI model coefficients 
  $\{\boldsymbol{\Xi}^{(i)}\}_{i\in \mathcal{N}_{\mathbb{D}_{v}}}$, where $\mathcal{N}_{\mathbb{D}_{v}}$ contains indices of parameters in $\mathbb{D}_{v}$
  
  \begin{algorithmic}[1]
        \State Set $v = 1$ \Comment{iteration counter}
        \State Set $w = 1$ \Comment{level counter for random-subset selection}
        \State Set epoch = 1 \Comment{training epoch counter}
        \While{epoch $\le N_{epoch}$} \Comment{gLaSDI training iterations}
            \State Update $\boldsymbol{\theta}_{enc}$, $\boldsymbol{\theta}_{dec}$, $\{\boldsymbol{\Xi}^{(i)}\}_{i\in \mathcal{N}_{\mathbb{D}_{v-1}}}$ 
            by minimizing the gLaSDI loss function in Eq.~\eqref{eq.total_loss} with $\mathcal{DB}_{v-1}$ and $\mathbb{D}_{v-1}$ 
            \If{epoch $\bmod N_{up}$ = 0} \Comment{greedy sampling}
                \State Obtain $\mathbb{D}_v, \mathcal{DB}_{v}, e_{v}^{max}, N_{subset}, w$ \Comment{algorithm \ref{alg:glasdi_sampling}}
                \State $v \gets v+1$
            \EndIf
            \If{$\big( e_{v}^{max} \le tol$ and $w = 2 \big)$  or $\big( |\mathbb{D}_{v}| > N_{\mu}^{max} \big)$ }
                \State \textbf{break} \Comment{terminate gLaSDI training}
            \EndIf
            \State epoch $\gets$ epoch + 1
        \EndWhile\label{epoch_loop}
        \State \textbf{return} $\mathbb{D}_v$, $\mathcal{DB}_{v}$, $\boldsymbol{\theta}_{enc}$, $\boldsymbol{\theta}_{dec}$, $\{\boldsymbol{\Xi}^{(i)}\}_{i\in \mathcal{N}_{\mathbb{D}_v}}$
  \end{algorithmic}
\end{algorithm}

\begin{algorithm}
  \caption{Greedy sampling with random-subset evaluation}\label{alg:glasdi_sampling}
  \textbf{Input}: 
  A parameter set $\mathbb{D}_{v-1} \subseteq \mathcal{D}^h$ and the associated database $\mathcal{DB}_{v-1}$; 
  a target tolerance of the maximum relative error $tol$; 
  a random-subset size $N_{subset}$;
  a random-subset level $w$; 
  the number of nearest neighbors $k$ for \RB{k-NN} convex interpolation
  
  \textbf{Output}: 
  updated parameter set $\mathbb{D}_{v}$; 
  database $\mathcal{DB}_{v}$;
  estimated maximum relative error $e_{v}^{max}$;
  random-subset size $N_{subset}$;
  random-subset level $w$
  
  \begin{algorithmic}[1]
        \State Select a random subset of parameters $\mathbb{D}^{'} \in\mathcal{D}^h$ with a size of $N_{subset}$
        % \State Set $\hat{\mathcal{DB}} = \{\}$ 
        \For{$\boldsymbol{\mu} \in \mathbb{D}^{'}$} \Comment{gLaSDI predictions}
            \State Obtain $\hat{\mathbf{U}}(\boldsymbol{\mu})$ from model evaluation \Comment{algorithm \ref{alg:glasdi_eval}}
            % \State $\hat{\mathcal{DB}} \gets \{\hat{\mathcal{DB}}, \hat{\mathbf{U}}(\boldsymbol{\mu})\}$
        \EndFor
        
        \State Compute $e^{res}(\hat{\mathbf{U}}(\boldsymbol{\mu})), \ \forall \boldsymbol{\mu} \in \mathbb{D}^{'}$ by Eq. (\ref{eq.residual_norm})
        \State Obtain $\boldsymbol{\mu}^{*}$ by solving Eq. (\ref{eq.optim_param})
        \State Obtain $\mathbf{U}(\boldsymbol{\mu}^{*})$ by solving Eq. (\ref{eq.govern_eqn}) numerically 
        \State $\mathcal{DB}_v \gets \{\mathcal{DB}_{v-1},
        \mathbf{U}(\boldsymbol{\mu}^{*})\}$\Comment{update database}
        \State $\mathbb{D}_v \gets \{\mathbb{D}_{v-1}, \boldsymbol{\mu}^{*}\}$\Comment{update the parameter set}
        \State Obtain $\{\hat{\mathbf{U}}(\boldsymbol{\mu})\}_{\boldsymbol{\mu}\in \mathbb{D}_v}$ \Comment{algorithm \ref{alg:glasdi_eval}}

        \State Compute $\mathbf{E}_v^{max}$ by Eq. (\ref{eq.max_rel_err_vector})
        \State Compute $\mathbf{E}_v^{res}$ by Eq. (\ref{eq.res_err_vector})
        % \State Obtain the correlation coefficient $(k^*,b^*)$ by solving Eq. (\ref{eq.corr_eqn})
        \State Compute $e_{v}^{max}$ by Eqs. (\ref{eq.corr_eqn})-(\ref{eq.est_emax})
        
        \If{$e_{v}^{max} \le tol$ and $w < 2$}
            \State $N_{subset} \gets 2 \times N_{subset}$ \Comment{update the random subset size}
            \State $w \gets w+1$
        \EndIf
    \State \textbf{return} $\mathbb{D}_v, \mathcal{DB}_{v}, e_{v}^{max}, N_{subset}, w$
  \end{algorithmic}
\end{algorithm}

% ===================================================================================================
% \subsection{gLaSDI on-line stage}
After the gLaSDI model is trained by Algorithm \ref{alg:glasdi_train}, the physics-informed greedy sampled parameter set $\mathbb{D}$, the autoencoder parameters, and a set of local DI model parameters are obtained. The trained gLaSDI model can then be applied to efficiently predict dynamical solutions given a testing parameter by using Algorithm \ref{alg:glasdi_eval}. The prediction accuracy can be evaluated by the maximum relative error with respect to high-fidelity solutions using Eq. (\ref{eq.max_relative_error}).

\begin{algorithm}
  \caption{Evaluation of the gLaSDI model}\label{alg:glasdi_eval}
  \textbf{Input}: 
  A testing parameter $\boldsymbol{\mu} \in \mathcal{D}^h$; 
  the gLaSDI sampled parameter set $\mathbb{D}$;
  the model coefficients $\boldsymbol{\theta}_{enc}$; $\boldsymbol{\theta}_{dec}$; $\{\boldsymbol{\Xi}^{(i)}\}_{i\in \mathcal{N}_{\mathbb{D}}}$; and 
  the number of nearest neighbors $k$ for \RB{k-NN} convex interpolation
  
  \textbf{Output}: 
  gLaSDI prediction $\hat{\mathbf{U}}(\boldsymbol{\mu})$
  
  \begin{algorithmic}[1]
        \State Search for \RB{k-NN} parameters based on the $L_2$ distance, $\mathcal{N}_k(\boldsymbol{\mu})$
        \State Compute \RB{k-NN} convex interpolation functions by Eqs. (\ref{eq.shep_shape}-\ref{eq.shep_kern})
        \State Obtain $\boldsymbol{\Xi}_{interp}$ for $\boldsymbol{\mu}$ by convex interpolation in Eq. (\ref{eq.convex_interp})
        \State Compute initial latent variables $\mathbf{z}_0 = \boldsymbol{\phi}_e(\mathbf{u}_0;\boldsymbol{\theta}_{\text{enc}})$ in Eq. (\ref{eq.autoencoder}a)
        \State Compute $\{\hat{\mathbf{z}}_n\}_{n=0}^{N_t}$ by Eq. (\ref{eq.dynamics_identification_discrete})
        \State Compute $\{\hat{\mathbf{u}}_n\}_{n=0}^{N_t}$ by Eq. (\ref{eq.autoencoder}b)
        % \State Obtain $\mathbf{u}(t;\boldsymbol{\mu})$ by solving Eq. (\ref{eq.backward_euler})
        % \State Evaluate $e^{max}(\mathbf{u}(t), \hat{\mathbf{u}}(t);\boldsymbol{\mu})$ by Eq. (\ref{eq.max_relative_error})
        \State $\hat{\mathbf{U}}(\boldsymbol{\mu}) \gets [\hat{\mathbf{u}}_0, ..., \hat{\mathbf{u}}_{N_t}]$
    \State \textbf{return} $\hat{\mathbf{U}}(\boldsymbol{\mu})$
  \end{algorithmic}
\end{algorithm}
\section{Numerical results}\label{sec:result}
The performance of gLaSDI is demonstrated by solving four numerical problems: one-dimensional (1D) Burgers' equation, two-dimensional (2D) Burgers' equation, nonlinear time-dependent heat conduction, and radial advection. The effects of the number of nearest neighbors $k$ on the model performance are discussed. 
In each of the numerical examples, the gLaSDI's performance is compared with that of LaSDI, that is the one without adaptive greedy sampling.
Note that the physics-informed adaptive greedy sampling can be performed on either a continuous parameter space ($\mathcal{D}$) or a discrete parameter space ($\mathcal{D}^h \subseteq \mathcal{D}$).
In the following examples, a discrete parameter space ($\mathcal{D}^h$) is considered. For \RA{all} numerical experiments, we always start with
the initial database $\mathcal{DB}_0$ and the associated parameter set $\mathbb{D}_0$ of four corner points of the parameter space. 

The training of gLaSDI are performed on a NVIDIA V100 (Volta) GPU of the Livermore Computing Lassen system at the Lawrence Livermore National Laboratory, with 3,168 NVIDIA CUDA Cores and 64 GB GDDR5 GPU Memory. 
The open-source TensorFlow library \cite{abadi2016tensorflow} and the Adam optimizer \cite{kingma2014adam} are employed for model training. 
\RB{The testing of gLaSDI and high-fidelity simulations are both performed on an IBM Power9 CPU with 128 cores and 3.5 GHz.}

% ===================================================================================================
\subsection{1D Burgers equation}\label{sec:result_1Dburger}
A 1D parameterized inviscid Burgers equation is considered
% \begin{subequations}\label{eq.1d_burger}
%     \begin{align}
%         \frac{\partial u(x,t;\boldsymbol{\mu})}{\partial t} + u(x,t;\boldsymbol{\mu}) \frac{\partial u(x,t;\boldsymbol{\mu})}{\partial x} & = 0, \quad x \in \Omega = [-3,3], \quad t \in [0,1], \\
%         u(3,t;\boldsymbol{\mu}) & = u(-3,t;\boldsymbol{\mu}).
%     \end{align}
% \end{subequations}
\begin{subequations}\label{eq.1d_burger}
    \begin{align}
        \frac{\partial u}{\partial t} + u \frac{\partial u}{\partial x} & = 0, \quad \Omega = [-3,3], \quad t \in [0,1], \\
        u(3,t;\boldsymbol{\mu}) & = u(-3,t;\boldsymbol{\mu}).
    \end{align}
\end{subequations}
Eq. (\ref{eq.1d_burger}b) is a periodic boundary condition. The initial condition is parameterized by the amplitude $a$ and the width $w$, defined as
\begin{equation}\label{eq.1d_burger_initial_condition}
    u(x,0;\boldsymbol{\mu}) = a e^{-\frac{x^2}{2w^2}},
\end{equation}
where $\boldsymbol{\mu} = \{a, w\}$. 
A uniform spatial discretization with 1,001 discrete points and nodal spacing as $dx=$6/1,000 is applied. The first order spatial derivative is approximated by the backward difference scheme. A semi-discertized system characterized by the ODE described in Eq. (\ref{eq.govern_eqn}) is obtained, which is solved by using the implicit backward Euler time integrator with a uniform time step of $\Delta t=$1/1,000 to obtain the full-order model solutions. \RB{Some solution snapshots are shown in \ref{appendix:1d_burger}.}

% ------------------------------------------------------------------------
\subsubsection{Case 1: Effects of the number of nearest neighbors k}\label{sec:result_1Dburger_case1}
In the first example, the effects of the number of nearest neighbors $k$ on model performance are investigated. 
The parameter space, $\mathcal{D}^h$, considered in this example is constituted by the parameters of the initial condition, including the width, $w \in [0.9,1.1]$, and the amplitude, $a \in [0.7,0.9]$, each with 21 evenly distributed discrete points in the respective parameter range, resulting in 441 parameter cases in total.
The gLaSDI model is composed of an autoencoder with an architecture of 1,001-100-5 and linear DI models. 
The greedy sampling is performed until the estimated maximum relative error of sampled parameter points is smaller than the target tolerance, $tol = 5\%$.
An initial random subset size $N_{subset}=64$ is used, around 20$\%$ of the size of $\mathcal{D}^h$. 
A two-level random-subset evaluation scheme is adopted, as described in Section \ref{sec:greedy_subset}. 
The maximum number of training epoch $N_{epoch}$ is set to be 50,000.

\paragraph{Adaptive greedy sampling with $k=1$} \hfill \break

\begin{figure}[htp]
    \centering
    \includegraphics[width=0.8\linewidth]{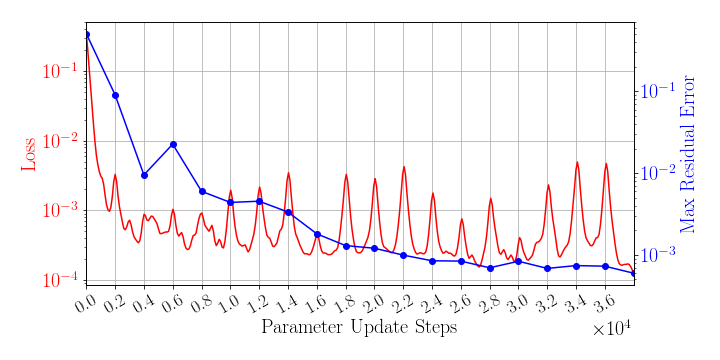}
    \caption{The history of the loss function and the maximum residual-based error of the sampled parameter points for the 1D Burgers problem. The \RB{k-NN} parameter, $k=1$, is used for greedy sampling procedure during training.}\label{fig.1d_burger_k1greedy_loss}
\end{figure}

The greedy sampling frequency $N_{up}$ is set to be 2,000. 
The training is performed by using Algorithm \ref{alg:glasdi_train}, where $k=1$ is used for greedy sampling procedure (Algorithm \ref{alg:glasdi_sampling}), which means the gLaSDI model utilizes the DI coefficient matrix of the existing parameter that is closest to the randomly selected parameter to perform dynamics predictions. 
Fig. \ref{fig.1d_burger_k1greedy_loss} shows the history of the loss function in a red solid line and the maximum residual-based error of the sampled parameter points in a blue solid line, demonstrating the convergence of the training. 
% The blue circles in Fig. \ref{fig.1d_burger_k1greedy_loss} indicate the epoch when greedy sampling is performed. 
In Fig. \ref{fig.1d_burger_k1greedy_loss}, the first blue point indicates the initial state of the model where four corner points of the parameter space are sampled, while the last blue point indicates the final state of the model that satisfies the prescribed termination criterion and no sampling is performed. 
\RB{The blue points in-between indicate the steps where new samples and associated DI models are introduced, corresponding to the spikes in the red curve (training loss history).}
At the end of the training, 22 parameter points are sampled and stored, including the initial 4 parameter points located at the four corners of the parameter space, which means 22 local DI models are constructed and trained in the gLaSDI model.

\begin{figure}[htp]
\centering
    \begin{subfigure}{0.495\textwidth}
        \centering
        \includegraphics[width=1\linewidth]{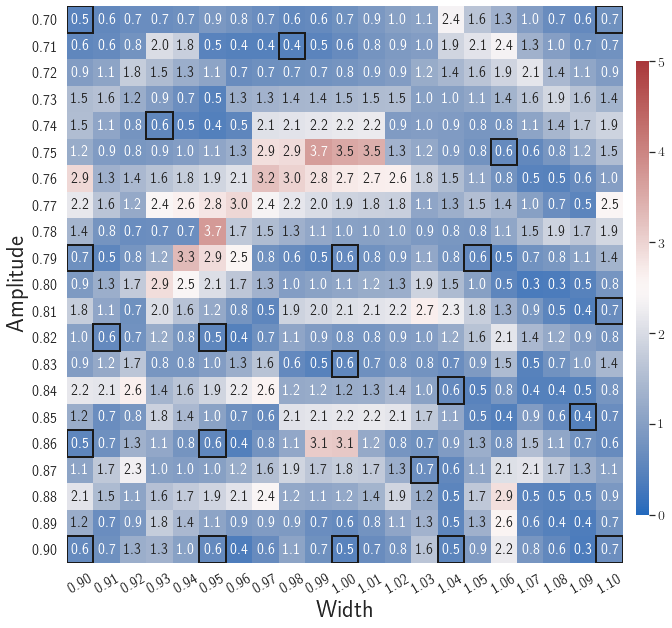}
        \caption{$k=1$ evaluation}
    \end{subfigure}
    \begin{subfigure}{0.495\textwidth}
        \centering
        \includegraphics[width=1\linewidth]{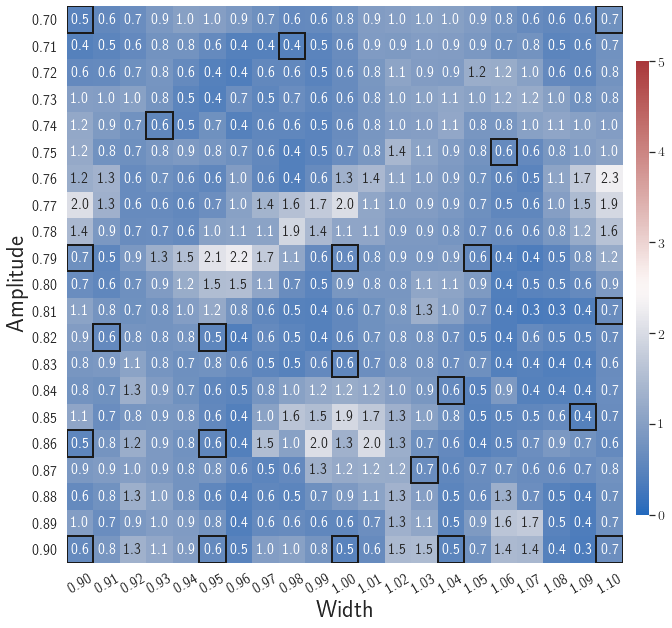}
        \caption{$k=3$ evaluation}
    \end{subfigure}
    \begin{subfigure}{0.495\textwidth}
        \centering
        \includegraphics[width=1\linewidth]{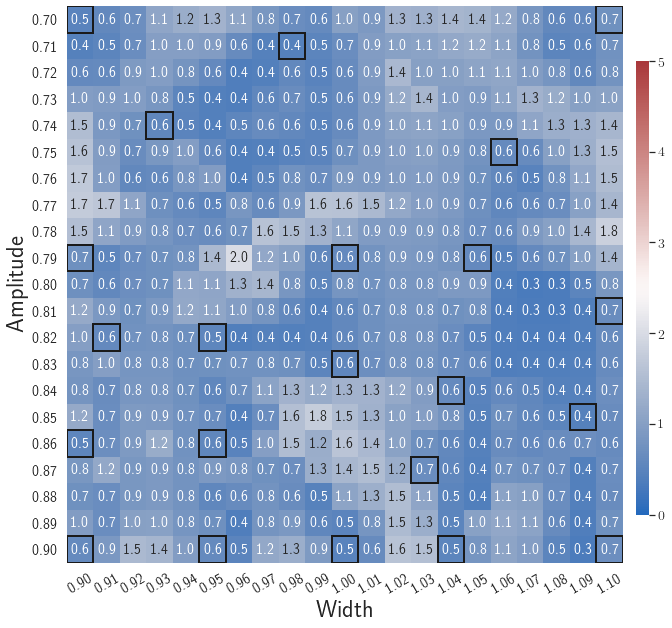}
        \caption{$k=4$ evaluation}
    \end{subfigure}
    \begin{subfigure}{0.495\textwidth}
        \centering
        \includegraphics[width=1\linewidth]{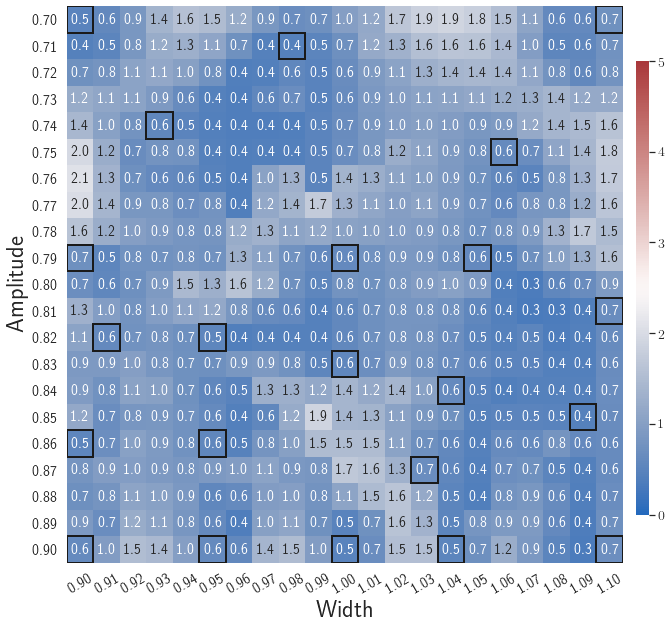}
        \caption{$k=5$ evaluation}
    \end{subfigure}
\caption{Maximum relative errors in the parameter space $\mathcal{D}^h$ evaluated with different values of $k$ for \RB{k-NN} convex interpolation of the DI coefficient matrices of the testing parameters for 1D Burgers problem: (a) $k=1$, (b) $k=3$, (c) $k=4$, (d) $k=5$. The number on each box denotes the maximum relative error of the associated parameter case. The black square boxes indicate the locations of the parameter points sampled from training. $k=1$ is used for greedy sampling procedure during training.}\label{fig.1d_burger_k1greedy_mre}
\end{figure}

After training, the gLaSDI model is applied for predictions by Algorithm \ref{alg:glasdi_eval}, where different values of $k$ are used for \RB{k-NN} convex interpolation of the DI coefficient matrix of the testing parameter. Fig. \ref{fig.1d_burger_k1greedy_mre} shows the maximum relative errors in the parameter space $\mathcal{D}^h$ evaluated with different values of $k$. 
The number on each box denotes the maximum relative error of the associated parameter case. 
The black square boxes indicate the locations of the sampled parameter points. 
The distance between the sampled parameter points located in the interior domain of $\mathcal{D}^h$ is relatively larger than that near the domain corners/boundaries. 
That means the interior sampled parameter points tend to have a larger \textit{trust region} within which the model prediction accuracy is high. 
It can also be observed that model evaluation with $k>1$ produces higher accuracy with around $2\%$ maximum relative error in $\mathcal{D}^h$, smaller than the target tolerance $tol = 5\%$, which is contributed by the \RB{k-NN} convex interpolation that exploits \textit{trust region} of local DI models. 
\RB{Compared with the high-fidelity simulation (in-house Python code) that has an around $2\%$ maximum relative error with respect to the high-fidelity data used for gLaSDI training, the gLaSDI model achieves 89$\times$ speed-up.
More information about the speed-up performance of gLaSDI for the 1D Burgers problem can be found in \ref{appendix:1d_burger:speedup}.}

\paragraph{Adaptive greedy sampling with $k=4$} \hfill \break

\begin{figure}[htp]
    \centering
    \includegraphics[width=0.8\linewidth]{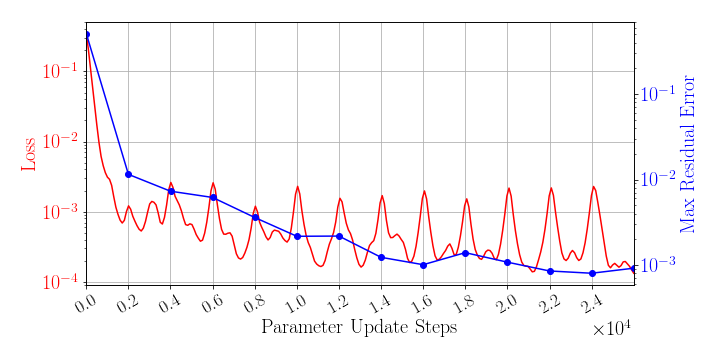}
    \caption{The history of the loss function and the maximum residual-based error of the sampled parameter for 1D Burgers problem. The \RB{k-NN} parameter $k=4$ is used for greedy sampling procedure during training.}\label{fig.1d_burger_k4greedy_loss}
\end{figure}

Fig. \ref{fig.1d_burger_k4greedy_loss} shows the history of the loss function and the maximum residual-based error of the gLaSDI model trained with $k=4$ for greedy sampling procedure. 
At the end of the training, 16 parameter points are sampled, including the initial four parameter points located at the 4 corners of the parameter space, which means 16 local DI models are constructed and trained in the gLaSDI model. 
Compared to the gLaSDI model trained with $k=1$ as in Fig. \ref{fig.1d_burger_k1greedy_loss}, the training with $k=4$ terminates faster with a sparser sampling to achieve the target tolerance of the maximum relative error, $tol = 5\%$, implying more efficient training. 
It is because a small $k$ for greedy sampling procedure leads to a more conservative gLaSDI model with more local DIs, while a large $k$ results in a more aggressive gLaSDI model with fewer local DIs as the trust region of local DI models are exploited by the \RB{k-NN} convex interpolation during training. 

\begin{figure}[htp]
\centering
    \begin{subfigure}{0.495\textwidth}
        \centering
        \includegraphics[width=1\linewidth]{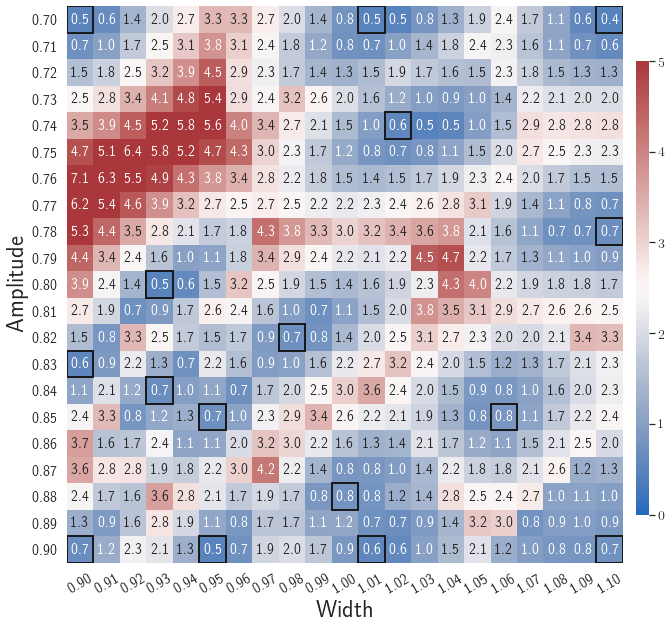}
        \caption{$k=1$ evaluation}
    \end{subfigure}
    \begin{subfigure}{0.495\textwidth}
        \centering
        \includegraphics[width=1\linewidth]{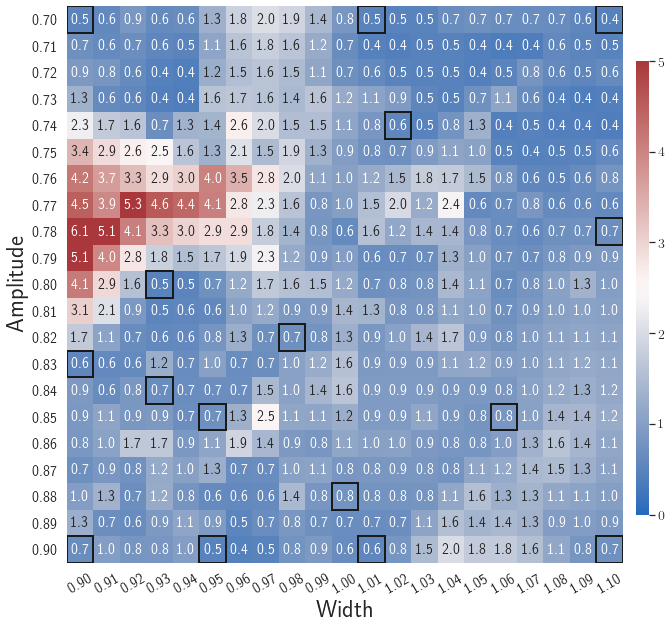}
        \caption{$k=3$ evaluation}
    \end{subfigure}
    \begin{subfigure}{0.495\textwidth}
        \centering
        \includegraphics[width=1\linewidth]{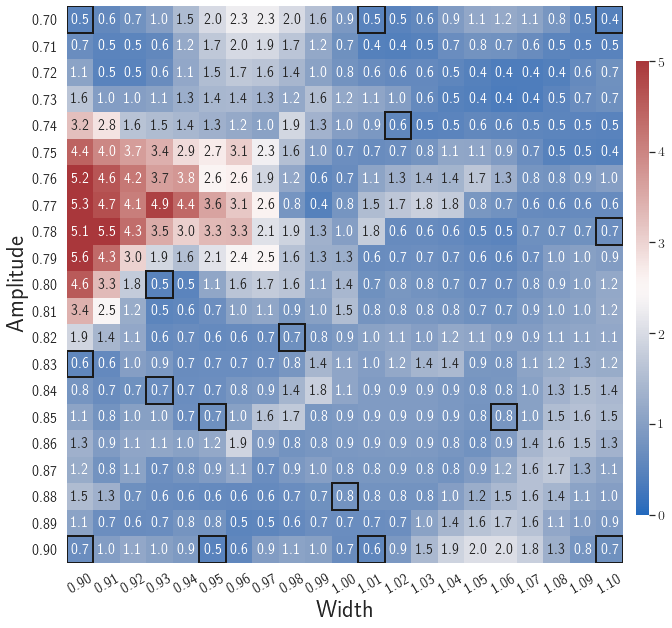}
        \caption{$k=4$ evaluation}
    \end{subfigure}
    \begin{subfigure}{0.495\textwidth}
        \centering
        \includegraphics[width=1\linewidth]{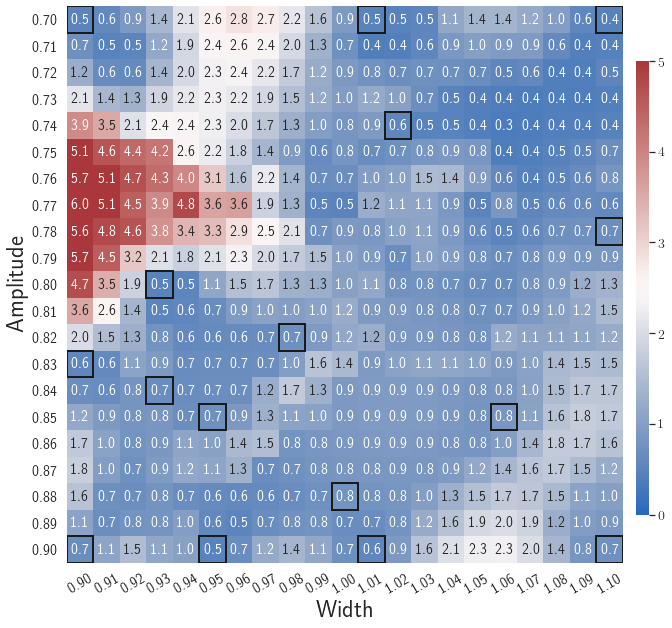}
        \caption{$k=5$ evaluation}
    \end{subfigure}
\caption{Maximum relative errors in the parameter space $\mathcal{D}^h$ evaluated with different values of $k$ for \RB{k-NN} convex interpolation of the DI coefficient matrices of the testing parameter for 1D Burgers problem: (a) $k=1$, (b) $k=3$, (c) $k=4$, (d) $k=5$. The number on each box denotes the maximum relative error of the associated parameter case. The black square boxes indicate the locations of the parameter points sampled from training. $k=4$ is used for greedy sampling procedure during training.}\label{fig.1d_burger_k4greedy_mre}
\end{figure}

Fig. \ref{fig.1d_burger_k4greedy_mre} shows the maximum relative errors in the parameter space $\mathcal{D}^h$ evaluated with different values of $k$. 
Similar to the results shown in Fig. \ref{fig.1d_burger_k1greedy_mre}, a larger $k$ results in higher model accuracy. 
It is noted that a few violations of the target tolerance $tol = 5\%$ exist even when $k>1$ is used for model evaluation. 
It shows that greedy sampling with $k>1$ results in a more aggressive gLaSDI model with fewer local DIs and higher training efficiency at the cost of model accuracy.
We also observed that the trained gLaSDI model achieved the best testing accuracy with $k=4$, which implies there exists a certain $k>1$ for optimal model performance.

The comparison of these two tests shows that a small $k$ for greedy sampling during training results in a more accurate gLaSDI model at the cost of training efficiency, and that using a $k>1$ for model evaluation (testing) improves generalization performance of gLaSDI.
In the following numerical examples, $k=1$ is used for greedy sampling procedure during training of gLaSDI. 
The trained gLaSDI models are evaluated by different $k>1$ and the optimal testing results are presented.

% \begin{itemize}
%     \item Interior local DIs tend to have larger \textit{trust region}, while local DIs closer to domain corners/boundaries tend to have smaller trust region.
%     \item Greedy training with $knn=1$ yields more accurate gLaSDI models than the ones obtained from using $knn>1$.
%     \item Model evaluation with $knn>1$ produces higher accuracy, which is contributed by the $knn$ convex interpolation that exploits \textit{trust region} of local DI models.
%     \item Given the same number of local DI models, gLaSDI models with $knn=1$ greedy training is more accurate than the ones with $knn>1$ greedy training
% \end{itemize}

% ------------------------------------------------------------------------
\subsubsection{Case 2: gLaSDI vs. LaSDI}\label{sec:result_1Dburger_case2}
In the second test, the autoencoder with an architecture of 1,001-100-5 and linear DI models are considered. 
% The parameter space considered in this example is constituted by the parameters of the initial condition, including the width, $w \in [0.9,1.1]$, and the amplitude, $a \in [0.7,0.9]$, each with 21 evenly distributed discrete points in the respective parameter range. 
% Thus, there are 441 parameter points in the parameter space $\mathcal{D}$. 
The same parameter space with $21 \times 21$ parameter cases in total is considered.
The gLaSDI training with adaptive greedy sampling is performed until the total number of sampled parameter points reaches 25. 
To investigate the effects of adaptive greedy sampling on model performances, a LaSDI model with the same architecture of the autoencoder and DI models is trained using 25 predefined training points uniformly distributed in a $5 \times 5$ grid in the parameter space. 
The performance of gLaSDI and LaSDI are compared and discussed. 
% The hyperparameters, e.g., learning rates and regularization parameters, are tuned to yield the best performance of each model. 

\begin{figure}[htp]
\centering
    \begin{subfigure}{0.9\textwidth}
        \centering
        \includegraphics[width=1\linewidth]{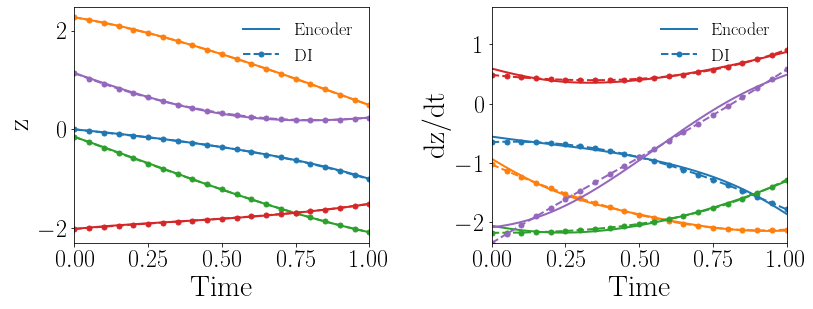}
        \caption{LaSDI}
    \end{subfigure}
    \begin{subfigure}{0.9\textwidth}
        \centering
        \includegraphics[width=1\linewidth]{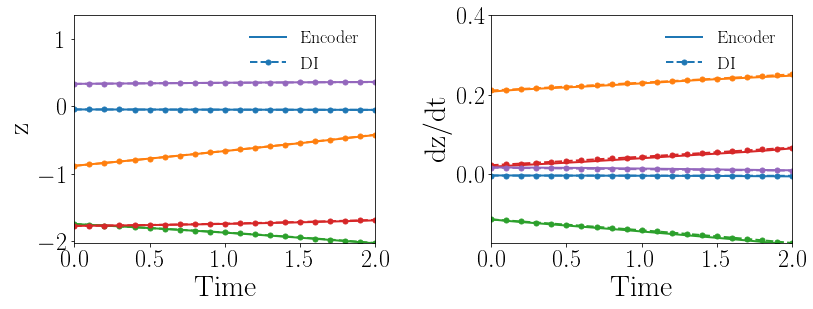}
        \caption{gLaSDI}
    \end{subfigure}
    \begin{subfigure}{0.495\textwidth}
        \centering
        \includegraphics[width=1\linewidth]{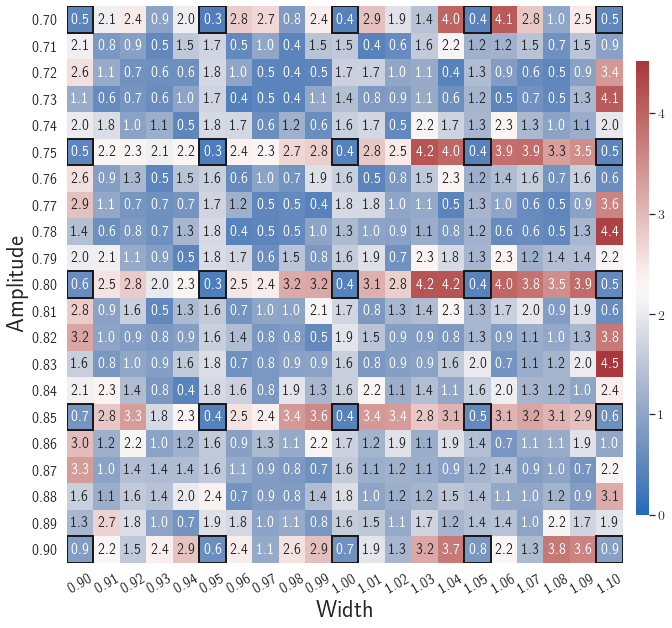}
        \caption{LaSDI}
    \end{subfigure}
    \begin{subfigure}{0.495\textwidth}
        \centering
        \includegraphics[width=1\linewidth]{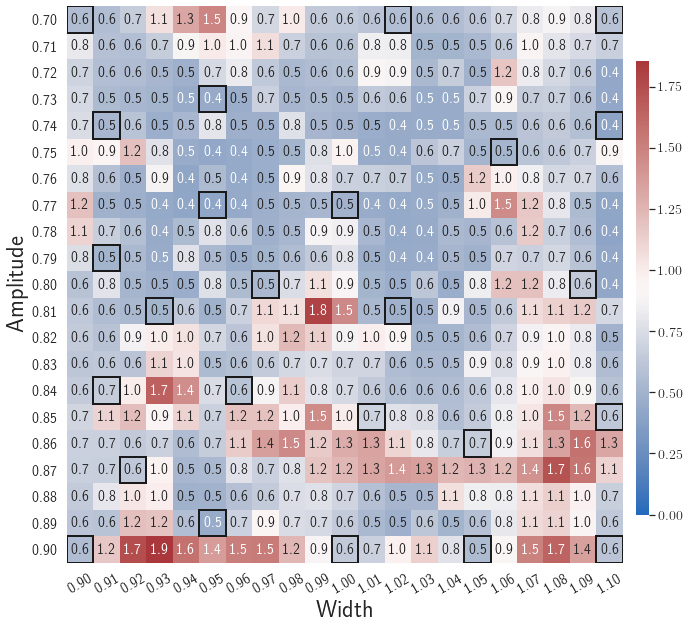}
        \caption{gLaSDI}
    \end{subfigure}
\caption{Comparison between LaSDI and gLaSDI with the same architecture of autoencoder (1,001-100-5) and dynamics identification models (linear) for 1D Burgers problem. The latent dynamics predicted by the trained encoder and the trained dynamics identification model from (a) LaSDI and (b) gLaSDI. The maximum relative errors in the parameter space $\mathcal{D}^h$ from (c) LaSDI with $k=4$ and (d) gLaSDI with $k=3$ for \RB{k-NN} convex interpolation during evaluation. The number on each box denotes the maximum relative error of the associated parameter case. The black square boxes indicate the locations of the sampled training parameter points. The \RB{k-NN} parameter $k=1$ is used for greedy sampling procedure during training of gLaSDI.}\label{fig.1d_burger_case2}
\end{figure}

Fig. \ref{fig.1d_burger_case2}(a-b) show the latent-space dynamics predicted by the trained encoder and the DI model from LaSDI and gLaSDI, respectively. The latent-space dynamics from gLaSDI is more linear and simpler than that from LaSDI. gLaSDI also achieves a better agreement between the encoder and the DI predictions, which is attributed by the interactive learning of gLaSDI. Note that the sequential training of the autoencoder and the DI models of the LaSDI could lead to more complicated and nonlinear latent-space dynamics because of the lack of interaction between the latent-space dynamics learned by the autoencoder and the DI models. As a consequence, it could pose challenges for the subsequent training of DI models to capture the latent-space dynamics learned by the autoencoder and cause higher prediction errors.

Fig. \ref{fig.1d_burger_case2}(c-d) show the maximum relative errors of LaSDI and gLaSDI predictions in the prescribed parameter space, respectively, where the black square boxes indicate the locations of the sampled parameter points. 
For LaSDI, the training parameter points are pre-selected and the associated high-fidelity solutions are obtained before training starts, whereas for gLaSDI, the training parameter points are sampled adaptively on the fly during the training, within which the greedy sampling algorithm is combined with the physics-informed residual-based error indicator, as introduced in Section \ref{sec:greedy}, which allows the points with the maximum error to be selected. Thus gLaSDI enhances the accuracy with less number of sampled parameter points than LaSDI.
It can be observed that gLaSDI tends to have denser sampling in the lower range of the parameter space.
Fig. \ref{fig.1d_burger_case2}(d) shows that gLaSDI achieves the maximum relative error of 1.9$\%$ in the whole parameter space, which is much lower than 4.5$\%$ of LaSDI in Fig. \ref{fig.1d_burger_case2}(c). 

% ------------------------------------------------------------------------
\subsubsection{Case 3: Effects of the loss terms} \label{sec:result_1Dburger_case3}
As mentioned in Sections~\ref{sec:autoencoder} and \ref{sec:dyn_ident}, there are three distinct terms in the loss function of the gLaSDI training. Each term includes a different set of trainable parameters. For example, the reconstruction term $\mathcal{L}_{recon}$ includes $\boldsymbol{\theta}_{\text{enc}}$ and $\boldsymbol{\theta}_{\text{dec}}$; the latent-space dynamics identification term $\mathcal{L}_{\dot{\mathbf{z}}}$ includes $\boldsymbol{\theta}_{\text{enc}}$ and $\{\boldsymbol{\Xi}^{(i)}\}_{i \in \mathbb{N}(N_{\mu})}$; and $\mathcal{L}_{\dot{\mathbf{u}}}$ includes all the trainable parameters, i.e., $\boldsymbol{\theta}_{\text{enc}}$,  $\boldsymbol{\theta}_{\text{dec}}$, and $\{\boldsymbol{\Xi}^{(i)}\}_{i \in \mathbb{N}(N_{\mu})}$.
Therefore, in this section, we demonstrate the effects of these loss terms.

\begin{figure}[htp]
\centering
    \begin{subfigure}{0.9\textwidth}
        \centering
        \includegraphics[width=1\linewidth]{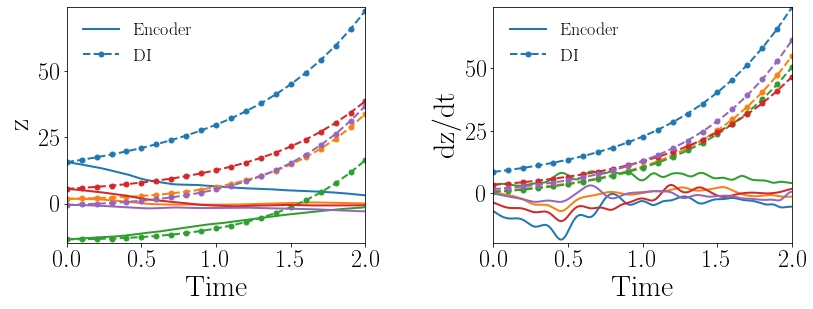}
        \caption{$\mathcal{L} = \mathcal{L}_{\dot{\mathbf{u}}}$}
    \end{subfigure}
    \begin{subfigure}{0.9\textwidth}
        \centering
        \includegraphics[width=1\linewidth]{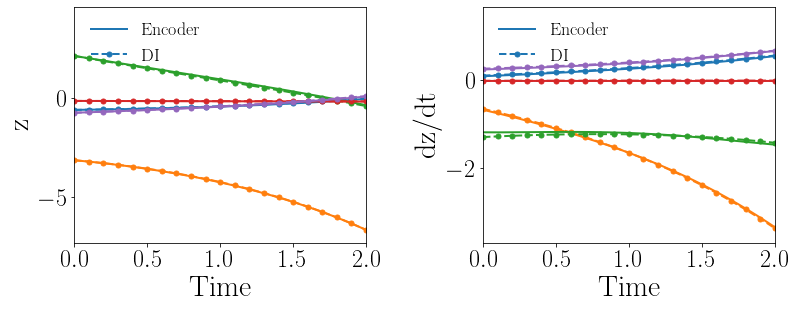}
        \caption{$\mathcal{L} = \mathcal{L}_{recon} + \beta_1 \mathcal{L}_{\dot{\mathbf{z}}}$}
    \end{subfigure}
    \begin{subfigure}{0.495\textwidth}
        \centering
        \includegraphics[width=1\linewidth]{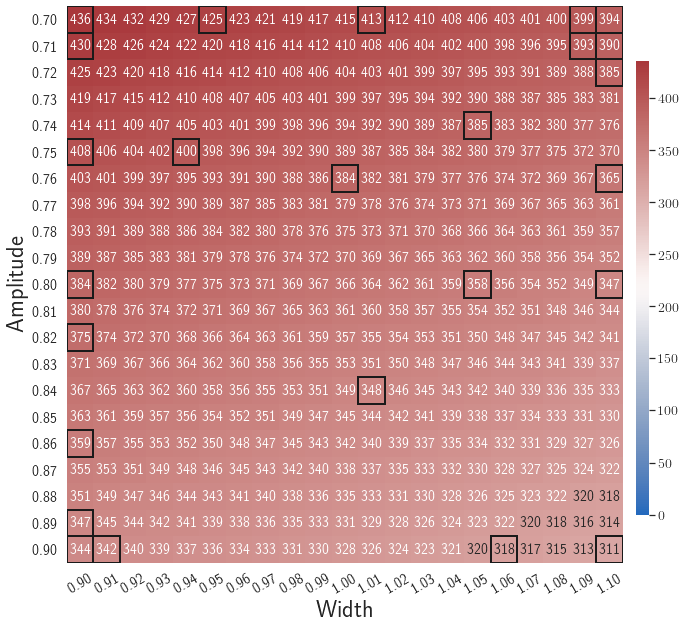}
        \caption{$\mathcal{L} = \mathcal{L}_{\dot{\mathbf{u}}}$}
    \end{subfigure}
    \begin{subfigure}{0.495\textwidth}
        \centering
        \includegraphics[width=1\linewidth]{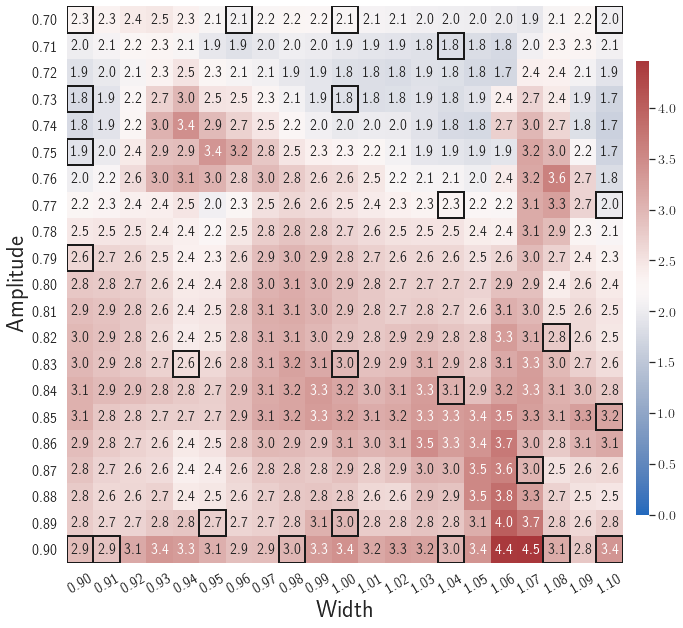}
        \caption{$\mathcal{L} = \mathcal{L}_{recon} + \beta_1 \mathcal{L}_{\dot{\mathbf{z}}}$}
    \end{subfigure}
\caption{Results of gLaSDI trained with different loss terms for 1D Burgers problem. An autoencoder of (1,001-100-5) and linear dynamics identification models are used. The latent dynamics predicted by the encoder and the dynamics identification model from the gLaSDI trained by (a) $\mathcal{L} = \mathcal{L}_{\dot{\mathbf{u}}}$ and (b) $\mathcal{L} = \mathcal{L}_{recon} + \beta_1 \mathcal{L}_{\dot{\mathbf{z}}}$, with $\beta_1=10^{-2}$. The maximum relative errors (evaluated with $k=4$) in the parameter space $\mathcal{D}^h$ from the gLaSDI trained by (c) $\mathcal{L} = \mathcal{L}_{\dot{\mathbf{u}}}$ and (d) $\mathcal{L} = \mathcal{L}_{recon} + \beta_1 \mathcal{L}_{\dot{\mathbf{z}}}$, with $\beta_1=10^{-2}$. The number on each box denotes the maximum relative error of the associated parameter case. The black square boxes indicate the locations of the sampled training parameter points. The \RB{k-NN} parameter $k=1$ is used for greedy sampling procedure during training of gLaSDI. }\label{fig.1d_burger_case3}
\end{figure}

Given the same settings of the gLaSDI model and the parameter space considered in Section \ref{sec:result_1Dburger_case2}, the gLaSDI is trained by only one loss term, $\mathcal{L}_{\dot{\mathbf{u}}}$ (Eq.~\eqref{eq.loss_udot}), that involves all the trainable parameters.
The predicted latent-space dynamics and maximum relative errors in the parameter space are shown in Fig. \ref{fig.1d_burger_case3}(a) and (c), respectively. 
The large deviation in the predicted latent-space dynamics between the encoder and the DI model leads to large prediction errors in the physical dynamics. 
It is also observed that the identified latent-space dynamics is more nonlinear, compared with that shown in Fig. \ref{fig.1d_burger_case2}(b). 
It implies that $\mathcal{L}_{\dot{\mathbf{u}}}$ cannot impose sufficient constraints on the trainable parameters to identify simple latent-space dynamics although it includes all the trainable parameters.

In the second test, the gLaSDI is trained by $\mathcal{L} = \mathcal{L}_{recon} + \beta_1 \mathcal{L}_{\dot{\mathbf{z}}}$ with different values of the regularization parameter $\beta_1$. 
The motivation is to see if we can achieve as good accuracy with only first two loss terms as the one with all three loss terms.
The maximum relative errors in the parameter space corresponding to different values of $\beta_1$ are summarized in Table \ref{table:effects_loss_term}. 
Using $\beta_1 = 10^{-2}$ yields the best model.
Compared with the latent-space dynamics of gLaSDI trained with all three loss terms, as shown in Fig. \ref{fig.1d_burger_case2}(b), the gLaSDI trained without $\mathcal{L}_{\dot{\mathbf{u}}}$ produces relatively more nonlinear latent-space dynamics, as shown in Fig. \ref{fig.1d_burger_case3}(b). 
The comparison also shows that the constraint imposed by $\mathcal{L}_{\dot{\mathbf{u}}}$ on model training enhances the generalization performance of gLaSDI, reducing the maximum relative error in the whole parameter space from 4.5$\%$ to 1.9$\%$, as shown in Fig. \ref{fig.1d_burger_case2}(d) and \ref{fig.1d_burger_case3}(d).

\begin{table}[!h]
\centering
\caption{The maximum relative errors in the parameter space of the gLaSDI trained by $\mathcal{L} = \mathcal{L}_{recon} + \beta_1 \mathcal{L}_{\dot{\mathbf{z}}}$ with different values of $\beta_1$.}
\begin{tabular}{ c c c c c c c c c } 
     \hline
     $\beta_1$ & $10^{-4}$ & $10^{-3}$ & $10^{-2}$ & $10^{-1}$ & $1$ & $10^{1}$ & $10^{2}$ & $10^{3}$ \\ 
     \hline
     $e^{max}$ ($\%$) & 11.5 & 8.5 & 4.5 & 4.7 & 4.6 & 14.5 & 20.9 & 23.8 \\ 
     \hline
\end{tabular}
\label{table:effects_loss_term}
\end{table}

% ===================================================================================================
\subsection{2D Burgers equation}\label{sec:result_2Dburger}
A 2D parameterized inviscid Burgers equation is considered
\begin{subequations}\label{eq.2d_burger}
    \begin{align}
        % \frac{\partial \mathbf{u}(\mathbf{x},t;\boldsymbol{\mu})}{\partial t} + \mathbf{u}(\mathbf{x},t;\boldsymbol{\mu}) \cdot \nabla \mathbf{u}(\mathbf{x},t;\boldsymbol{\mu}) & = \frac{1}{Re} \Delta \mathbf{u}(\mathbf{x},t;\boldsymbol{\mu}), \quad \mathbf{x} \in \Omega = [-3,3]\times[-3,3], \quad t \in [0,1], \\
        % \mathbf{u}(\mathbf{x},t;\boldsymbol{\mu}) & = \mathbf{0} \quad \text{on} \quad \partial\Omega.
        \frac{\partial \mathbf{u}}{\partial t} + \mathbf{u} \cdot \nabla \mathbf{u} & = \frac{1}{Re} \Delta \mathbf{u}, \quad \Omega = [-3,3]\times[-3,3], \quad t \in [0,1], \\
        \mathbf{u}(\mathbf{x},t;\boldsymbol{\mu}) & = \mathbf{0} \quad \text{on} \quad \partial\Omega,
    \end{align}
\end{subequations}
where $Re=10,000$ is the Reynolds number. Eq. (\ref{eq.2d_burger}b) is an essential boundary condition. The initial condition is parameterized by the amplitude $a$ and the width $w$, defined as
\begin{equation}\label{eq.2d_burger_initial_condition}
    \mathbf{u}(\mathbf{x},0;\boldsymbol{\mu}) = a e^{-\frac{||\mathbf{x}||^2}{w^2}},
\end{equation}
where $\boldsymbol{\mu} = \{a, w\}$. 
A uniform spatial discretization with $60 \times 60$ discrete points is applied. The first order spatial derivative is approximated by the backward difference scheme, while the diffusion term is approximated by the central difference scheme. 
The semi-discertized system is solved by using the implicit backward Euler time integrator with a uniform time step of $\Delta t=1/200$ to obtain the full-order model solutions.
\RB{Some solution snapshots are shown in \ref{appendix:2d_burger:snapshot}.}

% ------------------------------------------------------------------------
\subsubsection{Case 1: Comparison between gLaSDI and LaSDI}\label{sec:result_2Dburger_case1}
In the first test, a discrete parameter space $\mathcal{D}^h$ is constituted by the parameters of the initial condition, including the width, $w \in [0.9,1.1]$, and the amplitude, $a \in [0.7,0.9]$, each with 21 evenly distributed discrete points in the respective parameter range.
The autoencoder with an architecture of 7,200-100-5 and quadratic DI models are considered. 
The gLaSDI training is performed until the total number of sampled parameter points reaches 36. 
A LaSDI model with the same architecture of the autoencoder and DI models is trained using 36 predefined training points uniformly distributed in a $6 \times 6$ grid in the parameter space. 
The performance of gLaSDI and LaSDI are compared and discussed. 

\begin{figure}[htp]
\centering
    \begin{subfigure}{1\textwidth}
        \centering
        \includegraphics[width=0.8\linewidth]{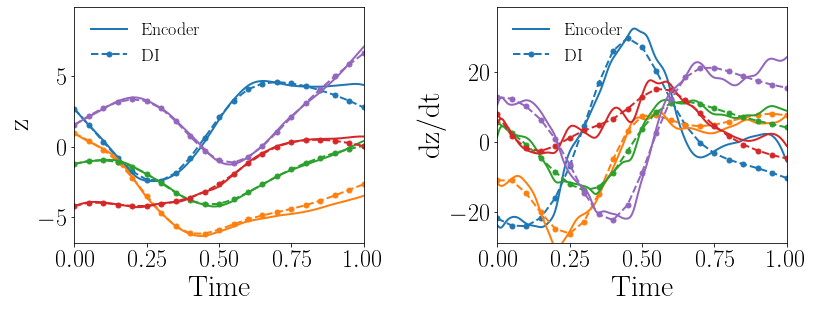}
        \caption{LaSDI}
    \end{subfigure}
    \begin{subfigure}{1\textwidth}
        \centering
        \includegraphics[width=0.8\linewidth]{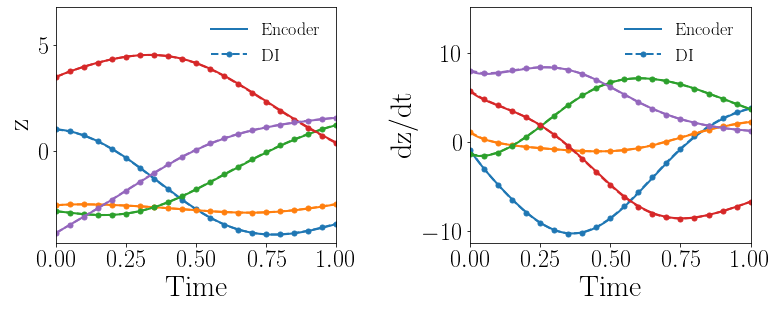}
        \caption{gLaSDI}
    \end{subfigure}
    \begin{subfigure}{0.495\textwidth}
        \centering
        \includegraphics[width=1\linewidth]{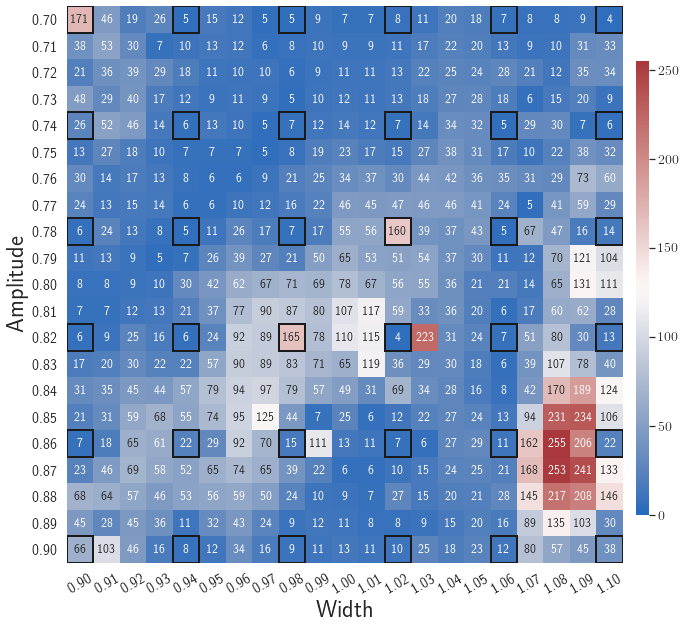}
        \caption{LaSDI}
    \end{subfigure}
    \begin{subfigure}{0.495\textwidth}
        \centering
        \includegraphics[width=1\linewidth]{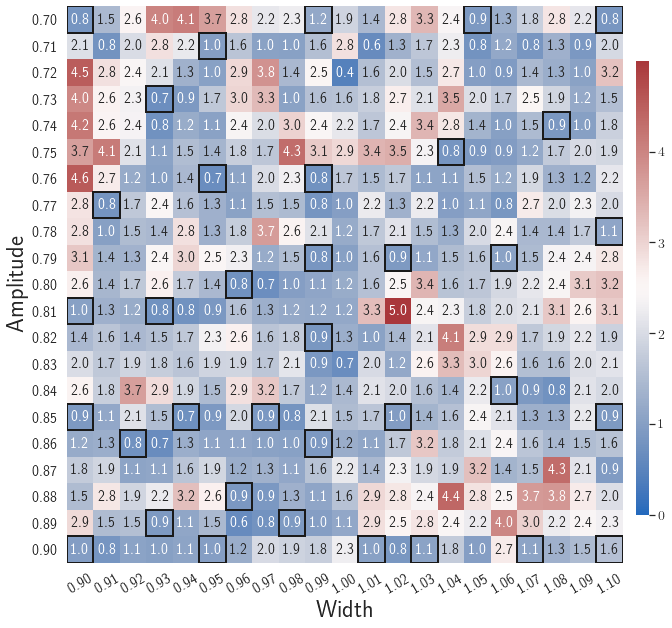}
        \caption{gLaSDI}
    \end{subfigure}
\caption{Comparison between LaSDI and gLaSDI with the same architecture of autoencoder (7,200-100-5) and dynamics identification models (\textbf{quadratic}) for 2D Burgers problem. The latent dynamics predicted by the trained encoder and the trained dynamics identification model from (a) LaSDI and (b) gLaSDI. The maximum relative errors in the parameter space $\mathcal{D}^h$ from (c) LaSDI with $k=4$ and (d) gLaSDI with $k=3$ for \RB{k-NN} convex interpolation during evaluation. The number on each box denotes the maximum relative error of the associated parameter case. The black square boxes indicate the location of the sampled training points. The \RB{k-NN} parameter $k=1$ is used for greedy sampling procedure for the training of gLaSDI.}\label{fig.2d_burger_case1}
\end{figure}

Fig. \ref{fig.2d_burger_case1}(a-b) show the latent-space dynamics predicted by the trained encoder and the DI model from LaSDI and gLaSDI, respectively. Again, the latent-space dynamics of gLaSDI is simpler than that of LaSDI, with a better agreement between the encoder and the DI predictions, which is attributed by the interactive learning of gLaSDI. 

Fig. \ref{fig.2d_burger_case1}(c-d) show the maximum relative error of LaSDI and gLaSDI predictions in the prescribed parameter space, respectively. 
The gLaSDI achieves the maximum relative error of 5$\%$ in the whole parameter space, much lower than 255$\%$ of LaSDI. 
The poor accuracy of LaSDI could be caused by the deviation between the DI predicted dynamics and the encoder predicted dynamics.
It is also observed that gLaSDI tends to have denser sampling in the lower range of the parameter space. This demonstrates the importance of the physics-informed greedy sampling procedure.
% Compared with the high-fidelity simulation based on an in-house Python code, the gLaSDI model achieves 1,740$\times$ speed-up.
\RB{Compared with the high-fidelity simulation (in-house Python code) that has an around $5\%$ maximum relative error with respect to the high-fidelity data used for gLaSDI training, the gLaSDI model achieves 871$\times$ speed-up.}

% ------------------------------------------------------------------------
\subsubsection{Case 2: Effects of the polynomial order in DI models and latent space dimension}\label{sec:result_2Dburger_case2}
In the second test, we want to see if simpler latent-space dynamics can be achieved by gLaSDI and how it affects the reduced-order modeling accuracy. The same parameter space with $21 \times 21$ parameter cases in total is considered.
The latent dimension is reduced from 5 to 3 and the polynomial order of the DI models is reduced from quadratic to linear. The autoencoder architecture becomes 7,200-100-3. 
The gLaSDI training is performed until the total number of sampled parameter \RA{points} reaches 36.
A LaSDI model with the same architecture of the autoencoder and DI models is trained using 36 predefined training parameter points uniformly distributed in a $6 \times 6$ grid in the parameter space.

\begin{figure}[htp]
\centering
    \begin{subfigure}{1\textwidth}
        \centering
        \includegraphics[width=0.8\linewidth]{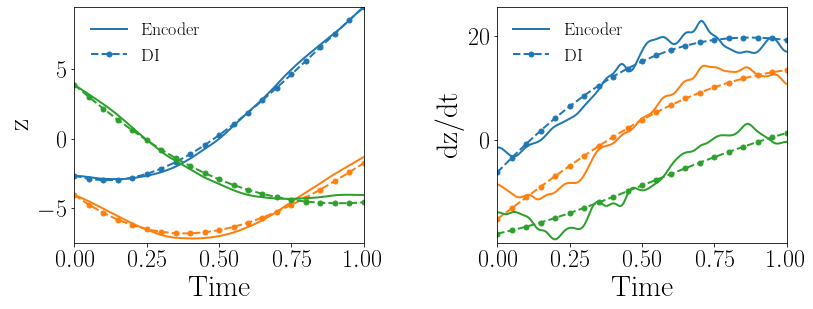}
        \caption{LaSDI}
    \end{subfigure}
    \begin{subfigure}{1\textwidth}
        \centering
        \includegraphics[width=0.8\linewidth]{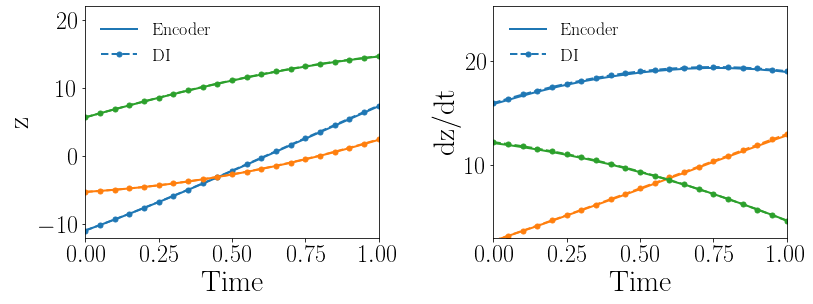}
        \caption{gLaSDI}
    \end{subfigure}
    \begin{subfigure}{0.495\textwidth}
        \centering
        \includegraphics[width=1\linewidth]{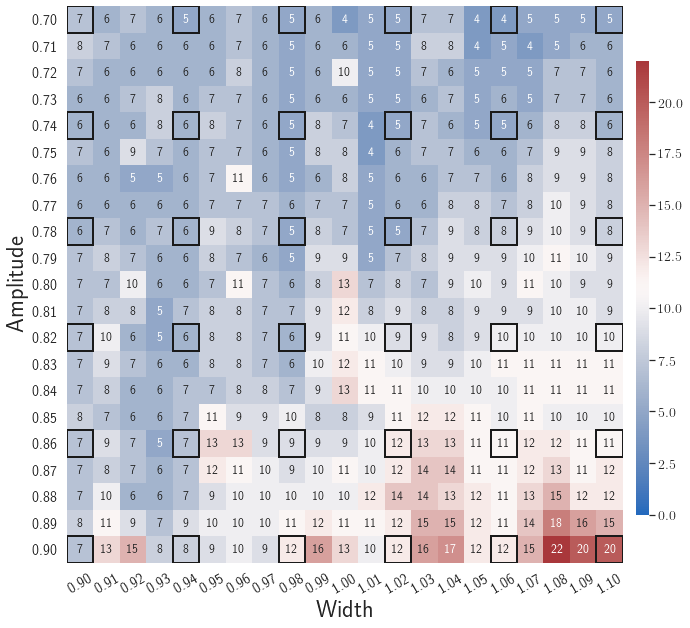}
        \caption{LaSDI}
    \end{subfigure}
    \begin{subfigure}{0.495\textwidth}
        \centering
        \includegraphics[width=1\linewidth]{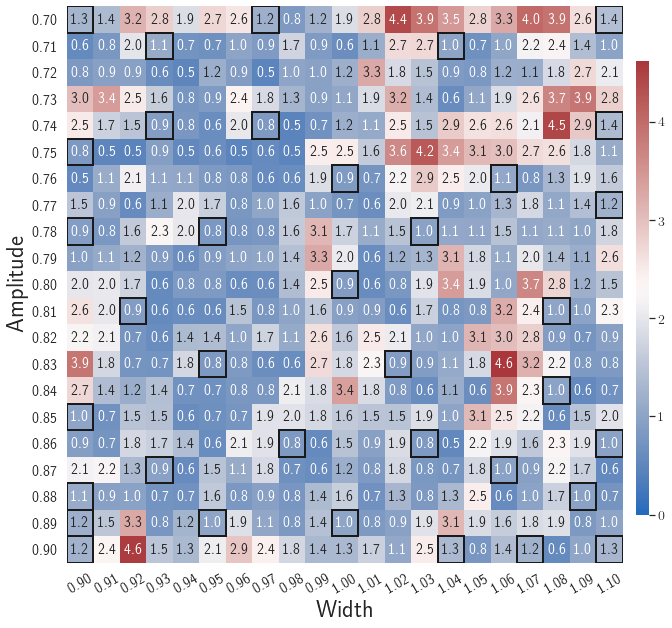}
        \caption{gLaSDI}
    \end{subfigure}
\caption{Comparison between LaSDI and gLaSDI with the same architecture of autoencoder (7,200-100-3) and dynamics identification models (\textbf{linear}) for 2D Burgers problem. The latent dynamics predicted by the trained encoder and the trained dynamics identification model from (a) LaSDI and (b) gLaSDI. The maximum relative errors in the parameter space $\mathcal{D}^h$ from (c) LaSDI with $k=3$ and (d) gLaSDI with $k=3$ for \RB{k-NN} convex interpolation during evaluation. The number on each box denotes the maximum relative error of the associated parameter case. The black square boxes indicate the location of the sampled training points. The \RB{k-NN} parameter $k=1$ is used for greedy sampling procedure during training of gLaSDI.}\label{fig.2d_burger_case2} 
\end{figure}

Fig. \ref{fig.2d_burger_case2}(a-b) show the latent-space dynamics predicted by the trained encoder and the DI model from LaSDI and gLaSDI, respectively.
The latent-space dynamics of gLaSDI is simpler than that of LaSDI, with a better agreement between the encoder and the DI predictions. 
It also demonstrates that gLaSDI could learn simpler latent-space dynamics, which enhances the reduced-order modeling efficiency. 
\RB{Compared with the high-fidelity simulation (in-house Python code) that has an around $5\%$ maximum relative error with respect to the high-fidelity data used for gLaSDI training, the gLaSDI model achieves 2,658$\times$ speed-up, which is 3.05 times the speed-up achieved by the gLaSDI model that has a latent dimension of 5 and quadratic DI models, as shown in Section \ref{sec:result_2Dburger_case1}.
More information about the speed-up performance of gLaSDI for the 2D Burgers problem can be found in \ref{appendix:2d_burger:speedup}.}

Fig. \ref{fig.2d_burger_case2}(c-d) show the maximum relative error of LaSDI and gLaSDI predictions on each parameter case in the prescribed parameter space, respectively. 
Compared with the example with a latent dimension of five and quadratic DI models, as shown in the previous subsection, both gLaSDI and LaSDI achieve lower maximum relative errors, reduced from 5.0$\%$ to 4.6$\%$ and from 255$\%$ to 22$\%$, respectively. 
It indicates that reducing the complexity of the latent-space dynamics allows the DI models of LaSDI to capture the encoder predicted dynamics more accurately although the error level of LaSDI is still larger than that of gLaSDI due to no interactions between the autoencoder and DI models during training.
Simplifying latent-space dynamics results in higher reduced-order modeling accuracy of both LaSDI and gLaSDI for this example. We speculate that the intrinsic latent space dimension for the 2D Burgers problem is close to three.
\RA{More results and discussion about the effects of latent dimension on model performance can be found in \ref{appendix:2d_burger:latent}.}

% ===================================================================================================
\subsection{Nonlinear time-dependent heat conduction}\label{sec:result_heat}
A 2D parameterized nonlinear time-dependent heat conduction problem is considered
\begin{subequations}\label{eq.heat_conduction}
    \begin{align}
        \frac{\partial u}{\partial t} & = \nabla \cdot (\kappa + \alpha u) \nabla u, \quad \Omega = [0,1]\times[0,1], \quad t \in [0,0.3], \\
        \frac{\partial u}{\partial \mathbf{n}} & = \mathbf{0} \quad \text{on} \quad \partial\Omega,
    \end{align}
\end{subequations}
where $\mathbf{n}$ denotes the unit normal vector. Eq. (\ref{eq.heat_conduction}b) is a natural insulating boundary condition. The coefficients $\kappa=0.5$ and $\alpha=0.01$ are adopted in the following examples. The initial condition is parameterized as
\begin{equation}\label{eq.heat_conduction_initial_condition}
    u(\mathbf{x},0;\boldsymbol{\mu}) = a \text{sin}(w||\mathbf{x}||_{L_2}) + a,
\end{equation}
where $\boldsymbol{\mu} = \{a, w\}$ denotes the \RA{parameters} of the initial condition. 
The spatial domain is discretized by first-order square finite elements constructed on a uniform grid of $33 \times 33$ discrete points. 
The implicit backward Euler time integrator with a uniform time step of $\Delta t = 0.005$ is employed. 
The conductivity coefficient is computed by linearizing the problem with the temperature field from the previous time step.
\RB{Some solution snapshots are shown in \ref{appendix:heat:snapshot}.}
\RA{Apart from the parameterization of the initial condition, the effectiveness of gLaSDI on the parameterization of the governing equations \eqref{eq.heat_conduction} is also demonstrated in \ref{appendix:heat:pde_parameterization}.}

% ------------------------------------------------------------------------
\subsubsection{Case 1: Comparison between gLaSDI and LaSDI}\label{sec:result_heat_case1}
In the first test, a parameter space $\mathcal{D}^h$ is constituted by the parameters of the initial condition, including the $w \in [4.0, 4.3]$ and $a \in [1.0, 1.4]$, each with 21 evenly distributed discrete points in the respective parameter range.
The autoencoder with an architecture of 1,089-100-3 and quadratic DI models are considered. 
The gLaSDI training is performed until the total number of sampled parameter points reaches 25. 
A LaSDI model with the same architecture of the autoencoder and DI models is trained using 25 predefined training parameter points uniformly distributed in a $5 \times 5$ grid in the parameter space. 
The performances of gLaSDI and LaSDI are compared and discussed. 

\begin{figure}[htp]
\centering
    \begin{subfigure}{1\textwidth}
        \centering
        \includegraphics[width=0.8\linewidth]{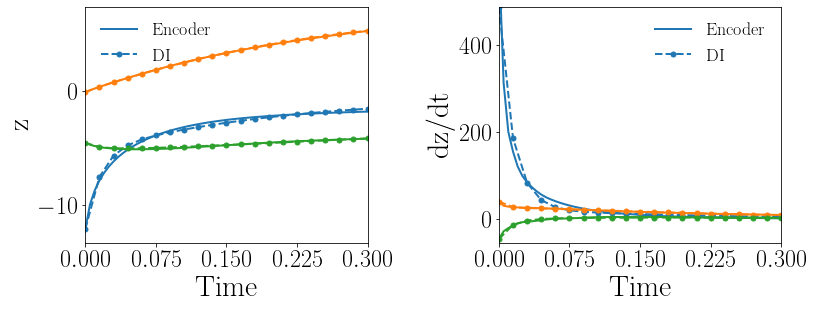}
        \caption{LaSDI}
    \end{subfigure}
    \begin{subfigure}{1\textwidth}
        \centering
        \includegraphics[width=0.8\linewidth]{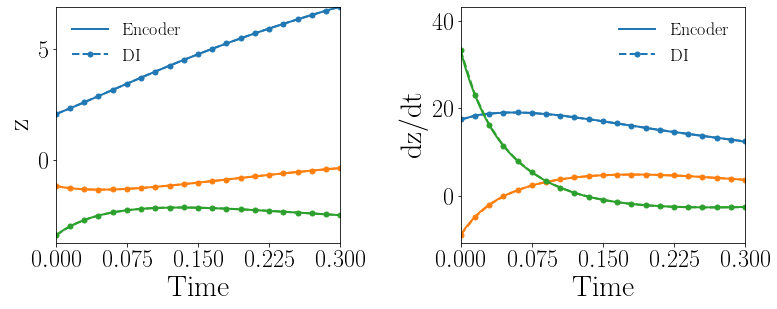}
        \caption{gLaSDI}
    \end{subfigure}
    \begin{subfigure}{0.495\textwidth}
        \centering
        \includegraphics[width=1\linewidth]{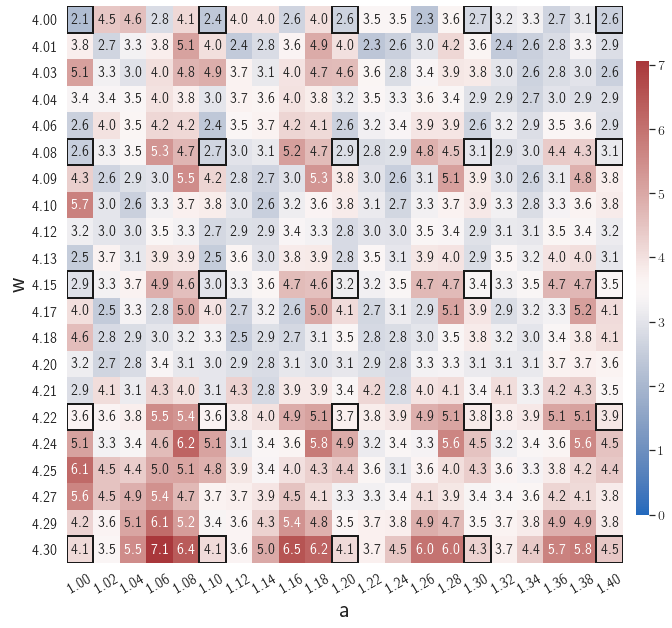}
        \caption{LaSDI}
    \end{subfigure}
    \begin{subfigure}{0.495\textwidth}
        \centering
        \includegraphics[width=1\linewidth]{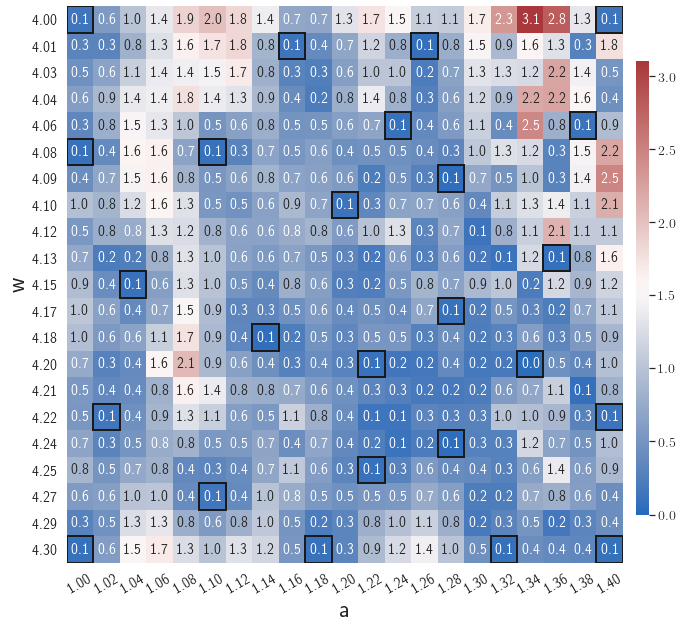}
        \caption{gLaSDI}
    \end{subfigure}
\caption{Comparison between LaSDI and gLaSDI with the same architecture of autoencoder (1,089-100-3) and dynamics identification models (\textbf{quadratic}) for the nonlinear time dependent heat conduction problem. The latent dynamics predicted by the trained encoder and the trained dynamics identification model from (a) LaSDI and (b) gLaSDI. The maximum relative errors in the parameter space $\mathcal{D}^h$ from (c) LaSDI with $k=4$ and (d) gLaSDI with $k=3$ for \RB{k-NN} convex interpolation during evaluation. The number on each box denotes the maximum relative error of the associated parameter case. The black square boxes indicate the locations of the sampled training points. The \RB{k-NN} parameter $k=1$ is used for greedy sampling procedure during training of gLaSDI.}\label{fig.heat_case1}
\end{figure}

Fig. \ref{fig.heat_case1}(a-b) show the latent-space dynamics predicted by the trained encoder and the DI model from LaSDI and gLaSDI, respectively. Again, gLaSDI achieves a better agreement between the encoder and the DI prediction than the ones for LaSDI.

Fig. \ref{fig.heat_case1}(c-d) show the maximum relative error of LaSDI and gLaSDI predictions in the prescribed parameter space, respectively. The gLaSDI achieves higher prediction accuracy than the LaSDI with the maximum relative error of 3.1$\%$ in the whole parameter space, compared to 7.1$\%$ of LaSDI. 
% Compared with the high-fidelity simulation based on MFEM \cite{anderson2021mfem}, the gLaSDI model achieves 66$\times$ speed-up.
\RB{Compared with the high-fidelity simulation (MFEM \cite{anderson2021mfem}) that has an around $3\%$ maximum relative error with respect to the high-fidelity data used for gLaSDI training, the gLaSDI model achieves 17$\times$ speed-up.}

% ------------------------------------------------------------------------
\subsubsection{Case 2: Effects of the polynomial order in DI models}\label{sec:result_heat_case2}
In the second test, we want to see if simpler latent-space dynamics can be achieved by gLaSDI and how it affects the reduced-order modeling accuracy. The same settings as the previous example are considered, including the parameter space, the autoencoder architecture (1,089-100-3), the number of training points (25), except that the polynomial order of the DI models is reduced from quadratic to linear. 

\begin{figure}[htp]
\centering
    \begin{subfigure}{1\textwidth}
        \centering
        \includegraphics[width=0.8\linewidth]{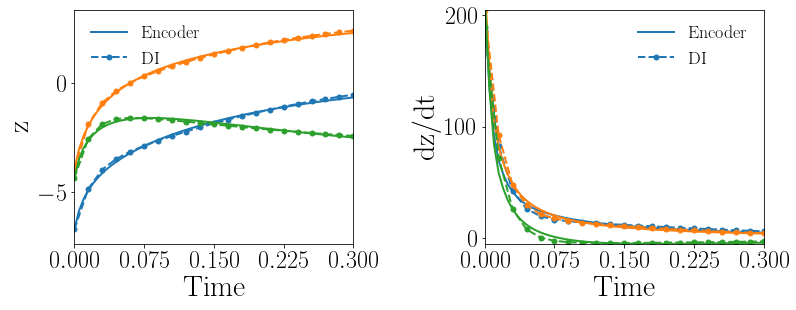}
        \caption{LaSDI}
    \end{subfigure}
    \begin{subfigure}{1\textwidth}
        \centering
        \includegraphics[width=0.8\linewidth]{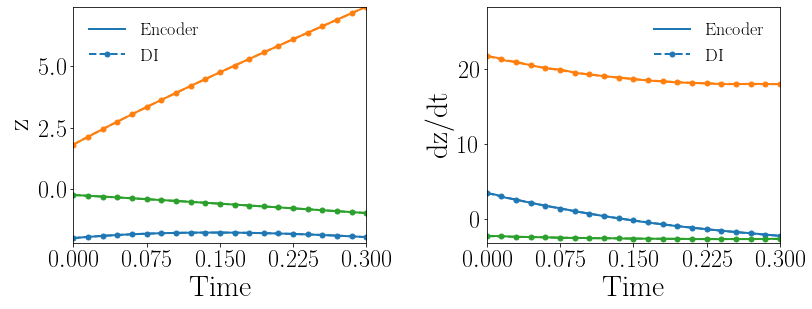}
        \caption{gLaSDI}
    \end{subfigure}
    \begin{subfigure}{0.495\textwidth}
        \centering
        \includegraphics[width=1\linewidth]{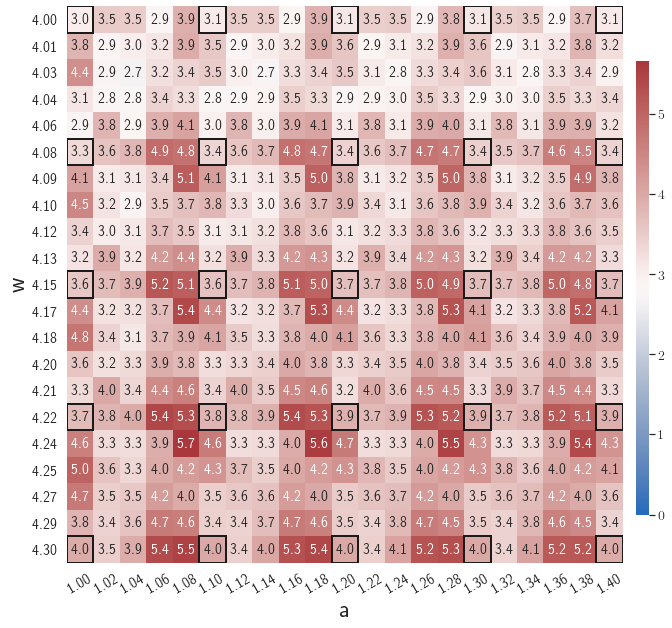}
        \caption{LaSDI}
    \end{subfigure}
    \begin{subfigure}{0.495\textwidth}
        \centering
        \includegraphics[width=1\linewidth]{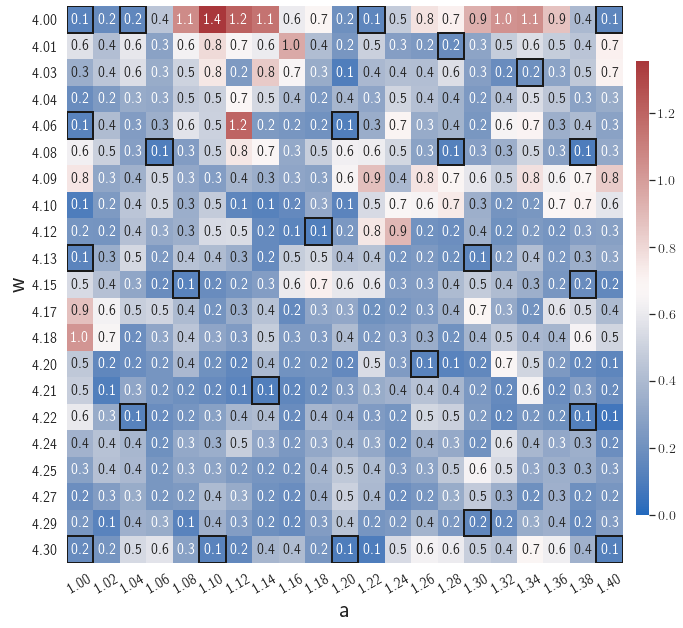}
        \caption{gLaSDI}
    \end{subfigure}
\caption{Comparison between LaSDI and gLaSDI with the same architecture of autoencoder (1,089-100-3) and dynamics identification models (\textbf{linear}) for the nonlinear time dependent heat conduction problem. The latent dynamics predicted by the trained encoder and the trained dynamics identification model from (a) LaSDI and (b) gLaSDI. The maximum relative errors in the parameter space $\mathcal{D}^h$ from (c) LaSDI with $k=4$ and (d) gLaSDI with $k=4$ for \RB{k-NN} convex interpolation during evaluation. The number on each box denotes the maximum relative error of the associated parameter case. The black square boxes indicate the locations of the sampled training points. The \RB{k-NN} parameter, $k=1$ is used for greedy sampling procedure during training of gLaSDI.}\label{fig.heat_case2}
\end{figure}

Fig. \ref{fig.heat_case2}(a-b) show the latent-space dynamics predicted by the trained encoder and the DI model from LaSDI and gLaSDI, respectively. 
Compared with the LaSDI's latent-space dynamics in the previous example, as shown in Fig. \ref{fig.heat_case1}(a-b), the agreement between the encoder and the DI predictions in this example improves, although not as good as that of gLaSDI. 
The dynamics learned by gLaSDI is simpler than the previous example with quadratic DI models. 
% Compared with the high-fidelity simulation based on MFEM \cite{anderson2021mfem}, the gLaSDI model achieves 205$\times$ speed-up, which is 3.11 times of the speed-up achieved by the gLaSDI model that has quadratic DI models, as shown in Section \ref{sec:result_heat_case1}.
\RB{Compared with the high-fidelity simulation (MFEM \cite{anderson2021mfem}) that has a similar maximum relative error with respect to the high-fidelity data used for gLaSDI training, the gLaSDI model achieves 58$\times$ speed-up, which is 3.37 times the speed-up achieved by the gLaSDI model that has quadratic DI models, as shown in Section \ref{sec:result_heat_case1}}
It further demonstrates that gLaSDI allows learning simpler latent-space dynamics, which could enhance the reduced-order modeling efficiency.

Fig. \ref{fig.heat_case2}(c-d) show the maximum relative error of LaSDI and gLaSDI predictions in the prescribed parameter space, respectively. 
Compared with the example with quadratic DI models, as shown in the previous subsection, the maximum relative error achieved by gLaSDI is reduced from 3.1$\%$ to 1.4$\%$, while that achieved by LaSDI is reduced from 7.1$\%$ to 5.7$\%$. Simplifying latent-space dynamics contributes to higher reduced-order modeling accuracy of both LaSDI and gLaSDI in this example. This implies that the higher order in DI model does not always help improving the accuracy.

% ===================================================================================================
\subsection{Time-dependent radial advection}\label{sec:result_advection}
A 2D parameterized time-dependent radial advection problem is considered
\begin{subequations}\label{eq.advection}
    \begin{align}
        \frac{\partial u}{\partial t} + \mathbf{v} \cdot \nabla  u & = 0, \quad \Omega = [-1,1]\times[-1,1], \quad t \in [0,3], \\
        u(\mathbf{x},t;\boldsymbol{\mu}) & = 0 \quad \text{on} \quad \partial\Omega,
    \end{align}
\end{subequations}
where Eq. (\ref{eq.advection}b) is a boundary condition and $\mathbf{v}$ denotes the fluid velocity, defined as
\begin{equation}\label{eq.advection_velocity}
    \mathbf{v} = \frac{\pi}{2} d [x_2, -x_1]^T,
\end{equation}
with $d = (1-x_1^2)^2(1-x_2^2)^2$.
The initial condition is defined as
\begin{equation}\label{eq.advection_initial_condition}
    u(\mathbf{x},0;\boldsymbol{\mu}) = \text{sin}(w_1 x_1) \text{sin}(w_2 x_2),
\end{equation}
where $\boldsymbol{\mu} = \{w_1, w_2\}$ denotes the paraemeters of the initial condition. 
The spatial domain is discretized by first-order periodic square finite elements constructed on a uniform grid of $96 \times 96$ discrete points. 
The fourth-order Runge-Kutta explicit time integrator with a uniform time step of $\Delta t = 0.01$ is employed. 
\RB{Some solution snapshots are shown in \ref{appendix:advection}.}

% ------------------------------------------------------------------------
\subsubsection{Case 1: Comparison between gLaSDI and LaSDI}
In the first test, a parameter space $\mathcal{D}^h$ is constituted by the parameters of the initial condition, including the $w_1 \in [1.5, 1.8]$ and $w_2 \in [2.0, 2.3]$, each with 21 evenly distributed discrete points in the respective parameter range.
The autoencoder with an architecture of 9,216-100-3 and linear DI models are considered. 
The gLaSDI training is performed until the total number of sampled parameter points reaches 25. 
A LaSDI model with the same architecture of the autoencoder and DI models is trained using 25 predefined training points uniformly distributed in a $5 \times 5$ grid in the parameter space. 
The performances of gLaSDI and LaSDI are compared and discussed. 

\begin{figure}[htp]
\centering
    \begin{subfigure}{1\textwidth}
        \centering
        \includegraphics[width=0.8\linewidth]{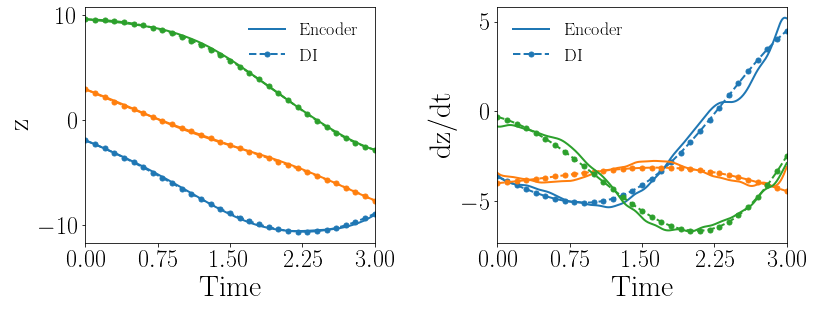}
        \caption{LaSDI}
    \end{subfigure}
    \begin{subfigure}{1\textwidth}
        \centering
        \includegraphics[width=0.8\linewidth]{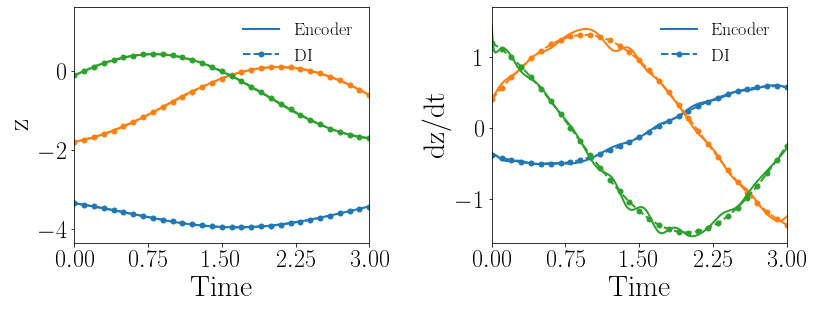}
        \caption{gLaSDI}
    \end{subfigure}
    \begin{subfigure}{0.495\textwidth}
        \centering
        \includegraphics[width=1\linewidth]{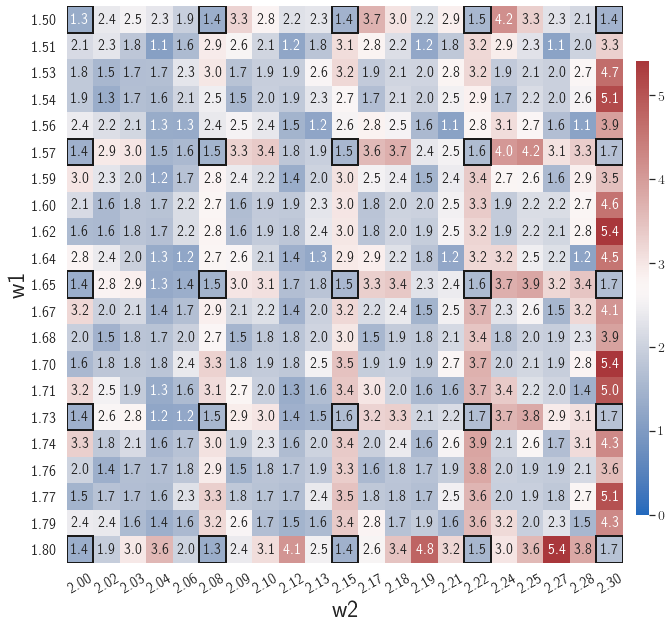}
        \caption{LaSDI}
    \end{subfigure}
    \begin{subfigure}{0.495\textwidth}
        \centering
        \includegraphics[width=1\linewidth]{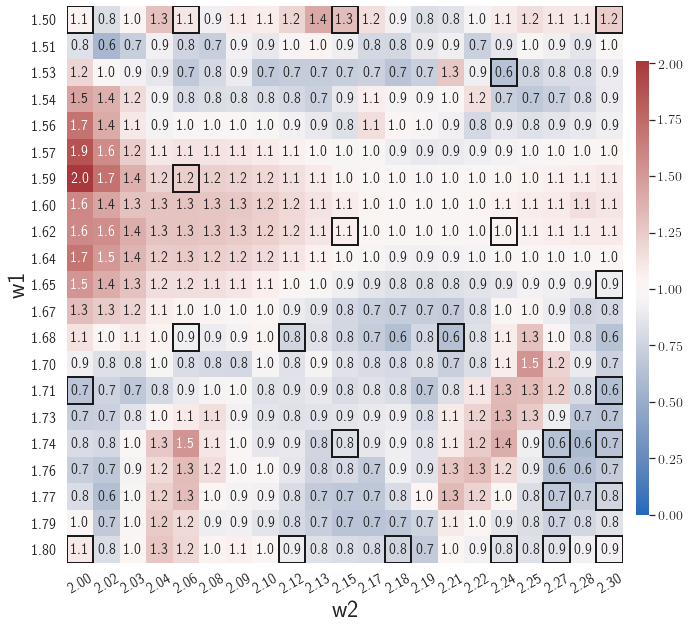}
        \caption{gLaSDI}
    \end{subfigure}
\caption{Comparison between LaSDI and gLaSDI with the same architecture of autoencoder (9,216-100-3) and dynamics identification models (\textbf{linear}) for the time dependent radial advection problem. The latent dynamics predicted by the trained encoder and the trained dynamics identification model from (a) LaSDI and (b) gLaSDI. The maximum relative errors in the parameter space $\mathcal{D}^h$ from (c) LaSDI with $k=4$ and (d) gLaSDI with $k=4$ for \RB{k-NN} convex interpolation during evaluation. The number on each box denotes the maximum relative error of the associated parameter case. The black square boxes indicate the locations of the sampled training points. The \RB{k-NN} parameter $k=1$ is used for greedy sampling procedure during training of gLaSDI.}\label{fig.advection_case1}
\end{figure}

Fig. \ref{fig.advection_case1}(a-b) show the latent-space dynamics predicted by the trained encoder and the DI model from LaSDI and gLaSDI, respectively. The gLaSDI achieves simpler time derivative latent-space dynamics than LaSDI, with a better agreement between the encoder and the DI prediction. 

Fig. \ref{fig.advection_case1}(c-d) show the maximum relative error of LaSDI and gLaSDI predictions in the prescribed parameter space, respectively. 
The gLaSDI achieves higher prediction accuracy than the LaSDI with the maximum relative error of 2.0$\%$ in the whole parameter space, compared to 5.4$\%$ of LaSDI. It is observed that gLaSDI tends to have denser sampling in the regime with higher parameter values, concentrating at the bottom-right corner of the parameter space, which implies that more vibrant change in dynamics is present for higher parameter values, requiring more local DI models there.
% Compared with the high-fidelity simulation based on MFEM \cite{anderson2021mfem}, the gLaSDI model achieves 200$\times$ speed-up.
\RB{Compared with the high-fidelity simulation (MFEM \cite{anderson2021mfem}) that has an around $2\%$ maximum relative error with respect to the high-fidelity data used for gLaSDI training, the gLaSDI model achieves 121$\times$ speed-up.}

% ------------------------------------------------------------------------
\subsubsection{Case 2: Effects of the size of parameter space}
In the second test, we want to see how the size of parameter space affect the model performances of LaSDI and gLaSDI. A larger parameter space $\mathcal{D}^h$ is considered and constituted by the parameters of the initial condition, including the $w_1 \in [1.5, 2.0]$ and $w_2 \in [2.0, 2.5]$, each with 21 evenly distributed discrete points in the respective parameter range.
Other settings remain the same as those used in the previous example, including the number of training points set 25.

\begin{figure}[htp]
\centering
    \begin{subfigure}{1\textwidth}
        \centering
        \includegraphics[width=0.8\linewidth]{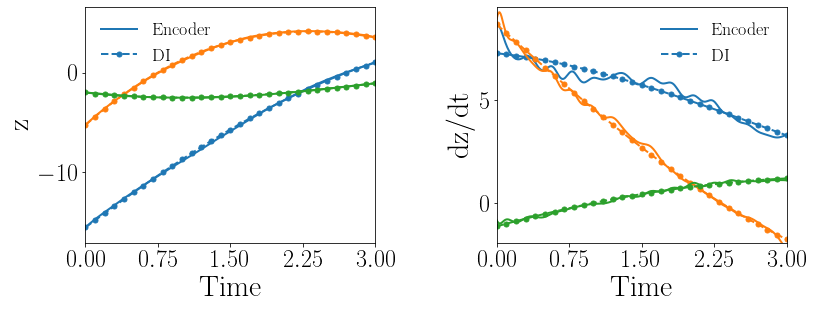}
        \caption{LaSDI}
    \end{subfigure}
    \begin{subfigure}{1\textwidth}
        \centering
        \includegraphics[width=0.8\linewidth]{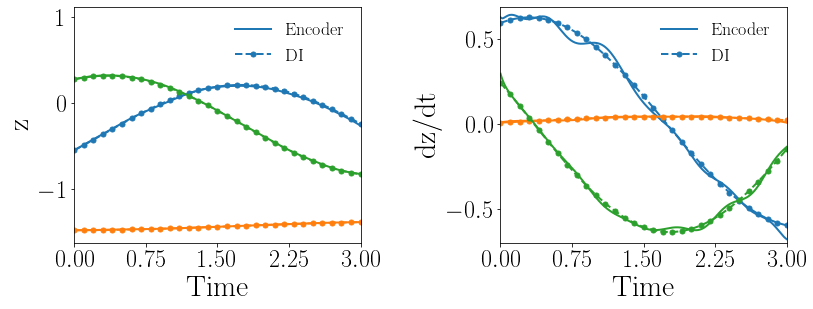}
        \caption{gLaSDI}
    \end{subfigure}
    \begin{subfigure}{0.495\textwidth}
        \centering
        \includegraphics[width=1\linewidth]{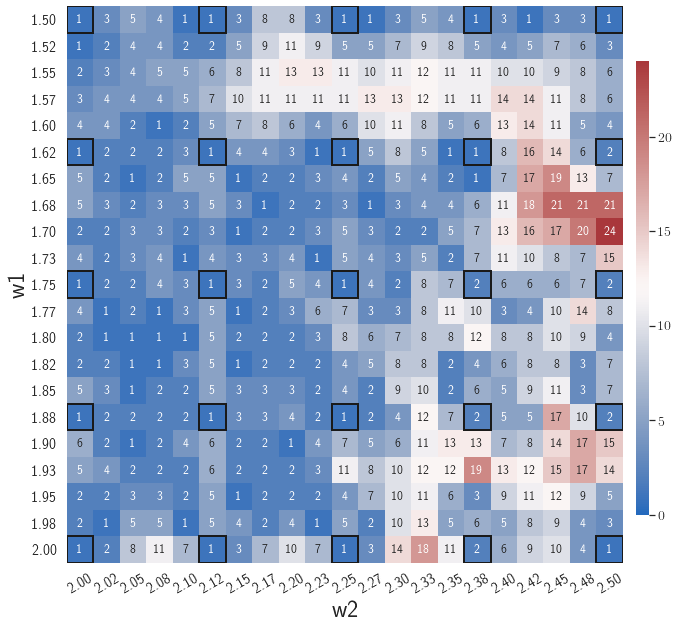}
        \caption{LaSDI}
    \end{subfigure}
    \begin{subfigure}{0.495\textwidth}
        \centering
        \includegraphics[width=1\linewidth]{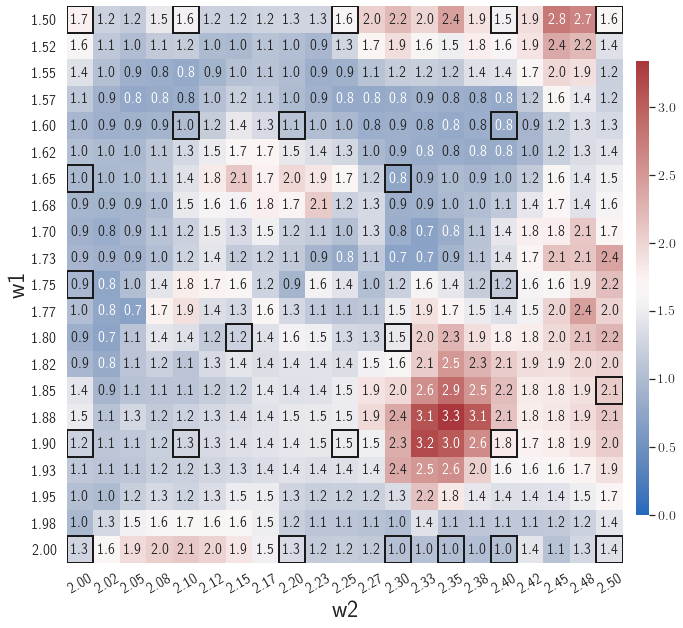}
        \caption{gLaSDI}
    \end{subfigure}
\caption{Comparison between LaSDI and gLaSDI with the same architecture of autoencoder (9,216-100-3) and dynamics identification models (\textbf{linear}) for the time dependent radial advection problem. The latent dynamics predicted by the trained encoder and the trained dynamics identification model from (a) LaSDI and (b) gLaSDI. The maximum relative errors in the parameter space $\mathcal{D}^h$ from (c) LaSDI with $k=4$ and (d) gLaSDI with $k=4$ for \RB{k-NN} convex interpolation during evaluation. The number on each box denotes the maximum relative error of the associated parameter case. The black square boxes indicate the locations of the sampled training points. The \RB{k-NN} parameter $k=1$ is used for greedy sampling procedure during training of gLaSDI. }\label{fig.advection_case2}
\end{figure}

Fig. \ref{fig.advection_case2}(a-b) show the latent-space dynamics predicted by the trained encoder and the DI model from LaSDI and gLaSDI, respectively. It again shows that gLaSDI learns smoother latent-space dynamics than LaSDI, with a better agreement between the encoder and the DI predictions than LaSDI.

Fig. \ref{fig.advection_case2}(c-d) show the maximum relative error of LaSDI and gLaSDI predictions in the prescribed parameter space, respectively. 
Compared with the example with a smaller parameter space, as shown in the previous subsection, the maximum relative error achieved by gLaSDI in the whole parameter space increases from 2.0$\%$ to 3.3$\%$, while that achieved by LaSDI increases from 5.4$\%$ to 24$\%$. 
It shows that gLaSDI maintains high accuracy even when the parameter space is enlarged, while LaSDI's error increases significantly due to non-optimal sampling and the mismatch between the encoder and DI predictions.
It is interesting to note that changing the parameter space affects the distribution of gLaSDI sampling.

\section{Conclusions}\label{sec:conclusion}
In this study, we introduced a physics-informed greedy parametric latent-space dynamics identification (gLaSDI) framework for accurate, efficient, and robust data-driven computing of high-dimensional nonlinear dynamical systems. 
The proposed gLaSDI framework is composed of an autoencoder that performs nonlinear compression of high-dimensional data and discovers intrinsic latent representations as well as dynamics identification (DI) models that capture local latent-space dynamics.
The autoencoder and DI models are trained interactively and simultaneously, enabling identification of simple latent-space dynamics for improved accuracy and efficiency of data-driven computing. 
To maximize and accelerate the exploration of the parameter space, we introduce an adaptive greedy sampling algorithm integrated with a physics-informed residual-based error indicator and random-subset evaluation to search for the optimal training samples on the fly.
Moreover, an efficient $k$-nearest neighbor convex interpolation scheme is employed for model evaluation to exploit local latent-space dynamics captured by the local DI models.

To demonstrate the effectiveness of the proposed gLaSDI framework, it has been applied to model various nonlinear dynamical problems, including 1D Burgers' equations, 2D Burgers' equations, nonlinear heat conduction, and time-dependent radial advection.
It is observed that greedy sampling with a small $k$ for model evaluation results in a more conservative gLaSDI model at the cost of training efficiency, and that the model testing with a large $k$ enhances generalization performance of gLaSDI.
Compared with LaSDI that depends on predefined uniformly distributed training parameters, gLaSDI with adaptive and sparse sampling can intelligently identify the optimal training parameter points to achieve higher accuracy with less number of training points than LaSDI.
Owning to interactive and simultaneous training of the autoencoder and DI models, gLaSDI is able to capture simpler and smoother latent-space dynamics than LaSDI that has sequential and decoupled training of the autoencoder and DI models.
In the radial advection problem, it is also shown that gLaSDI remains highly accurate as the parameter space increases, whereas LaSDI's performances could deteriorate tremendously.
In the numerical examples, compared with the high-fidelity models, gLaSDI achieves \RB{17 to 2,658}$\times$ speed-up, with 1 to 5$\%$ maximum relative errors in the prescribed parameter space, which reveals the promising potential of applying gLaSDI to large-scale physical simulations.

The proposed gLaSDI framework is general and not restricted \RB{to specific use of autoencoders, latent dynamics learning algorithms, or interpolation schemes for exploitation of localized latent-space dynamics learned by DI models.}
Depending on applications, various linear or nonlinear data compression techniques other than autoencoders could be employed. 
Further, latent-space dynamics identification could be performed by other system identification techniques or operator learning algorithms. 

The autoencoder architecture can be optimized to maximize generalization performance by integrating automatic \RB{neural architecture search \cite{elsken2019neural}} into the proposed framework.
The parameterization in this study considers the parameters from the initial conditions and the governing equations \RA{(\ref{appendix:heat:pde_parameterization})} of the problems. 
The proposed framework can be easily extended to account for other parameterization types, such as material properties, which will be useful for inverse problems.
\RB{As the training of gLaSDI proceeds, the training efficiency could decrease due to the increase in the training data. 
A combination of \textit{pre-training} and \textit{re-training} could potentially enhance the training efficiency. One could first \textit{pre-train} the DI model attached to the new sampled training parameter point using only its data rather than the aggregated training data that includes all training parameter points. Then, the gLaSDI model, consisting of an autoencoder and all DI models, is \textit{re-trained} by the combined training data to fine-tune the trainable parameters. 
More efficient training strategies will be investigated in future studies.}

% ===============================================================================
%%%%%%%%%%%%%%%%%%%%%%%%%%%%%%%%%%%%%%%%%%%%%%%%%%%%
%%%     End of body of article     %%%
%%%%%%%%%%%%%%%%%%%%%%%%%%%%%%%%%%%%%%%%%%%%%%%%%%%%
\section*{Acknowledgements}
This work was performed at Lawrence Livermore National Laboratory and partially funded by two LDRDs (21-FS-042 and 21-SI-006). Youngsoo was also supported for this work by the CHaRMNET Mathematical Multifaceted Integrated Capability Center (MMICC). Lawrence Livermore National Laboratory is operated by Lawrence Livermore National Security, LLC, for the U.S. Department of Energy, National Nuclear Security Administration under Contract DE-AC52-07NA27344 and LLNL-JRNL-834220.

\appendix
\section{1D Burgers equation}\label{appendix:1d_burger}
\RB{\subsection*{Solution snapshots}}\label{appendix:1d_burger:snapshot}
\RB{The solution fields at several time steps of the parameter case $(a=0.7, w=0.9)$ and the solution fields at the last time step of 4 different parameter cases, i.e., $(a=0.7, w=0.9)$, $(a=0.7, w=1.1)$, $(a=0.9, w=0.9)$, and $(a=0.9, w=1.1)$ are shown in Fig. \ref{fig.1d_burger_snapshot}.
\begin{figure}[htp]
\centering
    \begin{subfigure}{0.495\textwidth}
        \centering
        \includegraphics[width=1\linewidth]{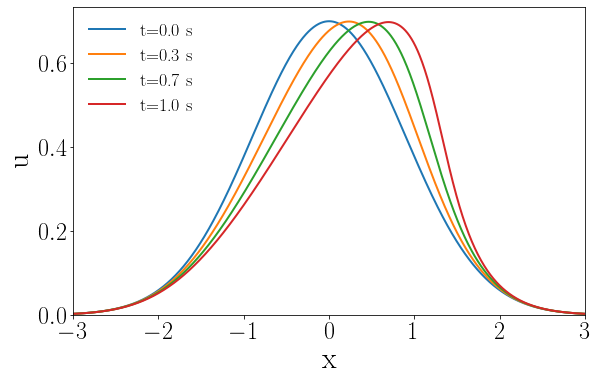}
        \caption{$a=0.7, w=0.9$}
    \end{subfigure}
    \begin{subfigure}{0.495\textwidth}
        \centering
        \includegraphics[width=1\linewidth]{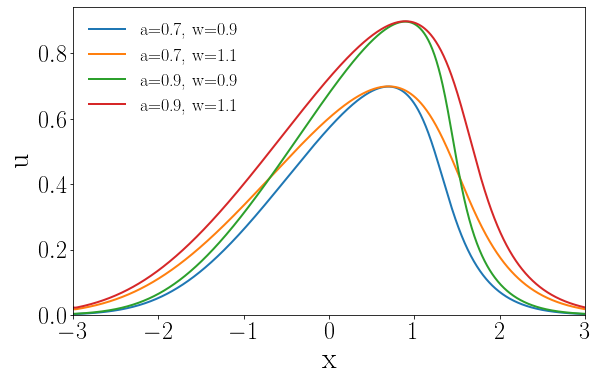}
        \caption{snapshots at the last time step}
    \end{subfigure}
\caption{\RB{Physical dynamics of the 1D Burgers problem: (a) solution fields at several time steps of the parameter case $(a=0.7, w=0.9)$; (b) solution fields at the last time step of 4 different parameter cases: $(a=0.7, w=0.9)$, $(a=0.7, w=1.1)$, $(a=0.9, w=0.9)$, and $(a=0.9, w=1.1)$.}}\label{fig.1d_burger_snapshot}
\end{figure}
}

\begin{figure}[htp]
    \centering
    \includegraphics[width=0.5\textwidth]{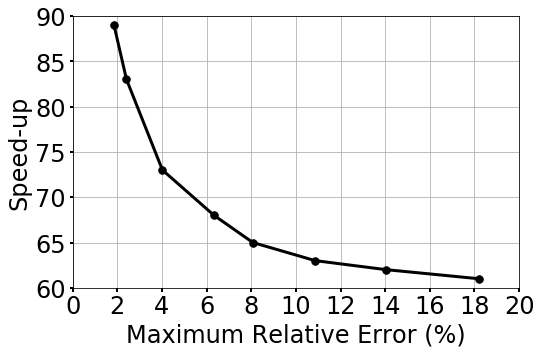}
    \caption{\RB{gLaSDI speed-up versus its maximum relative error (\%) in the parameter space for the 1D Burgers problem. The speed-up of gLaSDI is measured against the high-fidelity simulations that have similar maximum relative errors with respect to the high-fidelity data used for gLaSDI training.}}
    \label{fig.speedup_1dBurgers}
\end{figure}

\RB{\subsection*{Speed-up performance}}\label{appendix:1d_burger:speedup}
\RB{To further quantify the speed-up performance of gLaSDI, we have performed a series of tests using
the gLaSDI model and parameter space in case 2 (Section \ref{sec:result_1Dburger_case2}).
The gLaSDI model is trained until a prescribed target tolerance is reached by the maximum relative error estimated based on the residual error of the training parameter points (Eq. \eqref{eq.est_emax}).
The trained gLaSDI model is then evaluated in the parameter space prescribed for training and its maximum relative error is recorded.
Meanwhile, the high-fidelity simulations with similar maximum relative errors with respect to the high-fidelity data used for gLaSDI training are selected for speed-up comparison.
Fig. \ref{fig.speedup_1dBurgers} shows that the speed-up of gLaSDI increases as the maximum relative error decreases, which is expected as a lower error requires a higher dimension (resolution) of the high-fidelity solution that is much larger than the dimension of the latent space discovered by gLaSDI.}

% ===================================================================================================
\section{2D Burgers equation}\label{appendix:2d_burger}
\RB{\subsection*{Solution snapshots}\label{appendix:2d_burger:snapshot}
The solution fields of the first velocity component at several time steps of the parameter case $(a=0.7, w=0.9)$ are shown in Fig. \ref{fig.2d_burger_snapshot}.
\begin{figure}[htp]
\centering
    \begin{subfigure}{0.243\textwidth}
        \centering
        \includegraphics[width=0.8\linewidth]{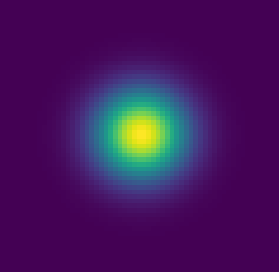}
        \caption{$t=0.0$ s}
    \end{subfigure}
    \begin{subfigure}{0.243\textwidth}
        \centering
        \includegraphics[width=0.8\linewidth]{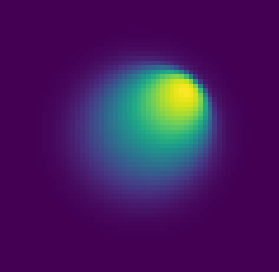}
        \caption{$t=0.3$ s}
    \end{subfigure}
    \begin{subfigure}{0.243\textwidth}
        \centering
        \includegraphics[width=0.8\linewidth]{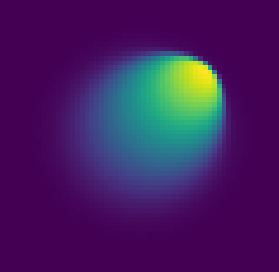}
        \caption{$t=0.7$ s}
    \end{subfigure}
    \begin{subfigure}{0.243\textwidth}
        \centering
        \includegraphics[width=0.96\linewidth]{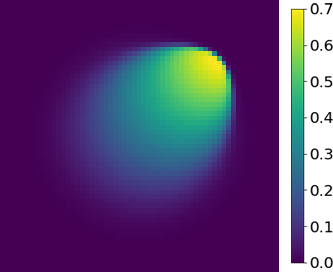}
        \caption{$t=1.0$ s}
    \end{subfigure}
\caption{\RB{Physical dynamics of the 2D Burgers problem. The solution fields of the first velocity component at several time steps of the parameter case $(a=0.7, w=0.9)$: (a) $t=0.0$ s, (b) $t=0.3$ s, (c) $t=0.7$ s, (d) $t=1.0$ s.}}\label{fig.2d_burger_snapshot}
\end{figure}
}

\RA{\subsection*{Effects of the latent dimension}\label{appendix:2d_burger:latent}}
\RA{The effects of the latent dimension on the ROM accuracy are further investigated. 
A series of tests are performed using the parameter space in case 1 (Section \ref{sec:result_2Dburger_case1}). 
The gLaSDI model has quadratic DI models and an autoencoder architecture of 7,200-100-$N_z$ with the latent dimension $N_z$ ranging from 2 to 7.
The gLaSDI is trained until the total number of sampled parameter points reaches 36.
For comparison, a LaSDI model with the same architecture of the autoencoder and DI models is trained using 36 predefined training parameter points uniformly distributed in a $6 \times 6$ grid in the parameter space.}

\begin{figure}[htp]
    \centering
    \includegraphics[width=0.5\textwidth]{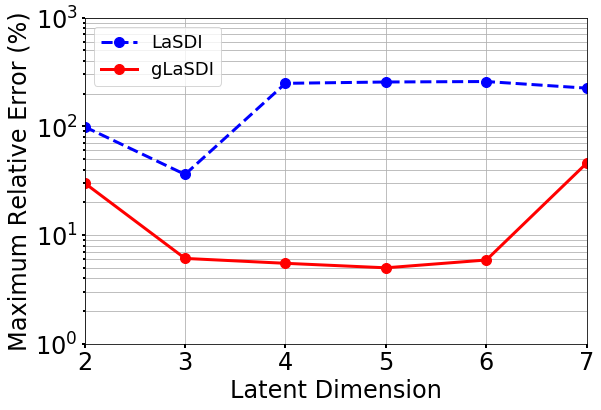}
    \caption{\RA{Maximum relative error (\%) of gLaSDI and LaSDI versus the latent dimension of gLaSDI and LaSDI.}}
    \label{fig.latent_dim_effects_2dBurgers}
\end{figure}

\RA{Fig. \ref{fig.latent_dim_effects_2dBurgers} shows that as the latent dimension increases from 2 to 3, the error LaSDI
decreases from around 99$\%$ to 36$\%$ and the error of gLaSDI decreases from around 30$\%$ to 6$\%$, which indicates that a latent dimension of 2 is insufficient for the autoencoder to capture all intrinsic features of the physical dynamics.
As the latent dimension further increases from 3 to 6, gLaSDI maintains a similar level of accuracy, around 5$\%$ error, while the error of LaSDI jumps significantly to around 250$\%$.
Due to strong nonlinearity and flexibility of the autoencoder, the complexity of the latent representation learned by the autoencoder increases with the latent dimension, posing more challenges for the subsequent DI training of LaSDI and therefore leading to large errors.
In contrast, the interactive autoencoder-DI training of gLaSDI provides additional constraints on the learned latent representation and contributes to a higher accuracy as well as more stable performance.
It is noticed that when the latent dimension is further increased to 7, the error of gLaSDI rises to around 46$\%$, which implies the constraints provided by the interactive training is insufficient to counteract the negative effects caused by the overly complex latent representations.
It shows that there exists a certain range of the latent dimension for optimal accuracy.
Note that a standard fully-connected autoencoder is applied in this study. 
The accuracy and robustness of gLaSDI could potentially be further improved by more advanced networks, such as convolutional autoencoders, and neural architecture search \cite{elsken2019neural}, which will be investigated in future studies.}

\RB{\subsection*{Speed-up performance}}\label{appendix:2d_burger:speedup}
\RB{The speed-up performance of gLaSDI is further investigated. A series of tests are performed using
the gLaSDI model and parameter space in case 2 (Section \ref{sec:result_2Dburger_case2}).
The gLaSDI model is trained until a prescribed target tolerance is reached by the maximum relative error estimated based on the residual error of the training parameter points (Eq. \eqref{eq.est_emax}).
The trained gLaSDI model is then evaluated in the parameter space prescribed for training and its maximum relative error is recorded.
Meanwhile, the high-fidelity simulations with similar maximum relative errors with respect to the high-fidelity data used for gLaSDI training are selected for speed-up comparison.
Similar to the observation in the speed-up analysis in \ref{appendix:1d_burger:speedup}, Fig. \ref{fig.speedup_2dBurgers} shows that the speed-up of gLaSDI increases as the maximum relative error decreases.}

\begin{figure}[htp]
    \centering
    \includegraphics[width=0.5\textwidth]{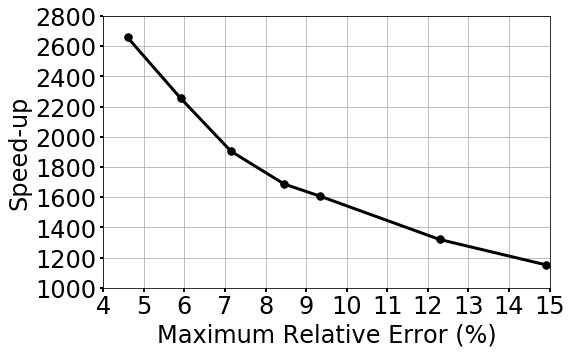}
    \caption{\RB{gLaSDI speed-up versus its maximum relative error (\%) in the parameter space for the 2D Burgers problem. The speed-up of gLaSDI is measured against the high-fidelity simulations that have similar maximum relative errors with respect to the high-fidelity data used for gLaSDI training.}}
    \label{fig.speedup_2dBurgers}
\end{figure}

% ===================================================================================================
\section{Nonlinear time-dependent heat conduction}\label{appendix:heat}
\RB{\subsection*{Solution snapshots}\label{appendix:heat:snapshot}
The solution fields at several time steps of the parameter case $(w=4, a=1)$ are shown in Fig. \ref{fig.heat_snapshot}.
\begin{figure}[htp]
\centering
    \begin{subfigure}{0.243\textwidth}
        \centering
        \includegraphics[width=0.8\linewidth]{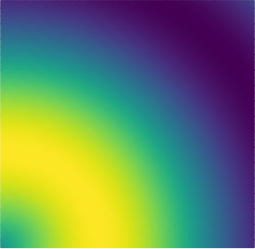}
        \caption{$t=0.0$ s}
    \end{subfigure}
    \begin{subfigure}{0.243\textwidth}
        \centering
        \includegraphics[width=0.8\linewidth]{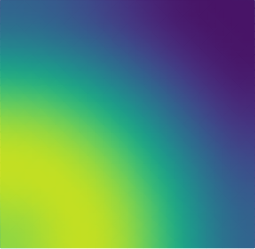}
        \caption{$t=0.03$ s}
    \end{subfigure}
    \begin{subfigure}{0.243\textwidth}
        \centering
        \includegraphics[width=0.8\linewidth]{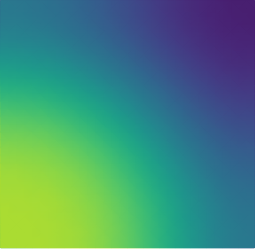}
        \caption{$t=0.07$ s}
    \end{subfigure}
    \begin{subfigure}{0.243\textwidth}
        \centering
        \includegraphics[width=0.96\linewidth]{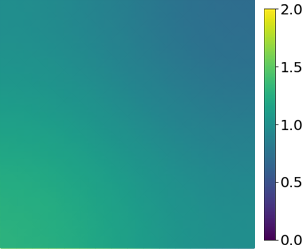}
        \caption{$t=0.3$ s}
    \end{subfigure}
\caption{\RB{Physical dynamics of the nonlinear heat conduction problem. The solution fields at several time steps of the parameter case $(w=4, a=1)$: (a) $t=0.0$ s, (b) $t=0.03$ s, (c) $t=0.07$ s, (d) $t=0.3$ s. The coefficients $\kappa = 0.5$ and $\alpha = 0.01$ are adopted in Eq. \eqref{eq.heat_conduction}.}}\label{fig.heat_snapshot}
\end{figure}
}

\RA{\subsection*{Parameterization of PDE}\label{appendix:heat:pde_parameterization}}
\RA{The effectiveness of the proposed gLaSDI framework on the parameterization of PDEs is investigated, where the coefficients $\kappa \in [0.3,0.7]$ and $\alpha \in [0.01,0.05]$ in Eq. \eqref{eq.heat_conduction} are considered to be the parameters that constitute the parameter space $\mathcal{D}^h$, each with 21 evenly distributed discrete points in the respective parameter range.
The parameters $w=4$ and $a=1$ are adopted in the initial condition.
The autoencoder with an architecture of 1,089-100-3 and linear DI models are considered. 
The gLaSDI training is performed until the total number of sampled parameter points reaches 25. 
}

\begin{figure}[htp]
\centering
    \begin{subfigure}{1\textwidth}
        \centering
        \includegraphics[width=0.8\linewidth]{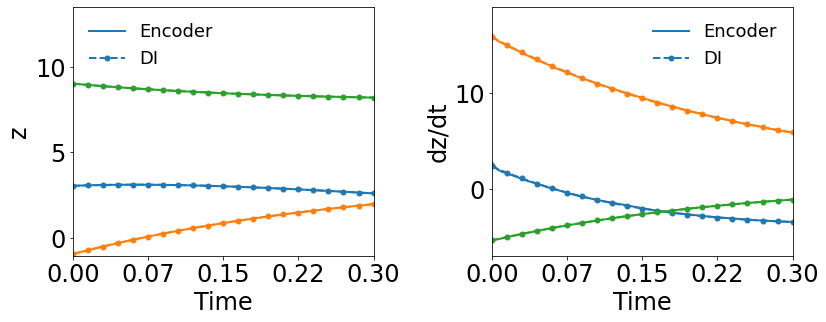}
        \caption{gLaSDI}
    \end{subfigure}
    \begin{subfigure}{0.6\textwidth}
        \centering
        \includegraphics[width=1\linewidth]{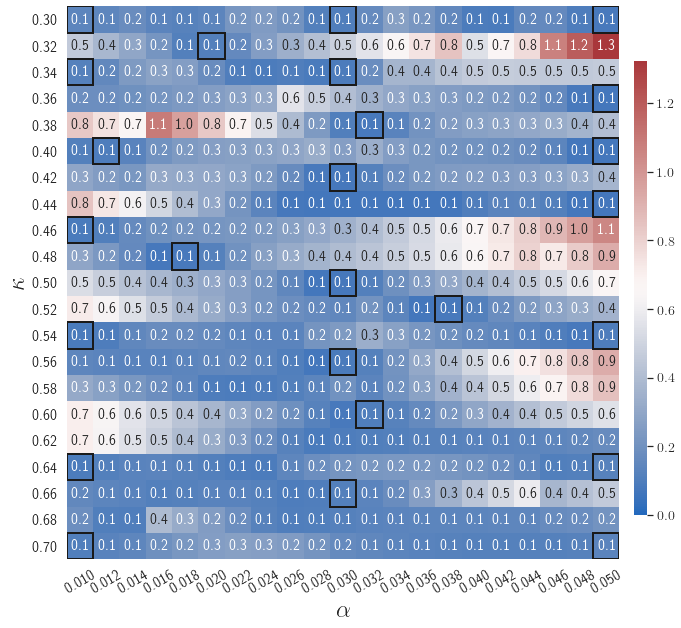}
        \caption{gLaSDI}
    \end{subfigure}
\caption{\RA{Results of gLaSDI with an autoencoder of 7,200-100-3 and linear dynamics identification (DI) models for the heat conduction problem with PDE parameterization: (a) latent dynamics predicted by the trained encoder and the trained DI model; (b) maximum relative errors in the parameter space with $k=3$ for k-NN convex interpolation during evaluation. The number on each box denotes the maximum relative error of the associated parameter case. The black square boxes indicate the location of the sampled training points. The k-NN parameter $k=1$ is used for greedy sampling procedure for the training of gLaSDI.}}\label{fig.heat_pde_parameterization}
\end{figure}

\RA{Fig. \ref{fig.heat_pde_parameterization}(a) shows that gLaSDI discovers simple latent-space dynamics with a good agreement between the predictions by the trained encoder and the DI model.
Fig. \ref{fig.heat_pde_parameterization}(b) shows that gLaSDI achieves a maximum relative error of 1.3$\%$ in the prescribed parameter space, which demonstrates the effectiveness of the proposed gLaSDI framework for reduced-order modeling with parameterization of PDEs.
}

% ===================================================================================================
\section{Time-dependent radial advection}\label{appendix:advection}
\RB{\subsection*{Solution snapshots}}\label{appendix:advection:snapshot}
\RB{The solution fields at several time steps of the parameter case $(w_1=1.5, w_2=2.0)$ are shown in Fig. \ref{fig.advection_snapshot}.
\begin{figure}[htp]
\centering
    \begin{subfigure}{0.243\textwidth}
        \centering
        \includegraphics[width=0.77\linewidth]{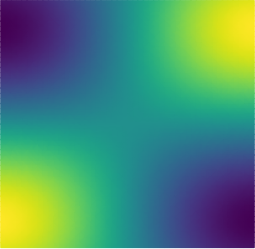}
        \caption{$t=0.0$ s}
    \end{subfigure}
    \begin{subfigure}{0.243\textwidth}
        \centering
        \includegraphics[width=0.77\linewidth]{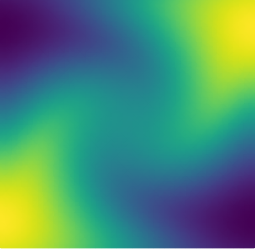}
        \caption{$t=0.7$ s}
    \end{subfigure}
    \begin{subfigure}{0.243\textwidth}
        \centering
        \includegraphics[width=0.77\linewidth]{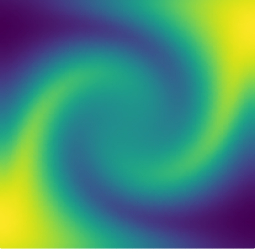}
        \caption{$t=1.7$ s}
    \end{subfigure}
    \begin{subfigure}{0.243\textwidth}
        \centering
        \includegraphics[width=1\linewidth]{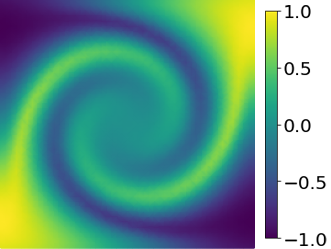}
        \caption{$t=3.0$ s}
    \end{subfigure}
\caption{\RB{Physical dynamics of the radial advection problem. The solution fields at several time steps of the parameter case $(w_1=1.5, w_2=2.0)$: (a) $t=0.0$ s, (b) $t=0.7$ s, (c) $t=1.7$ s, (d) $t=3.0$ s.}}\label{fig.advection_snapshot}
\end{figure}
}
% ===============================================================================

% \bibliography{reference}
\printbibliography

\end{document}